\documentclass[sn-mathphys-num]{sn-jnl}

\usepackage{graphicx}%
\usepackage{multirow}%
\usepackage{amsmath,amssymb,amsfonts}%
\usepackage{amsthm}%
\usepackage{mathrsfs}%
\usepackage[title]{appendix}%
\usepackage{xcolor}%
\usepackage{textcomp}%
\usepackage{manyfoot}%
\usepackage{booktabs}%
\usepackage{algorithm}%
\usepackage{algorithmicx}%
\usepackage{algpseudocode}%
\usepackage{listings}%
\usepackage{placeins}

\newcommand{\sig}{\ensuremath{\sigma}}
\newcommand{\sxx}{\ensuremath{\sigma_{xx}}}
\newcommand{\syy}{\ensuremath{\sigma_{yy}}}
\newcommand{\sxy}{\ensuremath{\sigma_{xy}}}
\newcommand{\syx}{\ensuremath{\sigma_{yx}}}

\begin{document}

\title[BISM review 2025]{
Tissue stress measurements with Bayesian Inversion Stress Microscopy}


\author[1]{\fnm{Lucas} \sur{Anger}}\email{lucas.anger@ijm.fr}
\equalcont{These authors contributed equally to this work.}
\author[1]{\fnm{Andreas} \sur{Schoenit}}\email{andreas.schonit@ijm.fr}
\equalcont{These authors contributed equally to this work.}
\author[1]{\fnm{Fanny} \sur{Wodrascka}}\email{fanny.wodrascka@ijm.fr}
\author[1,2]{\fnm{Carine} \sur{Rossé}}\email{carine.rosse@curie.fr}
\author[1]{\fnm{Ren\'e-Marc} \sur{M\`ege}}\email{rene-marc.mege@ijm.fr}
\author[1,3,4]{\fnm{Benoit} \sur{Ladoux}}\email{benoit.ladoux@ijm.fr}
\author[5]{\fnm{Philippe} \sur{Marcq}}\email{philippe.marcq@espci.fr}

\affil[1]{\orgdiv{Institut Jacques Monod}, \orgname{CNRS, Universit\'e Paris Cit\'e}, \orgaddress{\city{Paris}, \postcode{F-75013}, \country{France}}}

\affil[2]{Institut Curie, Paris Université Sciences et Lettres, Sorbonne Université, CNRS, UMR144, 75005 Paris, France}

\affil[3]{\orgdiv{Department of Physics}, \orgname{Friedrich-Alexander Universit\"at Erlangen-N\"urnberg}, \orgaddress{\city{Erlangen}, \country{Germany}}}

\affil[4]{\orgdiv{Max-Planck-Zentrum f\"ur Physik und Medizin}, \orgaddress{\city{Erlangen}, \country{Germany}}}

\affil[5]{\orgdiv{Physique et M\'ecanique des Milieux H\'et\'erog\`enes}, \orgname{CNRS, ESPCI Paris, Universit\'e PSL, Sorbonne Universit\'e, Universit\'e Paris Cit\'e}, \orgaddress{\city{Paris}, \postcode{F-75005}, \country{France}}}


\abstract{Cells within biological tissue are constantly subjected to dynamic mechanical forces. Measuring the internal stress of tissues has proven crucial for our understanding of the role of mechanical forces in fundamental biological processes like morphogenesis, collective migration, cell division or cell elimination and death. Previously, we have introduced Bayesian Inversion Stress Microscopy (BISM), which is relying on measuring cell-generated traction forces \textit{in vitro} and has proven particularly useful to measure absolute stresses in confined cell monolayers. We further demonstrate the applicability and robustness of BISM across various experimental settings with different boundary conditions, ranging from confined tissues of arbitrary shape to monolayers composed of different cell types. Importantly, BISM does not require assumptions on cell rheology. Therefore, it can be applied to complex heterogeneous tissues consisting of different cell types, as long as they can be grown on a flat substrate. Finally, we compare BISM to other common stress measurement techniques using a coherent experimental setup, followed by a discussion on its limitations and further perspectives.}

\keywords{Biophysics, Intercellular forces, Tissue mechanics, Stress measurements, Bayesian inference, Mechanobiology}


\maketitle

\begin{figure}[!t]
\includegraphics[width=1\linewidth]{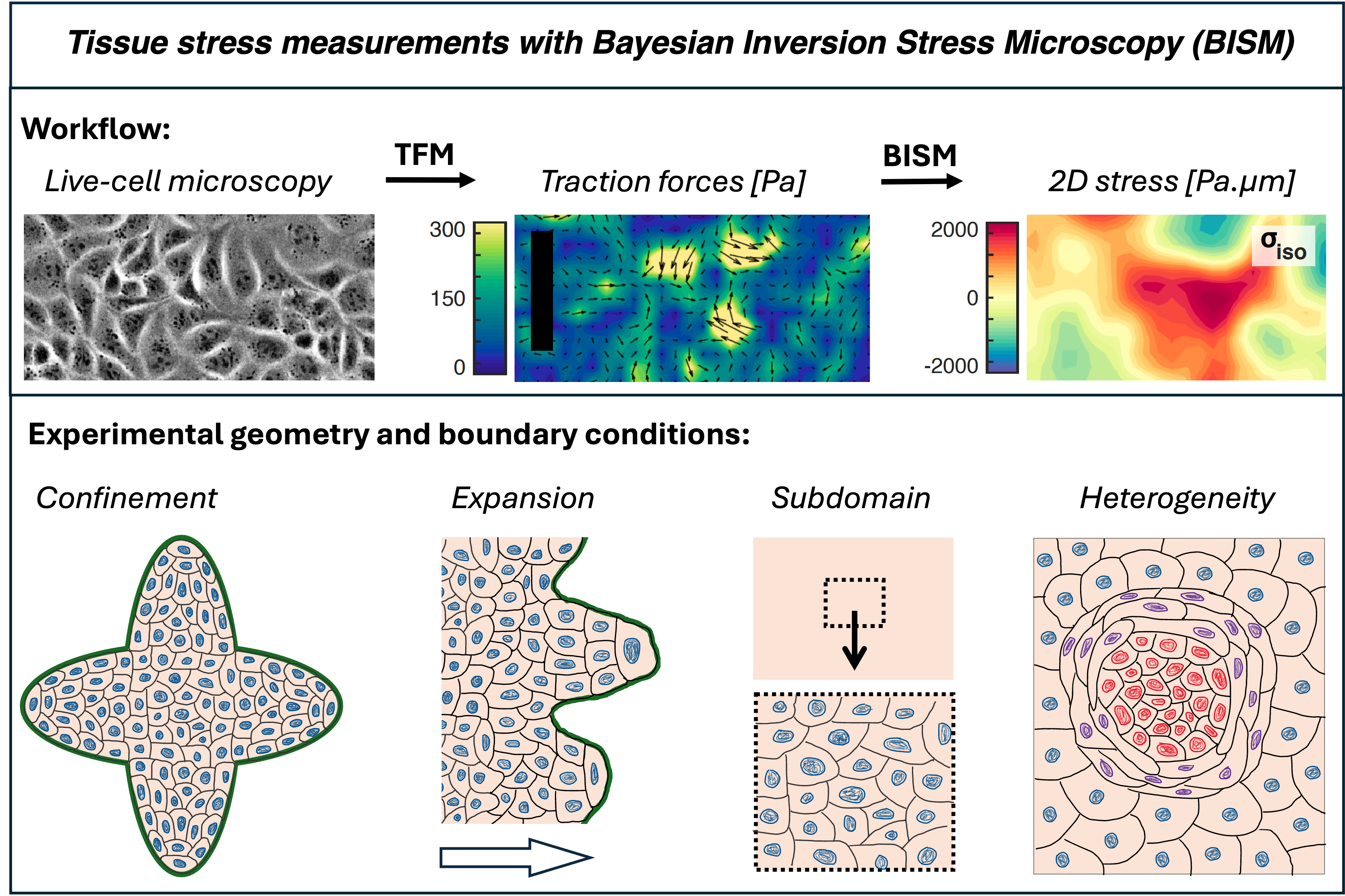}
\caption{
\textbf{Graphical abstract}}
\label{fig:graphical_abstract}
\end{figure}

\section{Introduction}
\label{itd}

Cells within biological tissues can be subjected to strong and dynamic cell-generated mechanical forces during development, in homeostasis and as a result of disease like cancer \cite{Fiore_2025, Heisenberg_2013, Lecuit_2007, Hayward_2021, Mao_2024, Ladoux_2017}. These forces can lead to deformations ranging from the sub-cellular to the tissue scale, which changes the local cellular mechanical environment. Extensively investigating cellular forces and stresses within \textit{in vivo} and \textit{in vitro} tissues has been crucial for and has greatly improved our understanding of multiple fundamental biological processes \cite{Ladoux_2017, Roca-Cusachs_2017, Polacheck_2016, Sugimura_2016}. Different methods allow probing tissue stresses on different levels, but the most straight-forward one is perhaps laser ablation, as it allows direct readout of tissue mechanical state by monitoring the ablated area \cite{rauzi_2008, Bonnet2012, Etournay_2015}. The recoil direction indicates tension or compression, and its velocity allows relative comparison of stress magnitudes. While laser ablation can be applied to any tissue with high spatial resolution, it can only provide a “snapshot” of the mechanical state.  Another approach is force inference based on cell shape \cite{Ishihara2012, Chiou_2012, Kong2019}. Assuming force balance, this approach requires the segmentation of cell-cell junctions, which in turn allows measuring relative tensions between cell junctions and relative pressures between cells (see \cite{Roffay_2021,Borges2024} for reviews). 

An absolute measurement of tissue stress requires a direct force measurement. 
Deformable objects like droplets \cite{Campas_2014}, pillars \cite{Roure2005} or bioprinted structures \cite{Maniou_2024} can be introduced into living tissues. Since the rheology of the force sensor is known, quantifying its deformation allows to estimate the stress exerted by the tissue, which in turn provides a tissue-scale readout of stresses in 3D and \textit{in vivo} tissue. In 2D, traction forces applied by the cells on their (elastic) substratum can be computed by measuring the displacement of fluorescent beads it contains \cite{Dembo1999, Butler_2002, Lekka_2021}. Tambe \textit{et al.} \cite{Tambe2011, Tambe2013} introduced Monolayer Stress Microscopy (MSM), which allows estimating the stress field within an elastic monolayer of cells, based on recording traction forces. With this non-invasive method, stresses can be continously monitored with high spatial and temporal resolution, \textit{e.g.} during collective migration \cite{Tambe2011} or cell division \cite{uroz_2018}. MSM assumes that the tissue behaves as a linear elastic body. 
However, the tissue of interest, when elastic, can consist of multiple cell types with different stiffnesses \cite{Luo_2016}, \textit{e.g.} organoids containing stem cells and differentiated cells \cite{Heo_2016}, normal and cancerous cells \cite{Alibert_2017} or heterogenous sub-populations within tumors \cite{Schoenit2025}. Stiffness fluctuations can even occur within a homogeneous tissue due to cell division \cite{Fischer-Friedrich_2016} or cell death \cite{Van_der_Meeren_2020}. Further, tissue rheology often deviates from the case treated by MSM (linear elasticity), whether due to nonlinearities observed for large enough deformations \cite{Latorre2018,Perros2024}, to viscoelastic behaviour \cite{Khalilgharibi2019,Bonfanti2020,Tlili2020}, or to both \cite{Mary2022}. Remarkably, living tissues may even dynamically tune their rheology through jamming \cite{Mongera2018} and fluidization \cite{Petridou2019,Menin2023} transitions.

Therefore, we have previously introduced Bayesian Inversion Stress Microscopy (BISM, Nier \textit{et al.} \cite{Nier2016}). BISM computes tissue stress based on tractions, but without making assumptions on tissue rheology. It provides an absolute measure of the stress. It has been extensively used to investigate internal tissue stresses in a variety of biological processes like collective migration \cite{Peyret2019, Ollech2020, Pricoupenko2024}, cell extrusion \cite{Saw2017, Balasubramaniam_2025}, hole formation \cite{Sonam2023}, mixed cultures \cite{Gauquelin2024} and cell competition \cite{Moitrier_2019, Schoenit2025}. BISM measured stresses robustly in these studies, for a panel of  geometries associated with different experimental boundary conditions.

Here, we provide the open-access code of BISM for \texttt{matlab}, accompanied by guidance on the concept, experimental setup and code implementation which can be found on the associated github pages. We discuss tested boundary conditions, with an emphasis on experimental validations and illustrations. These experimental systems range from confined and unconfined tissues over migrating monolayers to heterogenous tissues consisting of different cell types including \textit{ex vivo} tissue. We then use a consistent experimental condition to compare BISM to other tension measurement approaches including laser ablation, pressure inference based on cell shapes and MSM, revealing quantitative, cell-scale differences but a qualitative agreement in robustly reporting the mechanical state of cells.

\section{Bayesian inversion stress microscopy} 
\label{concept}

Throughout this work, we use the following terminology and notations.
Within a continuum approach, the two-dimensional mechanical stress
tensor $\sigma$ has three independent components $\sxx$, $\syy$ and $\sxy=\syx$
using cartesian coordinates.
It is customary to distinguish isotropic and deviatoric contributions
to the stress tensor:
$  \sigma_{ij} = \frac{1}{2}\sigma_{kk} \delta_{ij}  +
    \left( \sigma_{ij} - \frac{1}{2}\sigma_{kk} \delta_{ij}  \right)$
where $i,j \in \{x,y\}$, $\delta_{ij}$ is the Kronecker symbol,
and summation over repeated indices is implied. 
In matrix form, we find:
\begin{equation}
  \label{eq:iso:dev}
  \begin{pmatrix}
    \sxx & \sxy \\
    \sxy & \syy 
  \end{pmatrix}
  =
  \sigma_{\mathrm{iso}}
  \begin{pmatrix}
    1 & 0 \\
    0 & 1
  \end{pmatrix}
  +
    \begin{pmatrix}
    \sigma_{\mathrm{d}}  & \sxy \\
    \sxy & -\sigma_{\mathrm{d}} 
    \end{pmatrix}
\end{equation}
where  $\sigma_{\mathrm{iso}} = \frac{\sigma_{xx} + \sigma_{yy}}{2}$,
$\sigma_{xy}$ and $\sigma_{\mathrm{d}} = \frac{\sigma_{xx} - \sigma_{yy}}{2}$
respectively denote the isotropic, shear and deviatoric stresses. Of interest are also the maximum and minimum principal stresses, which provide a stress orientation
(see Table~\ref{tab1} for an overview of the different BISM outputs).
All those stresses may play important roles in biological systems, but since the trace of the stress $\sigma_{kk}$ is a tensor invariant,
isotropic stress has been the primary focus in many studies so far
\cite{gomez2020measuring}.
This quantity offers a straightforward mechanical interpretation:
positive isotropic stress corresponds to a state of tension,
while negative isotropic stress indicates compression.

Due to the relative smallness of the typical space and velocity scales
involved, inertial effects can be neglected when writing the force
balance equation of a cell monolayer:
\begin{equation}
  \label{eq:forcebalance}
\mathrm{div} \, \sig= \vec{t} \,,  
\end{equation}
where $\sigma$ and $\vec{t}$ respectively denote the intercellular tissue
stress field and the traction force field exerted by cells
on the underlying substrate. The stress field $\sigma$
generally contains both passive (\emph{e.g.} the viscous stress)
and active (\emph{e.g.} ATP-dependent) components.
Calculating stresses given the traction forces then amounts to solving
an underdetermined inversion problem, since the symmetric
tensor stress field has three independent components
($\sxx$, $\syy$ and $\sxy$) while the vector
traction force field has only two components ($t_x$ and $t_y$).
Bayesian inversion is a classical and reliable technique to solve
such problems \cite{kaipio2006statistical}, which is perhaps better
known in the Earth sciences \cite{tarantola2005inverse}
than in the Life sciences -- note however its early use in the context
of traction force microscopy to infer single cell traction forces from
substrate displacement data \cite{Dembo1996}.

\begin{figure}[!t]
\includegraphics[width=1\linewidth]{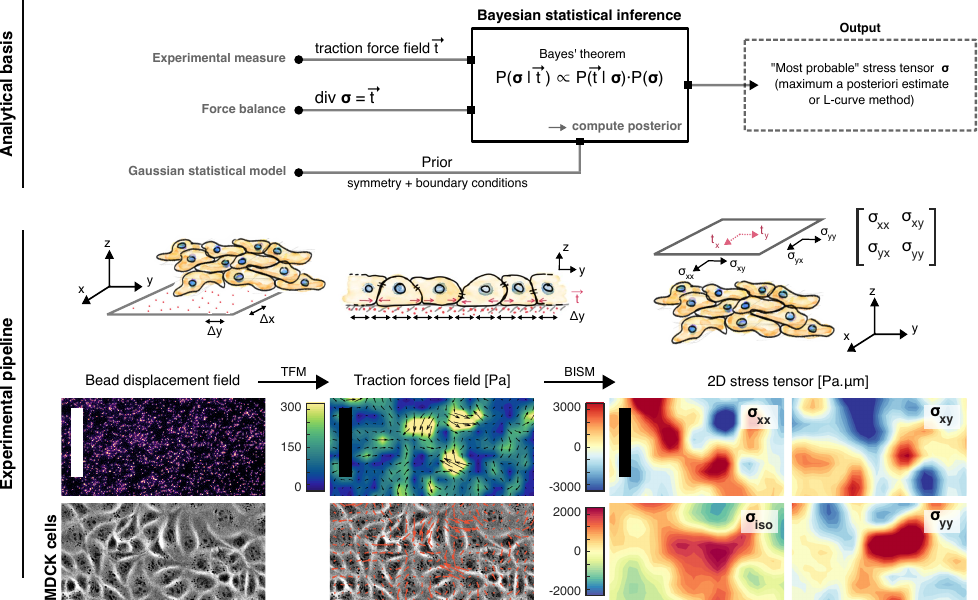}
\caption{
\textbf{Bayesian Inference Stress Microscopy workflow:} 
    Experimental and analytical pipeline of BISM. Top: Requirements and assumptions of the BISM algorithm. Force balance and experimental data from TFM allow computation of the most probable corresponding stress tensor. Middle: Cartoon illustrating the main steps of BISM implementation. The first step is monitoring the substrate displacement to calculate 2D traction force components. Then, BISM is applied to obtain the 2D stress tensor which includes different components. Bottom: Representative example of BISM analysis of a MDCK monolayer on a 15 kPa soft (PDMS) substrate. [Scale bar: $100 \,\mu$m.]
}
\label{fig:BISM}
\end{figure}

In \cite{Nier2016}, we implemented the Bayesian inversion of
Eq.~(\ref{eq:forcebalance}) using a Gaussian statistical model.
We refer the reader to the original article for technical details.
Both the algorithm and the experimental pipeline are described
graphically  in Fig.~\ref{fig:BISM}.
The likelihood function $P\left(\vec{t} \, | \sigma\right)$ imposes 
the force balance relationship Eq.~(\ref{eq:forcebalance})
up to an additive Gaussian noise of finite variance.
Within this Bayesian framework, the stress field is estimated
as the most likely value of the posterior distribution function 
$P\left(\sigma |\vec{t} \, \right)$,
equal in the Gaussian case to the posterior mean value,
which is readily computed from traction force data.
The prior distribution function $P\left(\sigma \right)$
acts as a regularization term,
and incorporates both the symmetry constraint imposed on the stress
field ($\sxy = \syx$) and, when appropriate, the stress-free
boundary conditions typical of confined systems
($\sigma_{ij} \, n_j=0$, where $\vec{n}$ denotes the vector normal to the
domain boundary). 
A dimensionless regularization parameter $\Lambda$,
characterizing the respective weight of the prior versus
that of the likelihood function, has to be determined from data.
In order to do so, it is possible to use either the Maximum a
Posteriori method \cite{kaipio2006statistical} or the
L-curve approach \cite{Hansen1992}, both yielding similar
quantitative results \cite{Nier2016} (typically $\Lambda \simeq 10^{-5}-10^{-6}$).
For simplicity, we set $\Lambda = 10^{-6}$ for all results presented
in this work.

Using numerical simulations of model tissues exchanging momentum
with an elastic substrate \cite{Nier2016}, we confirmed the accuracy
and robustness of Bayesian Inversion Stress Microscopy by varying
both the rheology of the tissue computational model and the ingredients of the
inference statistical model. While Bayesian inversion is designed to
analyse one snapshot at a time, we later implemented
a Kalman filter \cite{kaipio2006statistical} for dynamical stress
estimation from traction force \emph{movies} \cite{Nier2018}.
On note, the Kalman inversion algorithm yields stress estimates in good
agreement with those inferred by Bayesian inversion,
\emph{without} requiring the use of a prior distribution function.

In practice, both the temporal and spatial resolutions of
the output stress field inferred by BISM are identical
to those of the input traction force field, yielding high
temporal and spatial resolution.
Given a traction force field in kPa, the (2D) stress field will be 
in kPa.$\mu$m (see Eq.~(\ref{eq:forcebalance})). A mean (3D) stress field
in kPa is obtained by dividing the 2D stress values
by the cell monolayer height (see Sec.~\ref{methods:height}). 
When the error bar on the traction force measurement is known, 
the algorithm also calculates error bars on the inferred stress
using standard relationships valid for Gaussian probability
distribution functions. When analysing experimental data, a
coefficient of determination $R^2_t$ quantifies the similarity
between the ground truth $\vec{t}^{\mathrm{true}}$ and the
inferred traction force vector
$\vec{t}^{\mathrm{inf}} = \mathrm{div} \, \sig^{\mathrm{inf}}$.

\begin{figure}[!t]
    \begin{center}
      \hspace*{0.2cm}\textbf{a}\hspace*{0.4cm}
      \includegraphics[width=0.21\linewidth]{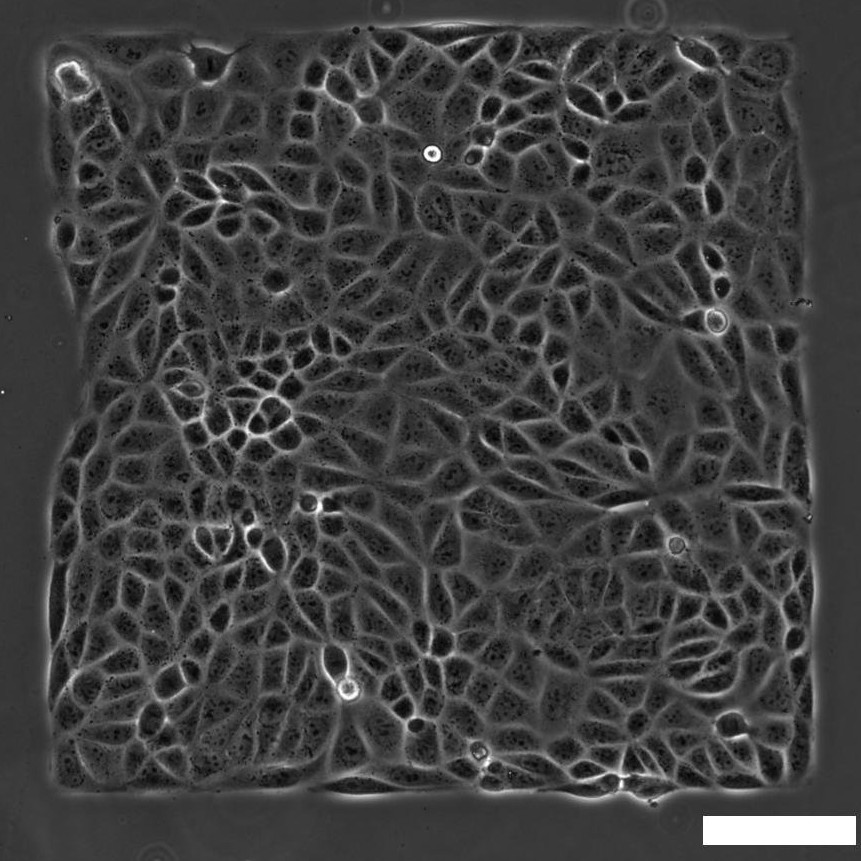}
   \hspace*{0.15cm} \textbf{b}
  \includegraphics[width=0.27\linewidth]{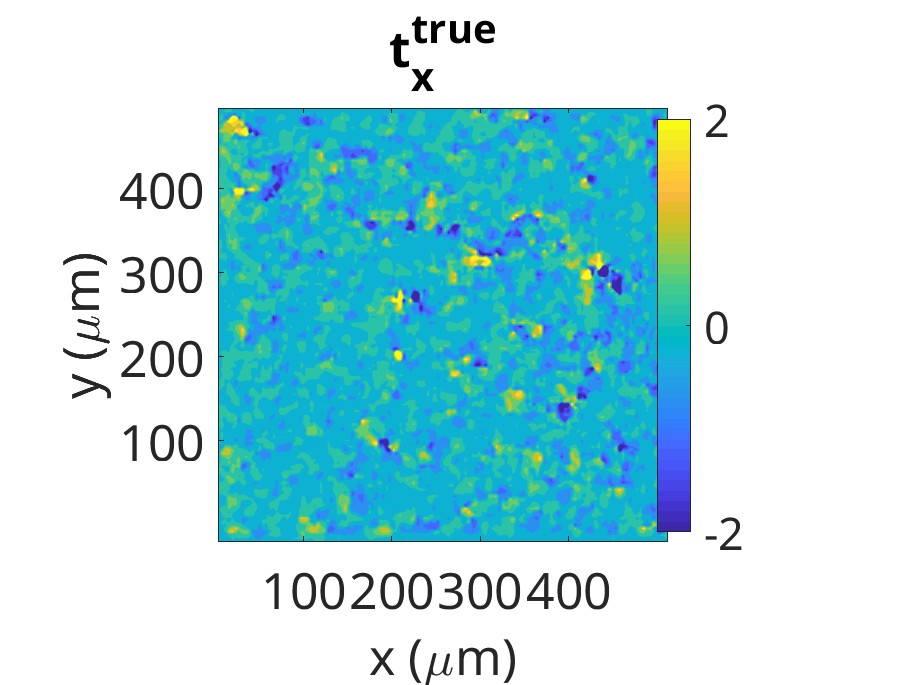} 
  \textbf{c}
  \includegraphics[width=0.27\linewidth]{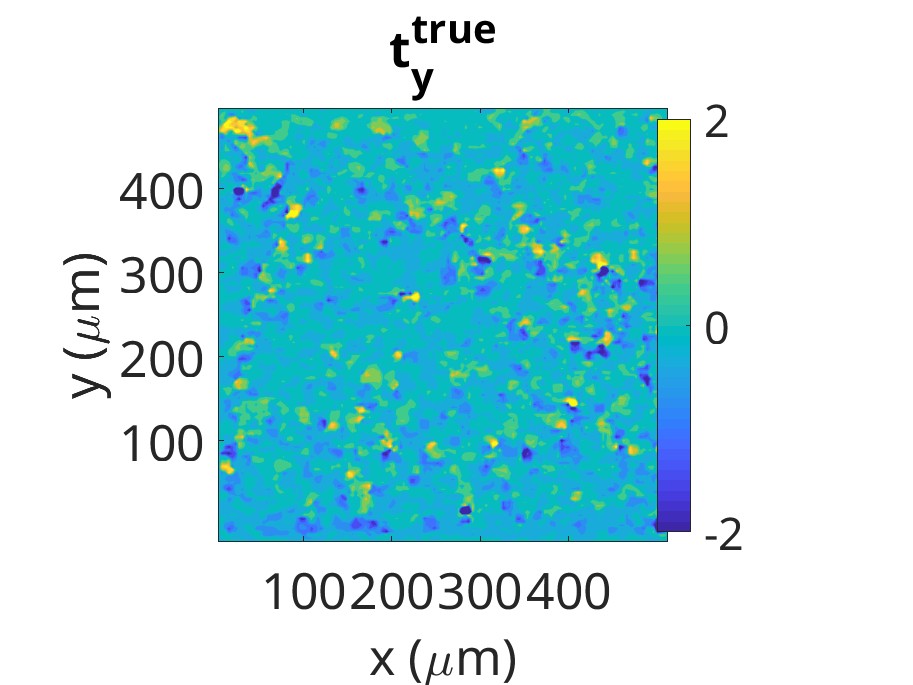}\\
  \medskip
  \textbf{d}
  \includegraphics[width=0.27\linewidth]{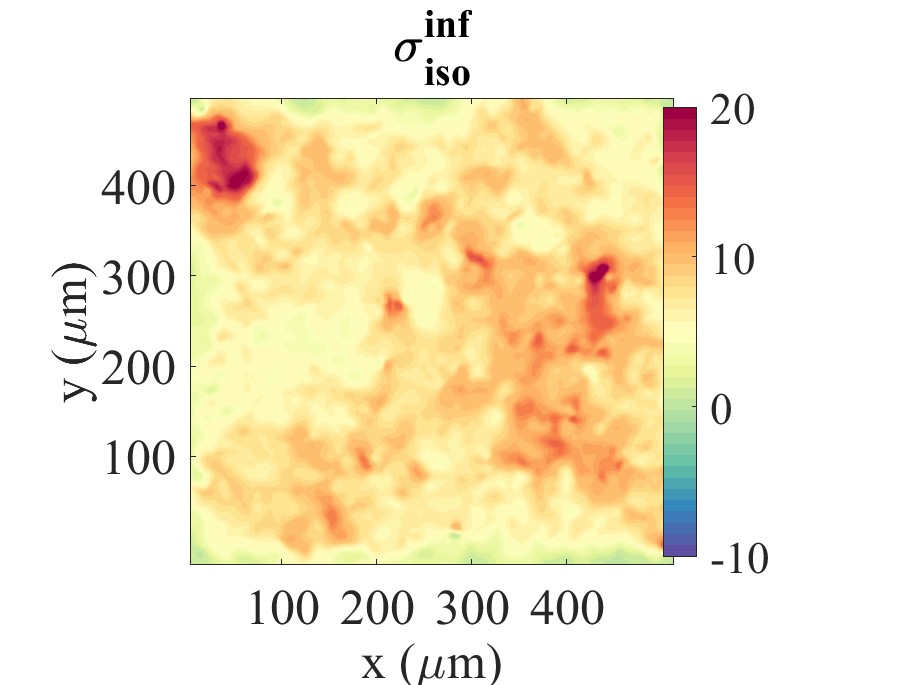} 
  \textbf{e}
  \includegraphics[width=0.27\linewidth]{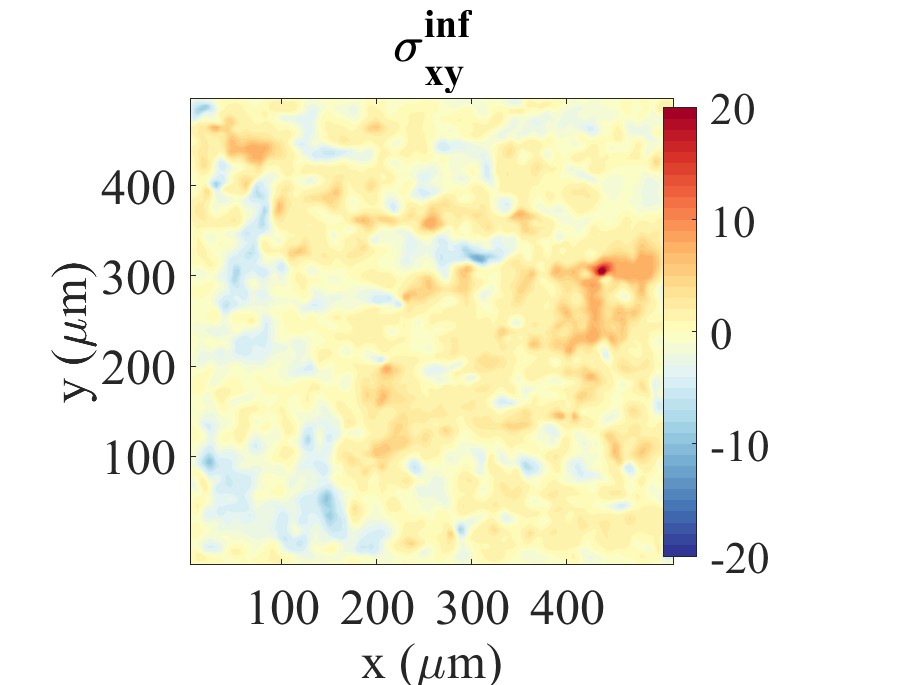}
  \textbf{f}
  \includegraphics[width=0.27\linewidth]{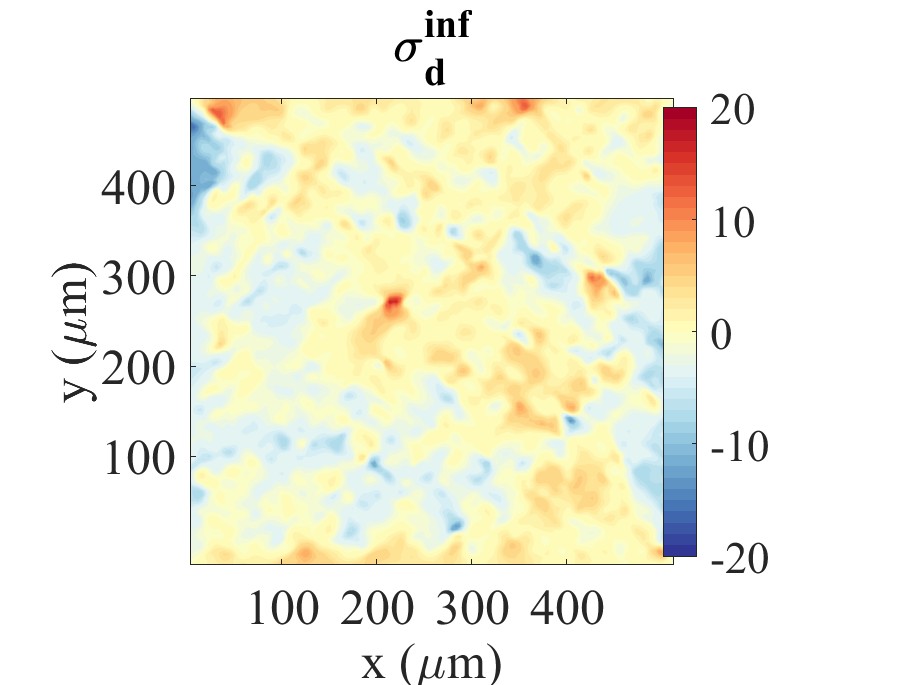} \\
  \smallskip
    \textbf{g}
  \includegraphics[width=0.27\linewidth]{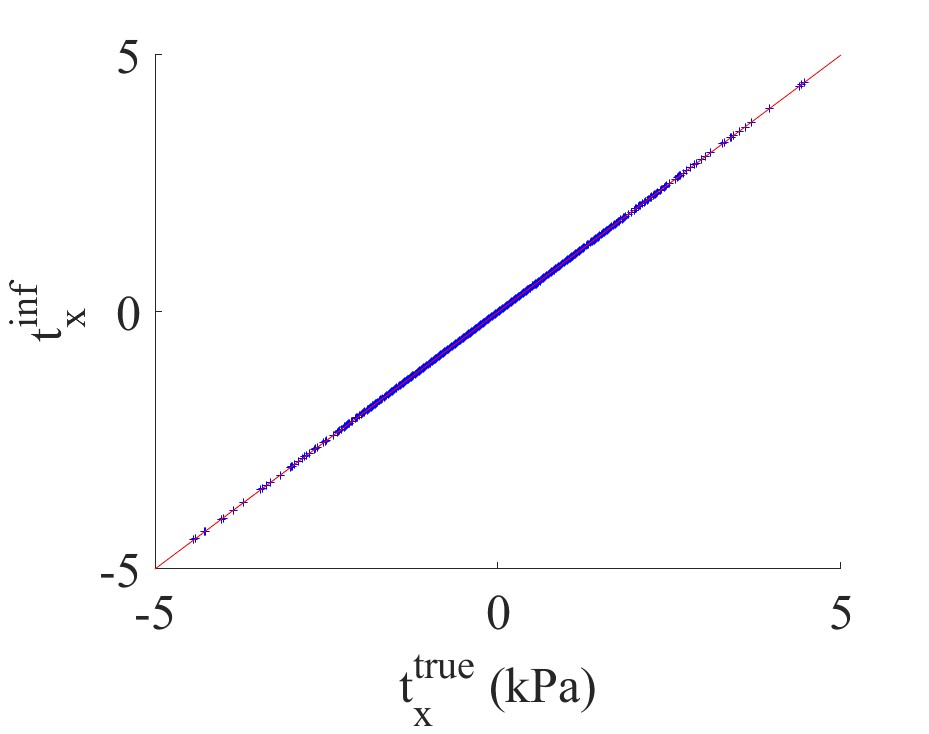}
  \textbf{h}
  \includegraphics[width=0.27\linewidth]{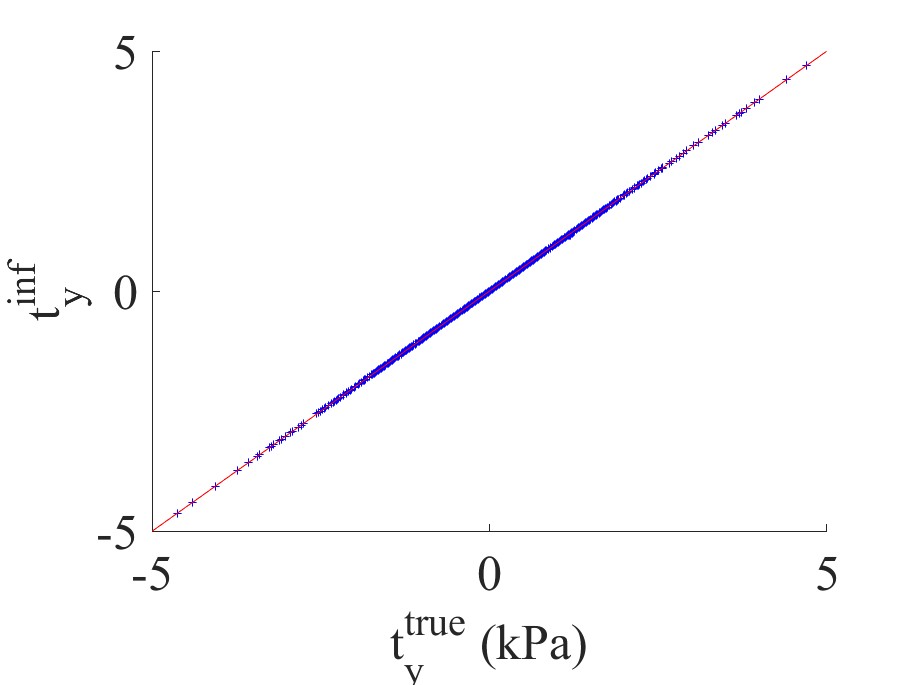} \\
    \smallskip
  \textbf{i}
  \includegraphics[width=0.27\linewidth]{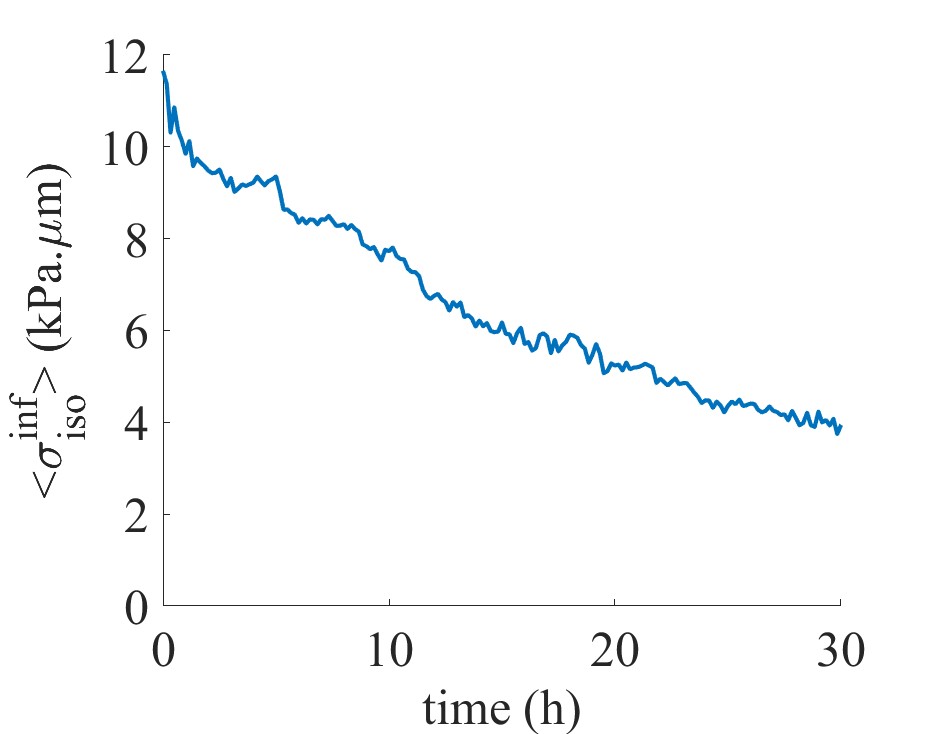} 
  \textbf{j}
  \includegraphics[width=0.27\linewidth]{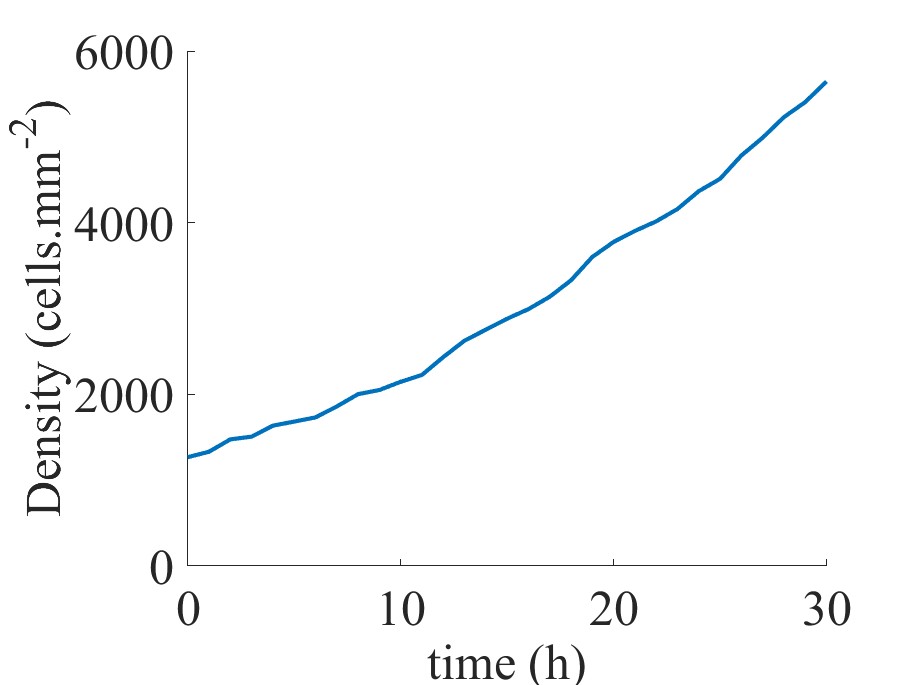}
  \textbf{k}
  \includegraphics[width=0.27\linewidth]{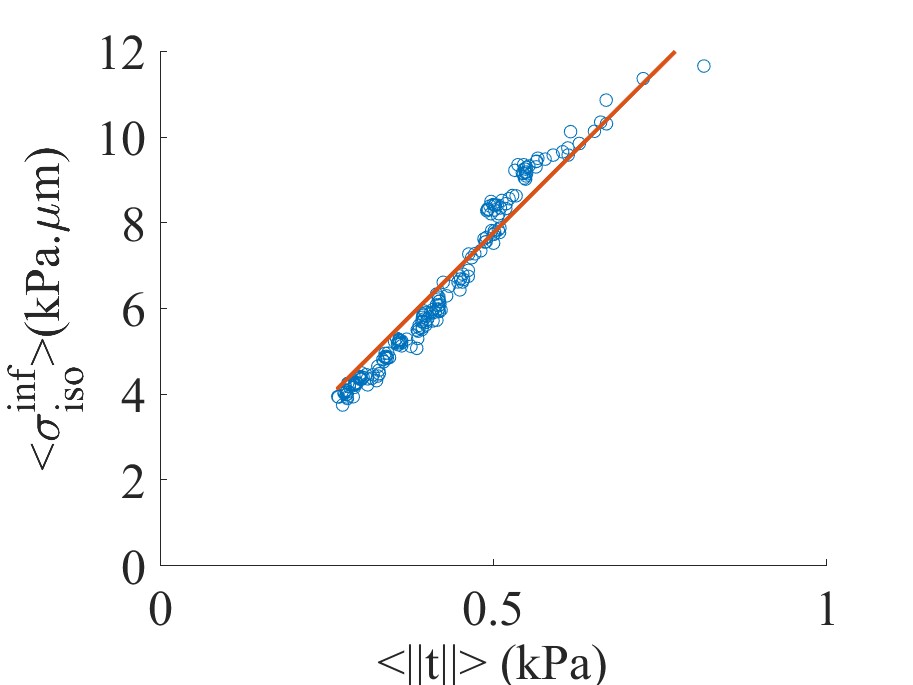} \\
  \end{center}
\caption{
\textbf{Stress inference in a confined system:} 
MDCK cell monolayer in a square domain of lateral extension
$L = 500 \,\mu$m \cite{Peyret2019}.
\textbf{a)} Phase contrast image of the confining domain.
\textbf{b, c)} Color maps of the
components of the traction force field
$t_x^{\mathrm{true}}$, $t_y^{\mathrm{true}}$ measured by TFM (in kPa).
\textbf{d, e, f)} Color maps of the inferred
isotropic stress $\sigma_{\mathrm{iso}}^{\mathrm{inf}}$ and
deviatoric stress components $\sigma_{\mathrm{xy}}^{\mathrm{inf}}$,
$\sigma_{\mathrm{d}}^{\mathrm{inf}}$ (in kPa.$\mu$m).
Here the zero-stress condition is imposed on the four boundaries.
\textbf{g, h)} Comparison of measured
($t^{\mathrm{true}}_x$, $t^{\mathrm{true}}_y)$
and inferred ($t^{\mathrm{inf}}_x$, $t^{\mathrm{inf}}_y$) values of
components of the traction force vectors (blue crosses),
computed from the inferred stress field
$\vec{t}^{\mathrm{inf}} = \mathrm{div} \, \sig^{\mathrm{inf}}$.
The bisectrix $y = x$ is plotted as a red line for comparison.
The coefficient of determination is $R^2_t = 1.0$.
We find excellent agreement between stress averages computed 
in the cell domain and the true values obtained from moments 
of the traction force field, thus confirming that BISM provides
an absolute measurement of stresses in confined domains: compare 
$  \langle \sigma_{\mathrm{iso}}^{\mathrm{true}} \rangle = 7.77 $ kPa
and
$  \langle \sigma_{\mathrm{iso}}^{\mathrm{inf}} \rangle = 7.76$ kPa;
then $  \langle \sigma_{\mathrm{xy}}^{\mathrm{true}} \rangle = 565 $ Pa
and
$  \langle \sigma_{\mathrm{xy}}^{\mathrm{inf}} \rangle = 568$ Pa;
and $  \langle \sigma_{\mathrm{d}}^{\mathrm{true}} \rangle = -420$ Pa
and
$  \langle \sigma_{\mathrm{d}}^{\mathrm{inf}} \rangle = -416$ Pa.
\textbf{i, j, k)} Time evolution of the monolayer.
We plot \textbf{i)} the mean tension; \textbf{j)} the cell density against time.
Panel \textbf{k)} shows the mean tissue tension as a function
of mean traction force norm (blue circles), for the same duration ($30$ h).
The red line is the linear regression of data points,
with a slope of $15.5\,\mu$m, of the order of a typical cell diameter.
[Scale bar: $100 \,\mu$m (a).]
}
\label{fig:confined:MDCK}
\end{figure}

For confined cell monolayers, and given the stress-free boundary condition 
$\sigma_{ij} \, n_j=0$, integrating by parts Eq.~(\ref{eq:forcebalance})
allows to compute the spatially-averaged value of each stress component
directly from traction force data. In particular, we use the identities 
\begin{align*}
 \langle \sigma_{\mathrm{iso}} \rangle & =
- \frac{1}{2} \left( \langle t_x \, x \rangle +
\langle t_y \, y \rangle \right) \,, \\
\langle \sigma_{\mathrm{xy}}  \rangle & =
 - \frac{1}{2} \left(  \langle t_x \, y \rangle
  + \langle t_y \, x \rangle \right)  \,,  \\
  \langle \sigma_{\mathrm{d}} \rangle & =
- \frac{1}{2} \left( \langle t_x \, x \rangle -
\langle t_y \, y \rangle \right) ,
\end{align*}
where brackets $\langle \, \rangle$ denote spatial averaging
over the cell domain (see \cite{Nier2016} for further details). 
Importantly, these classical calculations \cite{landau1975elasticity} allowed us 
to verify on experimental data that, unlike other relative stress inference methods 
which rely on cell shape or cell displacement, BISM indeed measures the absolute values 
of tissue stresses.

We include in Fig.~\ref{fig:confined:MDCK} an example of stress inference 
performed by BISM on the traction force field exerted by a Madin-Darby 
canine kidney (MDCK) cell monolayer, confined in a square box
using micropatterning.
Similar data obtained in a HaCaT skin cell monolayer is shown in
Fig.~\ref{fig:confined:HaCaT} (see \cite{Peyret2019}
for a detailed description of the experiments).
In addition to a snapshot of traction force and stress patterns,
Fig.~\ref{fig:confined:MDCK} displays several mechanical features 
revealed during  the time evolution of the cell monolayer over
a duration of $30$ h. 

Compressive tissue stresses are often assumed to be generated by a cell density increase. Here, measuring absolute stress values, we show that this is not necessarily the case. As cell density increases about fourfold within a constant area,
the mean tension decreases by a factor of three, but remains positive (Fig.~\ref{fig:confined:MDCK}ijk). Thus, the
monolayer does not reach a compressive state during this period. Although cells are confined in 2D, they can actively change their shape in 3D to accommodate the density increase, thereby remaining under tension.
Finally, we observe a linear relationship between
mean tension and mean traction force amplitude over the duration
of the experiment, as would perhaps be expected from the (linear) relationship
\eqref{eq:forcebalance}. At the beginning of the experiment,
larger traction force norms lead to a higher level
of tissue tension, while cell crowding later reduces cell activity,
leading to both lower mean traction force amplitude and lower mean tension. Furthermore, this relationship explains differences in tension magnitudes between different cell types, which is observed when comparing cells derived from kidney in Fig.~\ref{fig:confined:MDCK} and cells derived from skin in Fig.~\ref{fig:confined:HaCaT}. Differences in tension magnitude were also reported if traction force generation is altered through mutations \cite{Peyret2019, Gauquelin2024}. Thus, cell types which exert stronger traction forces generally generate higher tissue tension. 

Together, those measurements illustrate how BISM can be used to test common assumptions on tissue mechanics and to investigate general relationships between (external) traction forces and (internal) tissue stresses.

\section{Boundary conditions}
\label{BC}

As already noted in \cite{Nier2016}, solutions to the force balance
equation \eqref{eq:forcebalance} remain valid when adding
to the stress tensor arbitrary solutions of the homogeneous
equation $\mathrm{div} \,  \sigma = \vec{0}$.
BISM has originally been validated, both numerically and experimentally,
in the case of simple confined domains, such as ring or square
patterns \cite{Nier2016}, showing that enforcing stress-free
boundary conditions in the prior allows to lift this indeterminacy.
However, studying specific biological processes,
such as the influence of tissue boundary curvature or collective and
large-scale tissue dynamics, requires the use of more complex
geometries. This, in turn, necessitates modifications to
the experimental design and results in altered boundary conditions.
In this section, we examine the impact
of boundary conditions on stress inference. We show that BISM remains
reliable for measuring intercellular stress even when applied to
non-trivial geometries with modified boundary conditions. Furthermore,
we demonstrate that in many cases, relaxing constraints on the boundary
conditions is possible and still yields robust and consistent results.

\begin{figure}[!t]
  \begin{center}
      \hspace*{-0.1cm}
      \textbf{a.}
            \hspace*{0.5cm}
      \includegraphics[width=0.18\linewidth]{fig4a.jpg}
      \hspace*{0.5cm}
  \textbf{b.}
  \includegraphics[width=0.27\linewidth]{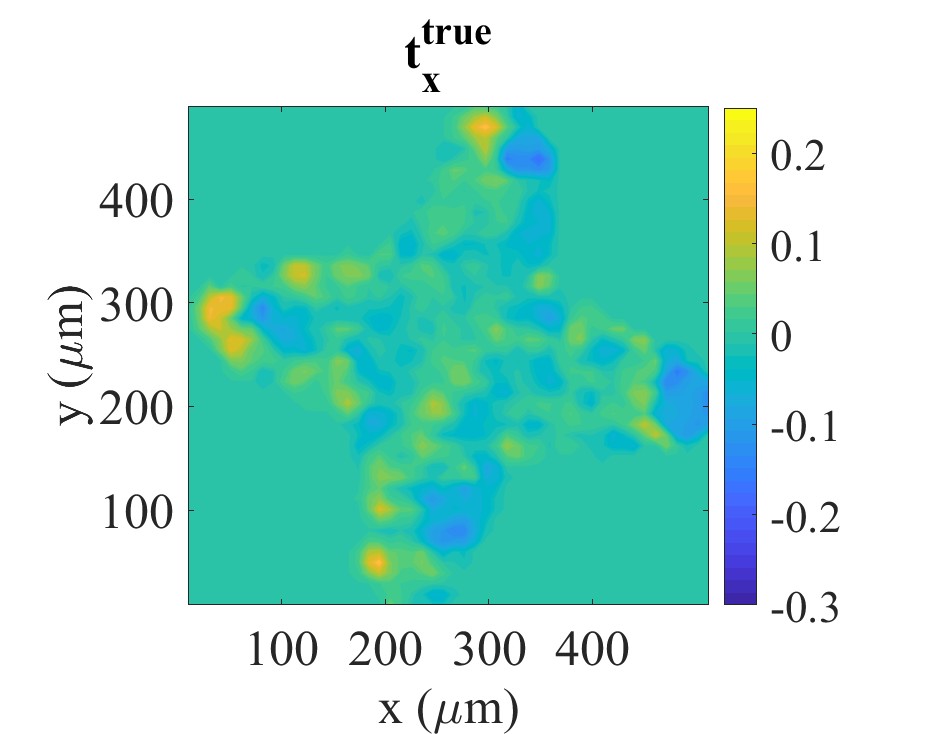} 
  \textbf{c.}
  \includegraphics[width=0.27\linewidth]{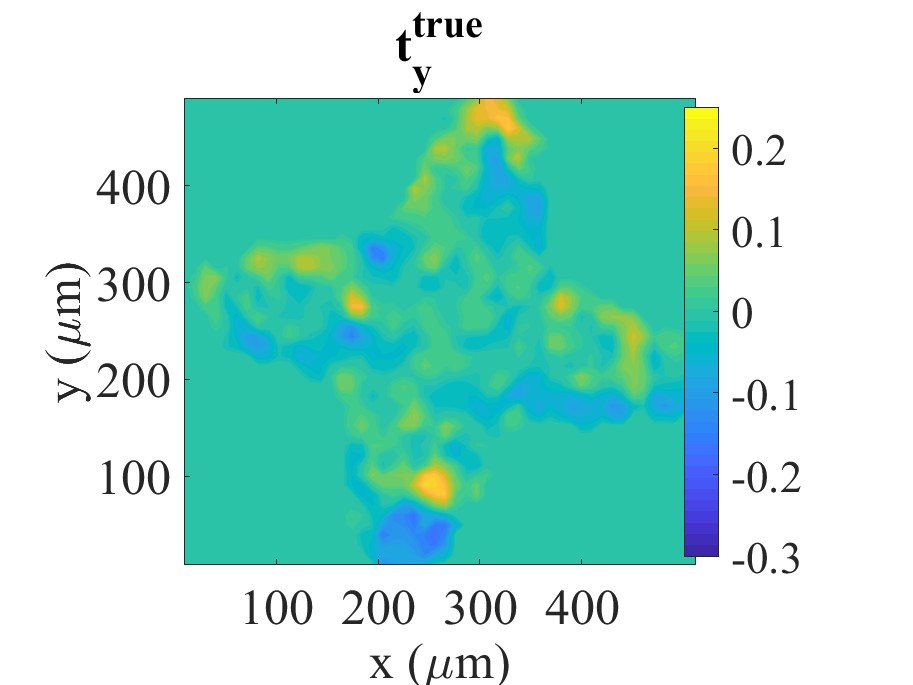}\\
  \textbf{d.}
  \includegraphics[width=0.27\linewidth]{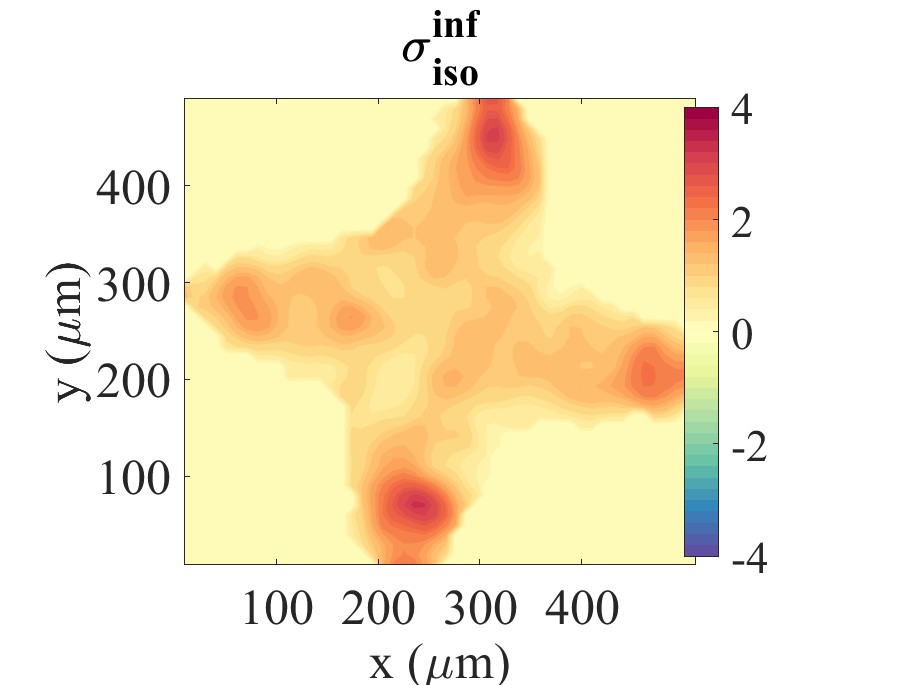} 
  \textbf{e.}
  \includegraphics[width=0.27\linewidth]{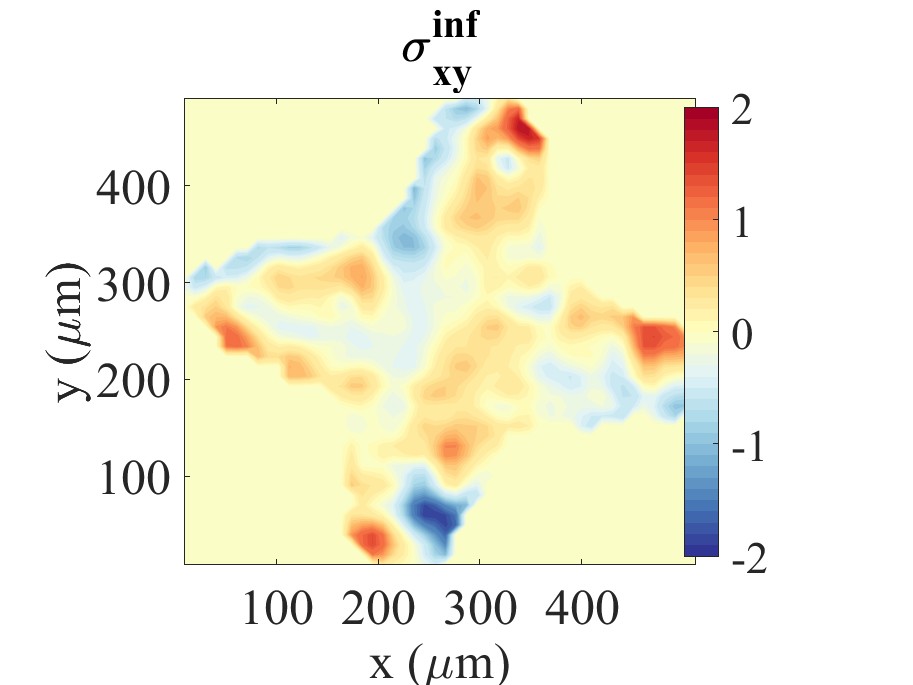}
  \textbf{f.}
  \includegraphics[width=0.27\linewidth]{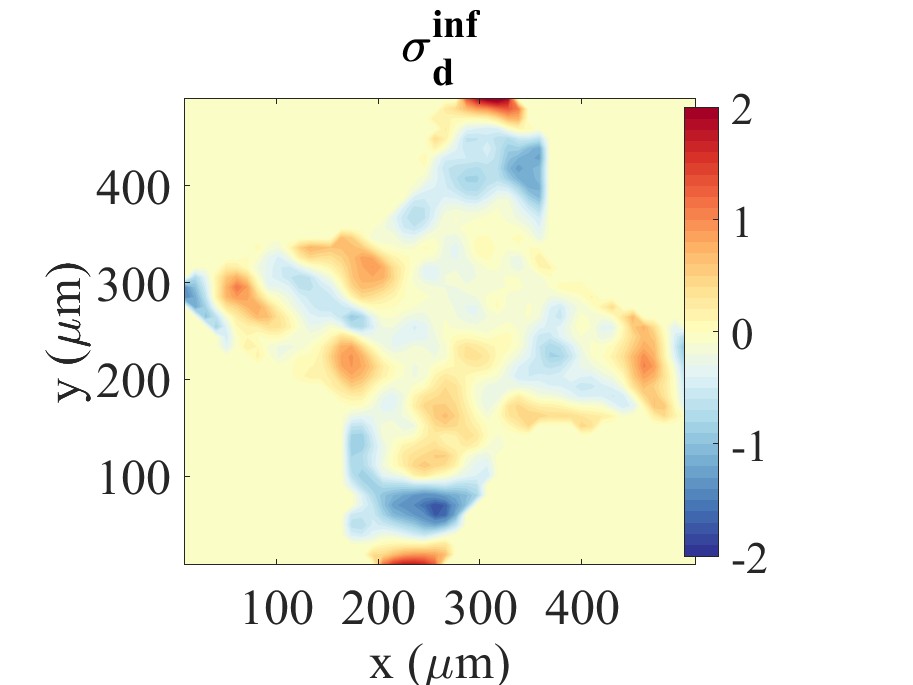} \\
    \textbf{g.}
  \includegraphics[width=0.27\linewidth]{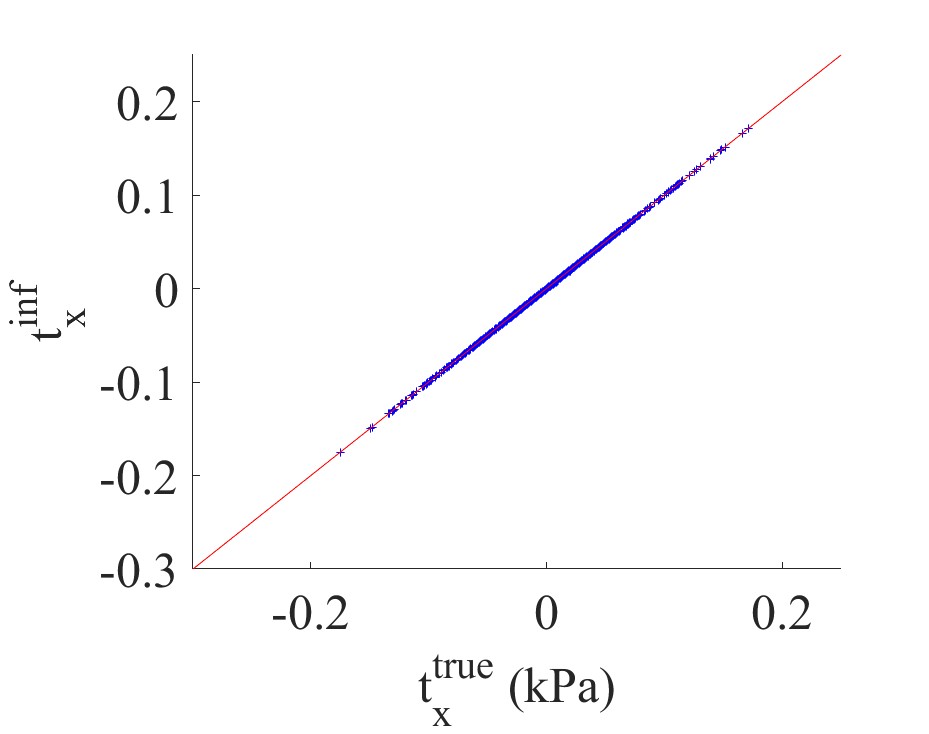}
  \textbf{h.}
  \includegraphics[width=0.27\linewidth]{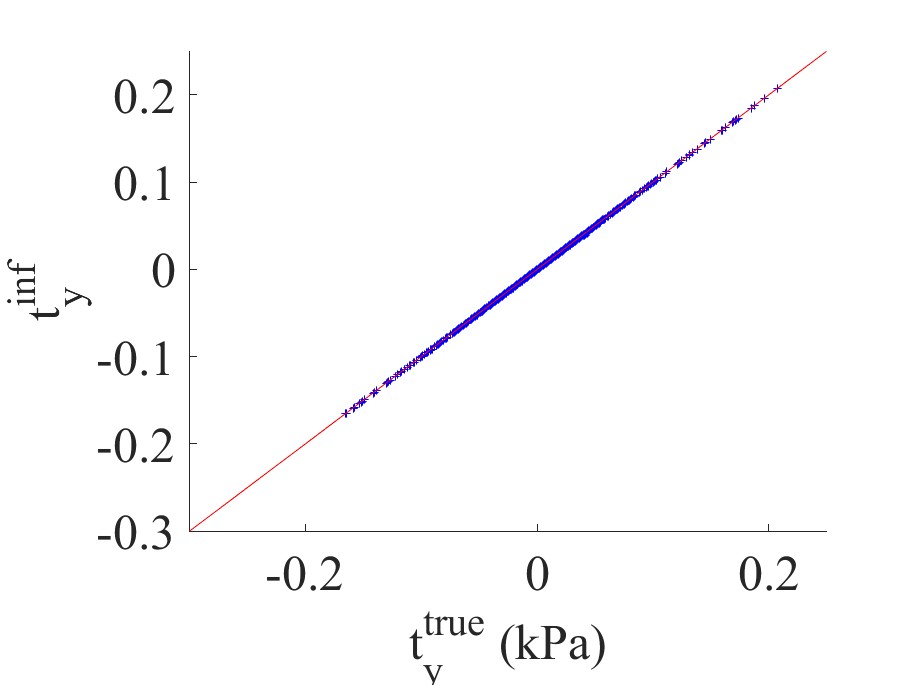}   
  \end{center}
  \caption{
\textbf{Stress inference in star-shaped tissue:} 
MDCK cell monolayer.
\textbf{a)} Phase contrast image of the cell island.
\textbf{b, c)} Color maps of the
components of the traction force field
$t_x^{\mathrm{true}}$, $t_y^{\mathrm{true}}$ measured by TFM (in kPa).
\textbf{d, e, f)} Color maps of the inferred
isotropic stress $\sigma_{\mathrm{iso}}^{\mathrm{inf}}$ and
deviatoric stress components $\sigma_{\mathrm{xy}}^{\mathrm{inf}}$,
$\sigma_{\mathrm{d}}^{\mathrm{inf}}$ (in kPa.$\mu$m).
Here the zero-stress condition
$\sigma_{ij} \, n_j=0$ is imposed on the four boundaries of the smallest
rectangular domain encompassing the cell domain.
\textbf{g, h)} Comparison of measured
($t^{\mathrm{true}}_x$, $t^{\mathrm{true}}_y)$
and inferred ($t^{\mathrm{inf}}_x$, $t^{\mathrm{inf}}_y$) values of
components of the traction force vectors (blue crosses),
computed from the inferred stress field
$\vec{t}^{\mathrm{inf}} = \mathrm{div} \, \sig^{\mathrm{inf}}$.
The bisectrix $y = x$ is plotted as a red line for comparison.
The coefficient of determination is $R^2_t = 1.0$.
As in Fig.~\ref{fig:confined:MDCK}, we find excellent agreement between
stress averages computed in the cell domain and the true values
obtained from moments of the traction force field: compare 
$  \langle \sigma_{\mathrm{iso}}^{\mathrm{true}} \rangle = 1.56 $ kPa
and
$  \langle \sigma_{\mathrm{iso}}^{\mathrm{inf}} \rangle = 1.57$ kPa;
$  \langle \sigma_{\mathrm{xy}}^{\mathrm{true}} \rangle = 58$ Pa
and
$  \langle \sigma_{\mathrm{xy}}^{\mathrm{inf}} \rangle = 69$ Pa;
$  \langle \sigma_{\mathrm{d}}^{\mathrm{true}} \rangle = -45$ Pa
and
$  \langle \sigma_{\mathrm{d}}^{\mathrm{inf}} \rangle = -47$ Pa.
[Scale bar 100 $\mu$m (a).] 
}
\label{fig:star}
\end{figure}

\subsection{Confined tissues of arbitrary shape}
\label{BC:island}

Geometry can play an important role in tissue morphogenesis. For example, high tissue curvature can promote cell extrusion \cite{Saw2017} and governs the self-organisation of intestinal organoids by localizing stem cells \cite{Gjorevski2022}. Microfabrication techniques allow generating tissues of arbitrary shape, \textit{e.g.} by restricting cell adhesion to micropatterned domains \cite{Ladoux_2017}. Thus, inferring stresses in such geometrically controlled experimental systems may provide important mechanistic insights into cell and tissue behaviour.

Applying BISM in the context of other geometries than the original rectangle, we found that the stress field within cell islands of arbitrary shape may be
inferred reliably by using as an effective domain the smallest rectangle encompassing
the island. This idea was originally implemented and validated numerically
in \cite{Pricoupenko2024} in the case of a circular domain.

Here, we present in Fig.~\ref{fig:star} results obtained by BISM
when applied on the smallest square encompassing a star-shaped MDCK cells island.
(The case of an elliptic cell island, studied in the same fashion, 
is presented in Fig.~\ref{fig:ellipse}.)
Stress-free boundary conditions are applied on the four boundaries of the square.
Spurious traction forces measured outside the star, but inside the rectangle, are set to zero.
Only stresses estimated within the cell island are plotted in Fig.~\ref{fig:star}.
The exact calculation of mean stress components, as introduced in Sec.~\ref{concept}, 
is expected to remain valid for confined domains of arbitrary shape. Indeed, 
we find excellent quantitative agreement 
between mean stress components and moments of the traction forces, where spatial averages 
are computed within the cell island (see the caption of Fig.~\ref{fig:star}
for numerical values). 
Those experiments demonstrate the applicability of BISM for tissues of arbitrary geometry, which opens up the possibility to study the relationship between the biological process in question, tissue shape and stresses resolved in space and time.

\subsection{One free boundary}
\label{BC:front}

\begin{figure}[!ht]
    \begin{center}
      \textbf{a}
      \hspace*{0.7cm}
      \includegraphics[width=0.12\linewidth]{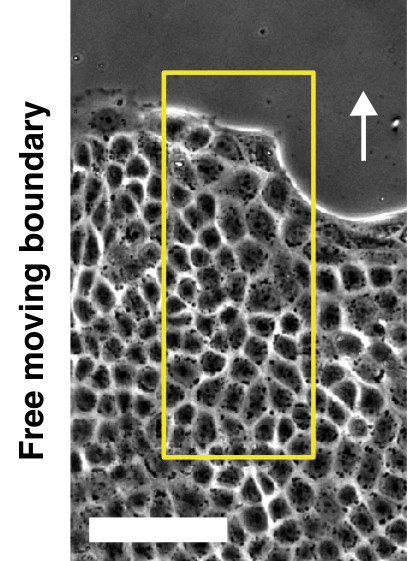}
      \hspace*{1.3cm}
      \textbf{b}
  \includegraphics[width=0.27\linewidth]{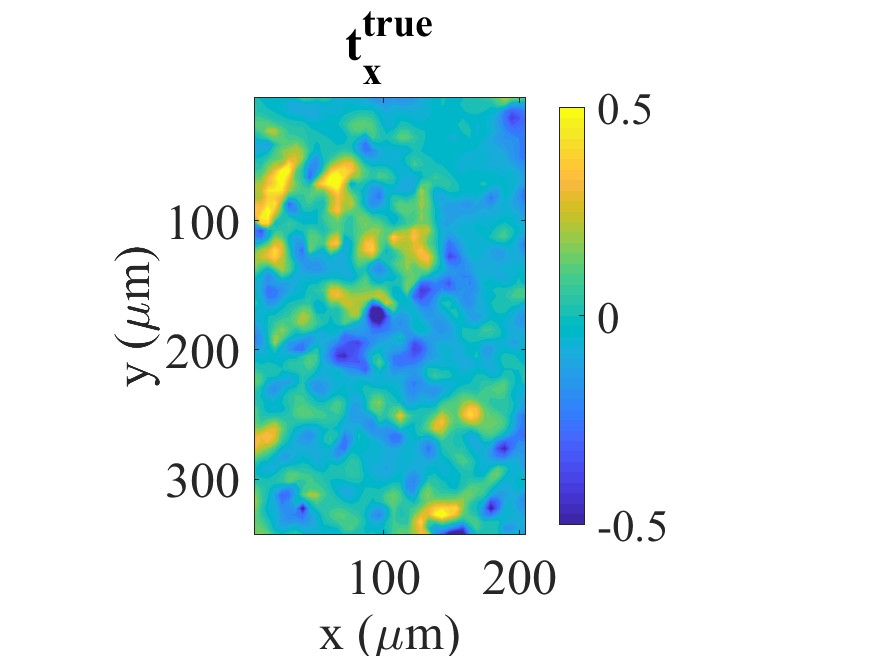} 
  \textbf{c}
  \includegraphics[width=0.27\linewidth]{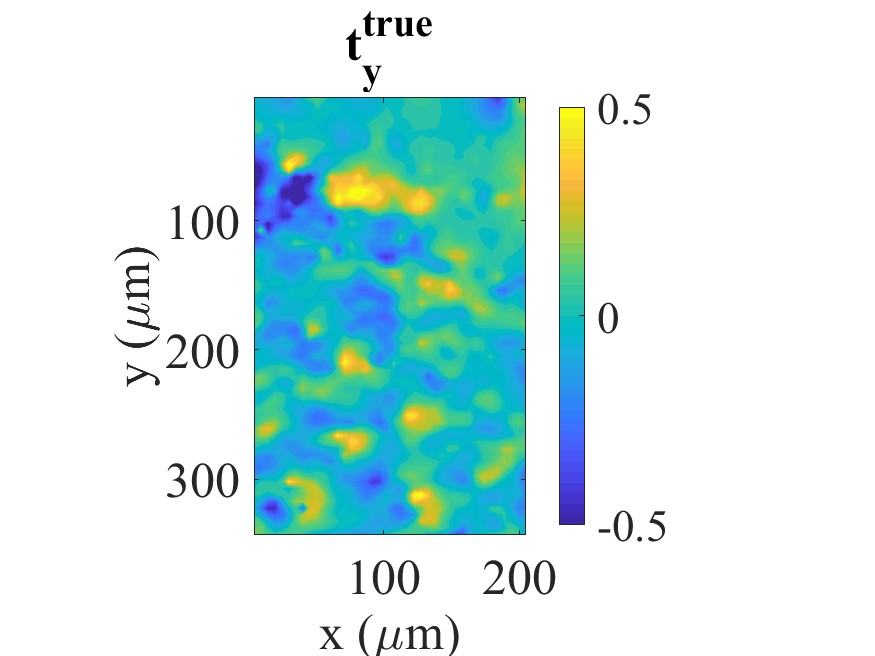}\\
  \medskip
  \textbf{d}
  \includegraphics[width=0.27\linewidth]{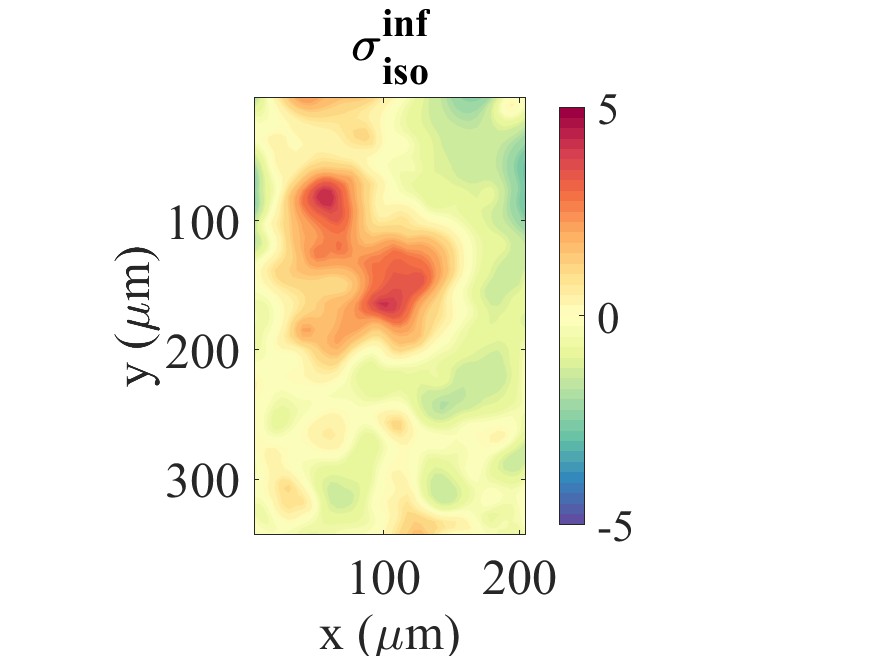} 
  \textbf{e}
  \includegraphics[width=0.27\linewidth]{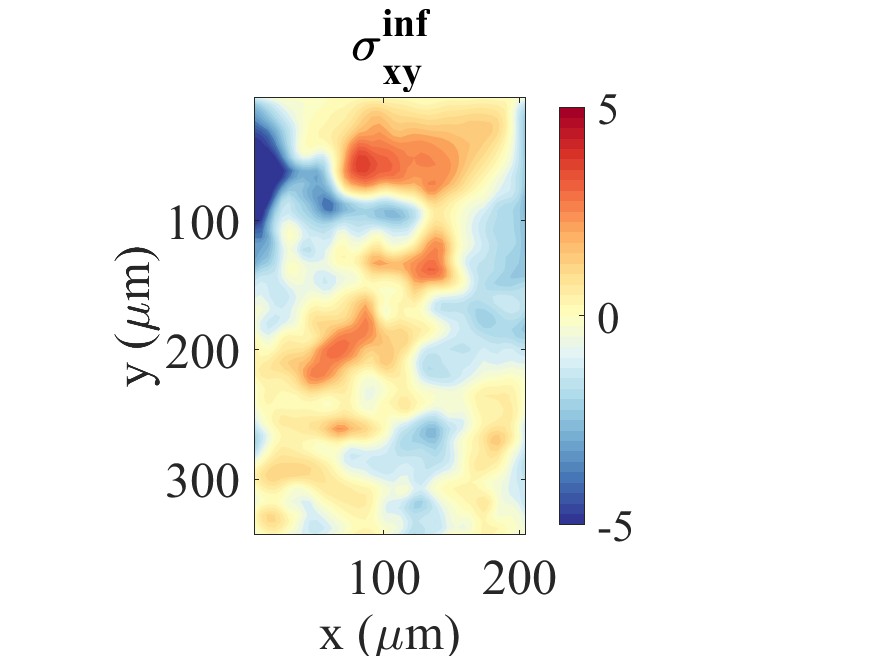}
  \textbf{f}
  \includegraphics[width=0.27\linewidth]{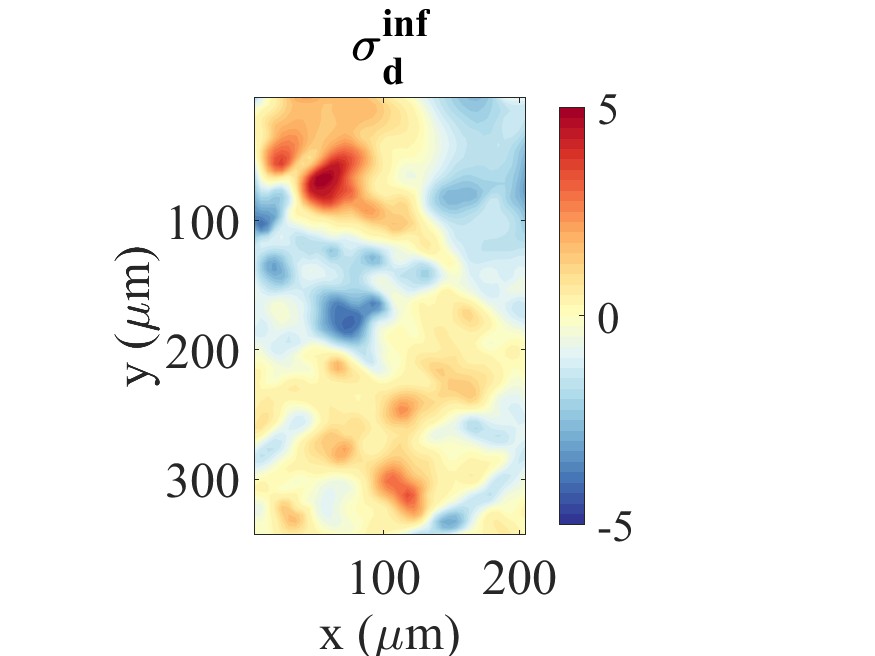} \\
  \smallskip
    \textbf{g}
  \includegraphics[width=0.27\linewidth]{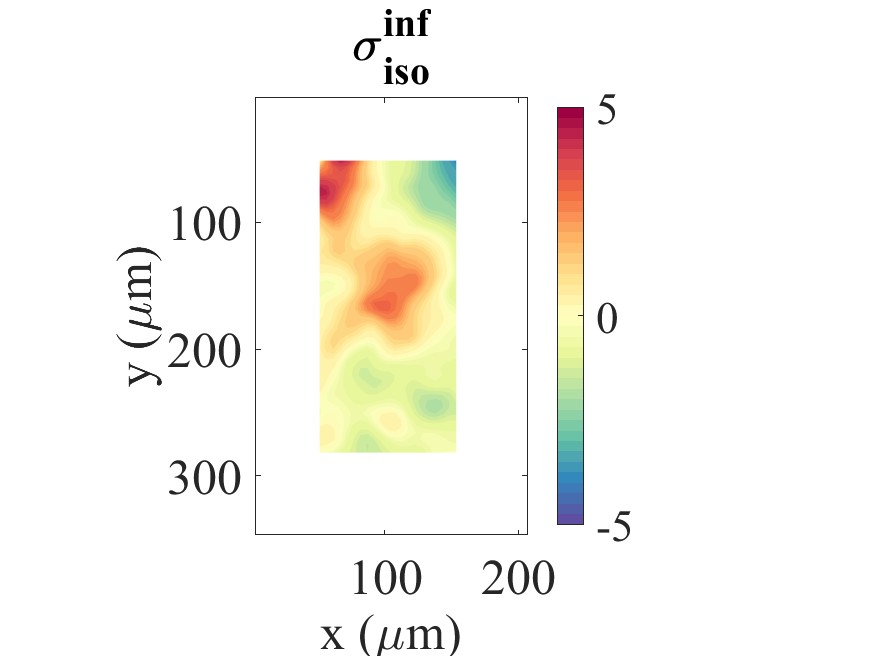}
  \textbf{h}
  \includegraphics[width=0.27\linewidth]{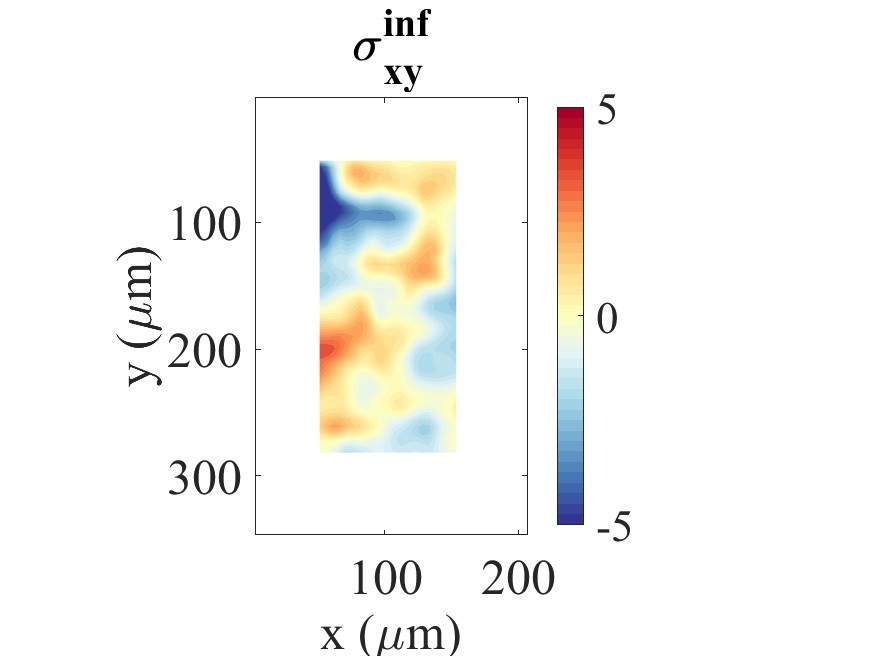}
    \textbf{i}
  \includegraphics[width=0.27\linewidth]{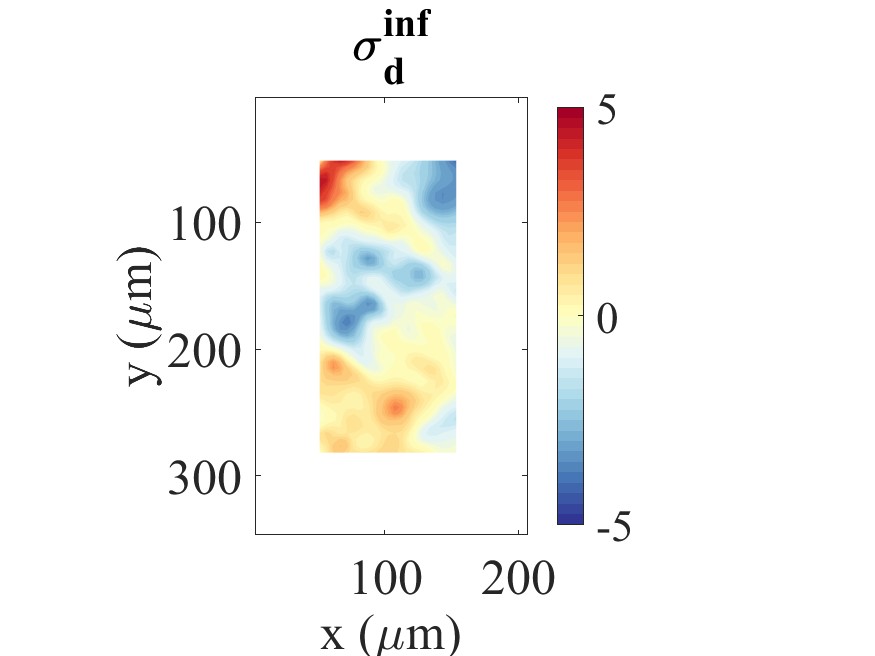} \\
    \smallskip
  \textbf{j}
  \includegraphics[width=0.27\linewidth]{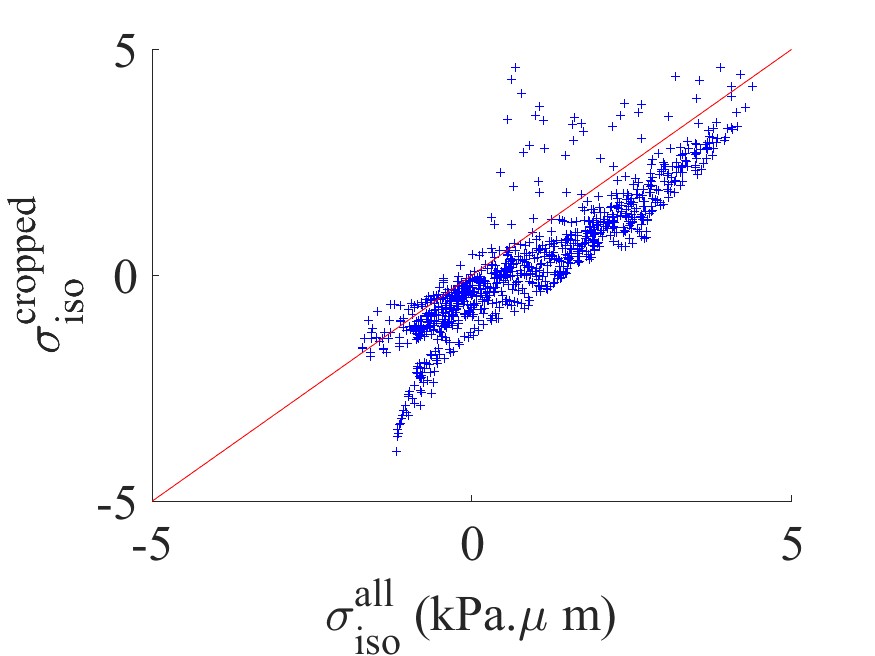} 
  \textbf{k}
  \includegraphics[width=0.27\linewidth]{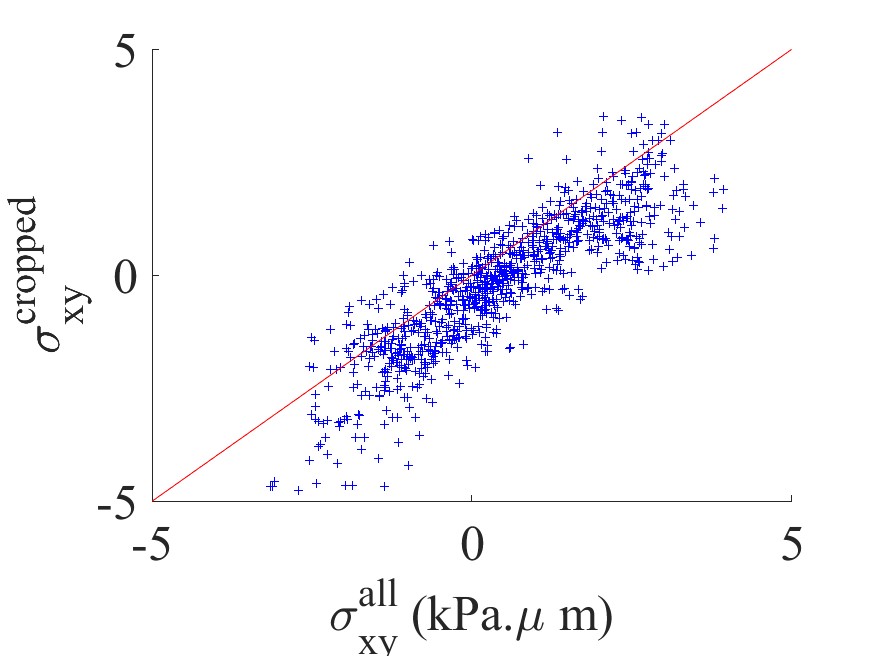}
  \textbf{l}
  \includegraphics[width=0.27\linewidth]{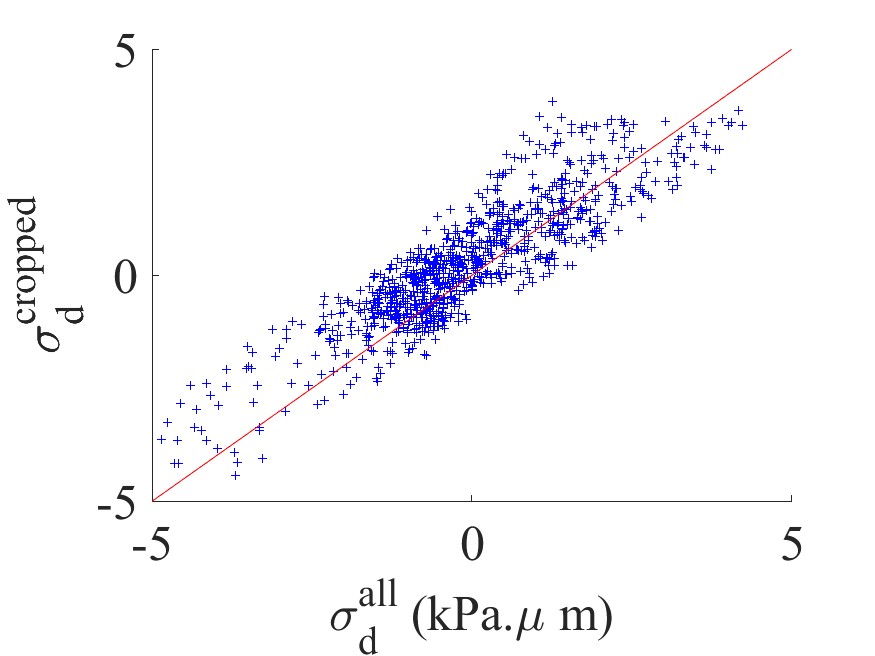} \\
  \end{center}
\caption{
\textbf{Stress inference in a system with a moving boundary.} 
\textbf{a})  Brightfield image representative of a MDCK WT cells wound healing assay. The free edge of the monolayer is moving in the direction of the white arrow.
\textbf{b, c)} Color maps of the
components of the traction force field
$t_x^{\mathrm{true}}$, $t_y^{\mathrm{true}}$ measured by TFM (in kPa).
\textbf{d, e, f)} Color maps of the inferred
isotropic stress $\sigma_{\mathrm{iso}}^{\mathrm{inf}}$ and
deviatoric stress components $\sigma_{\mathrm{xy}}^{\mathrm{inf}}$,
$\sigma_{\mathrm{d}}^{\mathrm{inf}}$ (in kPa.$\mu$m).
Here the zero-stress condition
$\sigma_{ij} \, n_j=0$ is imposed only on the top boundary,
while the three other edges are not subject to a boundary condition.
\textbf{g, h, i)} The same stress components are inferred under the same conditions, but within a subdomain that intersects the moving front.
\textbf{j, k, l)} Comparison of measured stress components
in the cropped region depending on whether it was computed
using the whole field of view (``all'') or only the cropped region
(``cropped''). The bisectrix $y = x$ is plotted as a red line for comparison.
    [Scale bar 80 $\mu$m (a).]
    }
\label{fig:front}
\end{figure}

Multiple biological processes during morphogenesis, but also wound healing or cancer invasion require cells to expand and migrate collectively \cite{friedl2009}. Letting an epithelium expand into empty space represents a minimal system to study collective migration. 
As tissue size constraints typically limit measuring all cell-exerted forces, stresses can often only be inferred in subdomains of the expanding tissue. Thus, the empty space generates one free moving boundary.

Depending on the chosen geometry and boundary conditions,
symmetry may be leveraged to estimate tissue stress.
A case in point is (directional) monolayer expansion.
Denoting $x$ and $y$ the coordinates perpendicular and
parallel to the moving front, translational invariance
may apply in  a statistical sense along $y$ (the boundary),
making an homogeneous system effectively 1D (along $x$),
provided that the spatial extension along $y$ is large enough.
However, Eq.~(\ref{eq:forcebalance}) is readily integrable
in one spatial dimension, yielding an absolute measure
of the (1D) tissue tension, which is highest close to free boundaries in MDCK monolayers \cite{Vincent2015}.

In \cite{Ollech2020}, we used this observation to validate Bayesian
inference in the case of an expanding monolayer, and checked
in particular that the average over the $y$-coordinate of
the inferred 2D stress agrees with the (exact) 1D stress.
Of note, this result was obtained using experimental data 
by imposing a stress-free condition
only on the side of the moving boundary, without setting specific
conditions on the three other boundaries of the domain
(see \cite{Ollech2020} for technical details).

This analysis underlines the applicability of BISM in the context of collective cell migration. In Fig.~\ref{fig:front}, we further illustrate this approach
and show that the accuracy of stress inference is unchanged when traction forces can only be measured on a small scale, as it may be the case in the context of high-resolution imaging.

\subsection{No boundary conditions}
\label{BC:unconfined}

\begin{figure}[!t]
    \begin{center}
      \textbf{a}
  \includegraphics[width=0.27\linewidth]{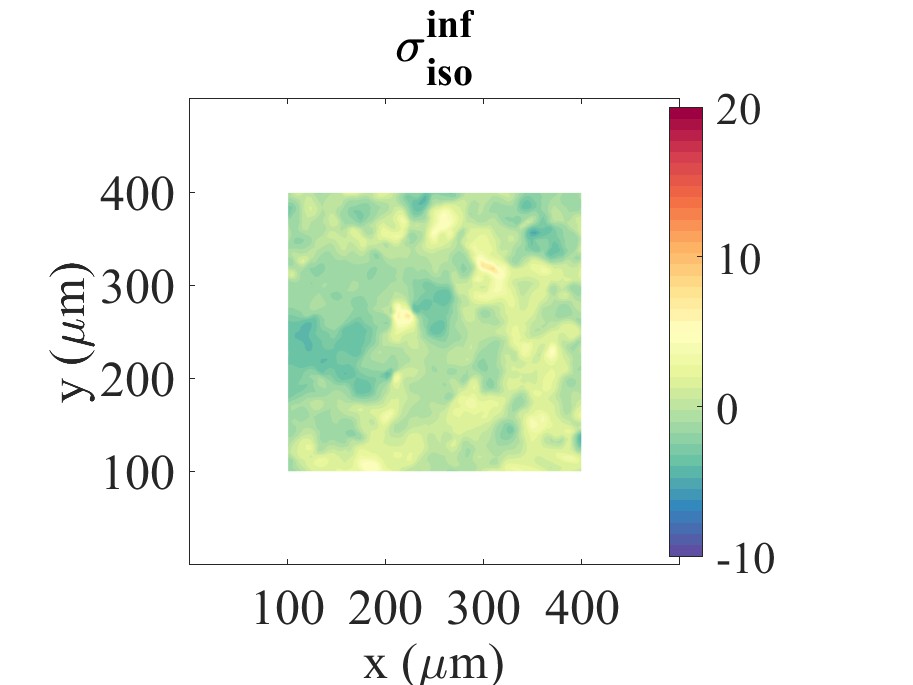}
  \textbf{b}
  \includegraphics[width=0.27\linewidth]{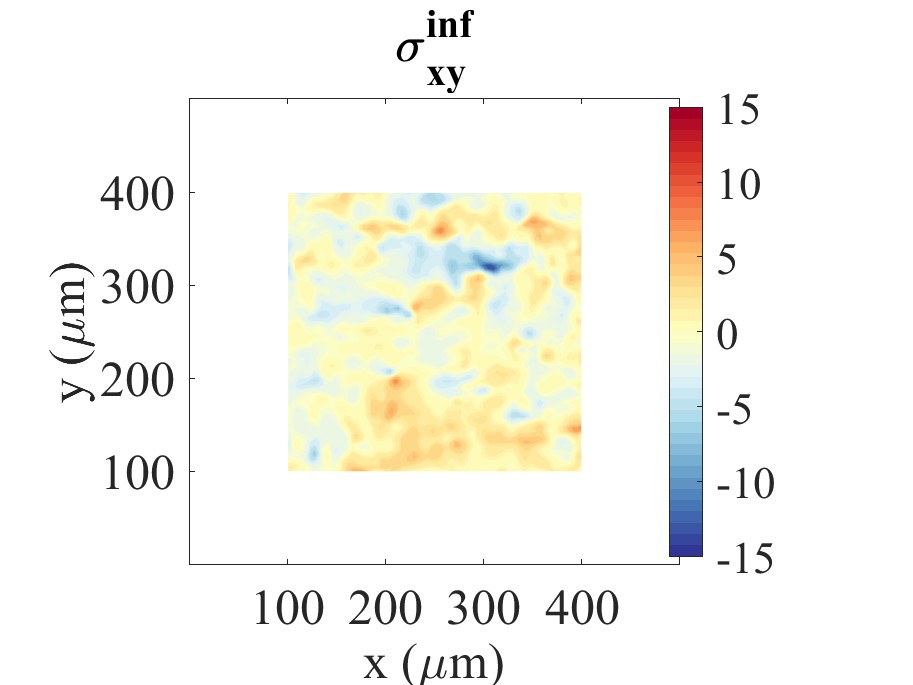} 
  \textbf{c}
  \includegraphics[width=0.27\linewidth]{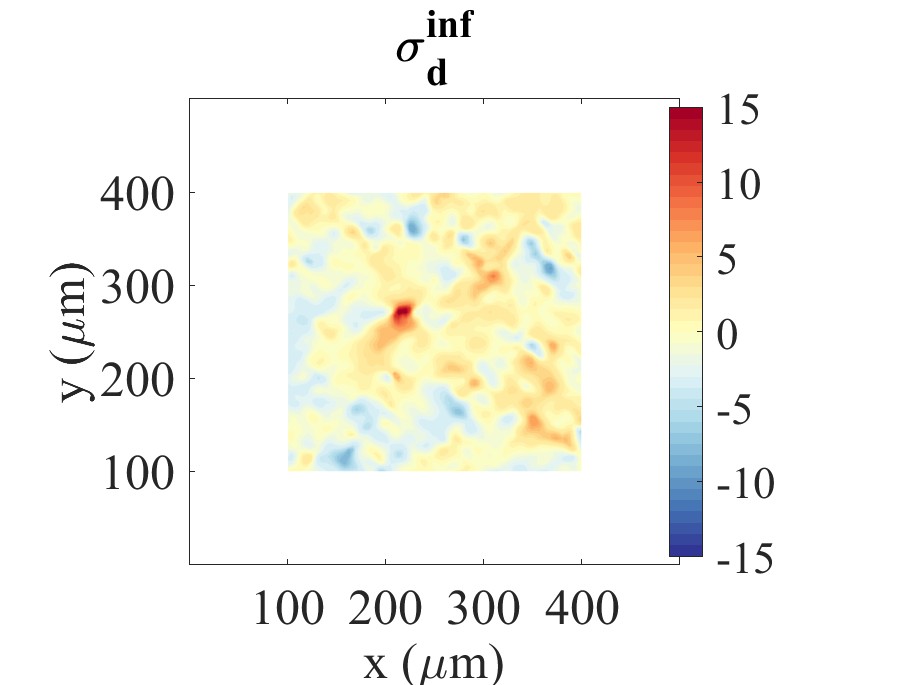}\\
  \medskip
  \textbf{d}
  \includegraphics[width=0.27\linewidth]{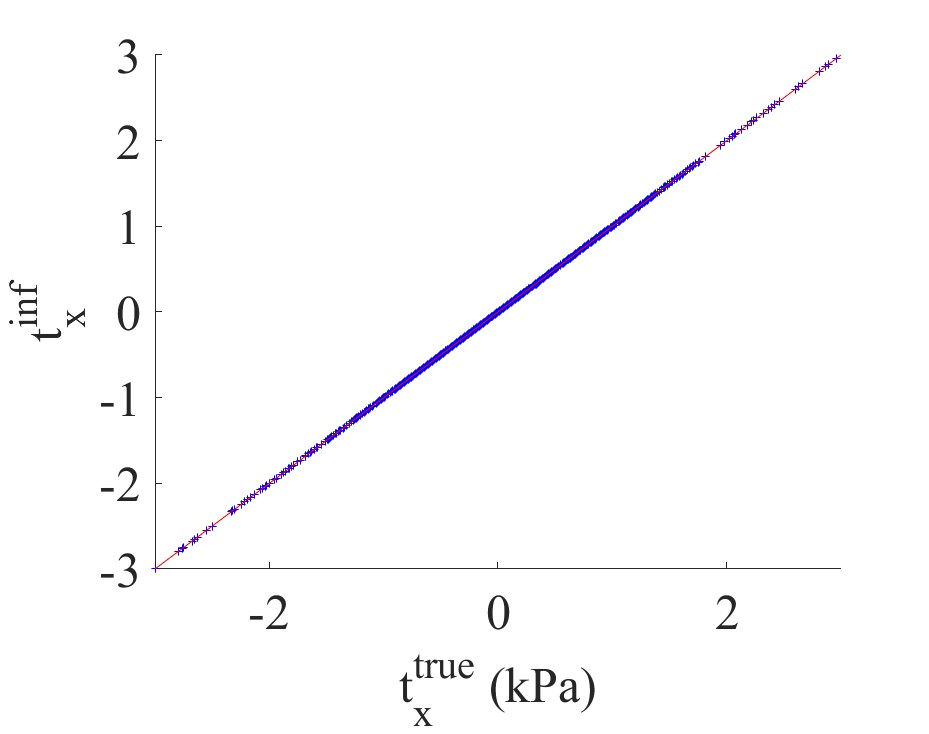} 
  \textbf{e}
  \includegraphics[width=0.27\linewidth]{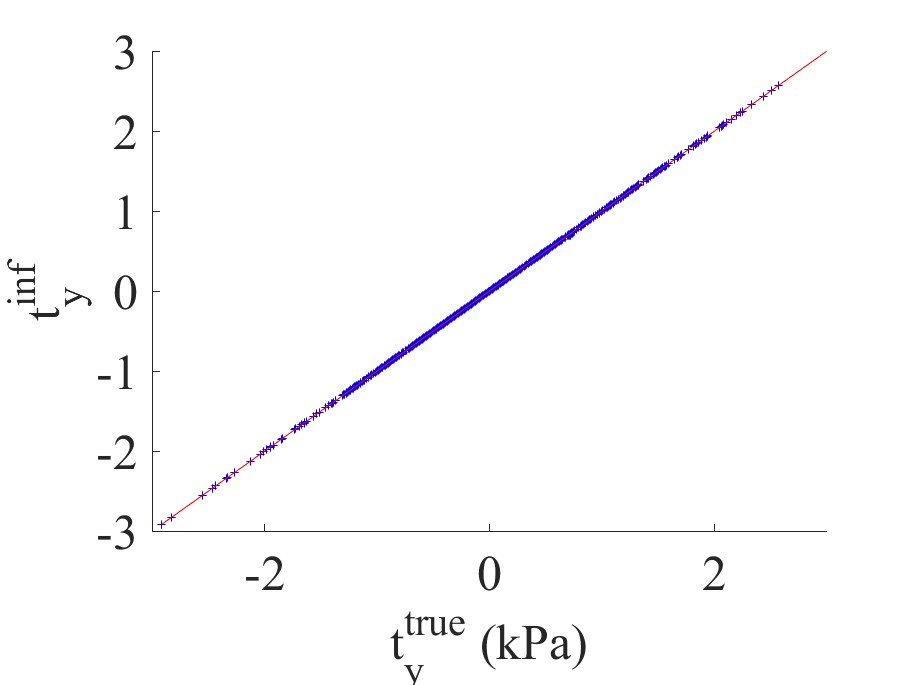}
    \textbf{f}
  \includegraphics[width=0.27\linewidth]{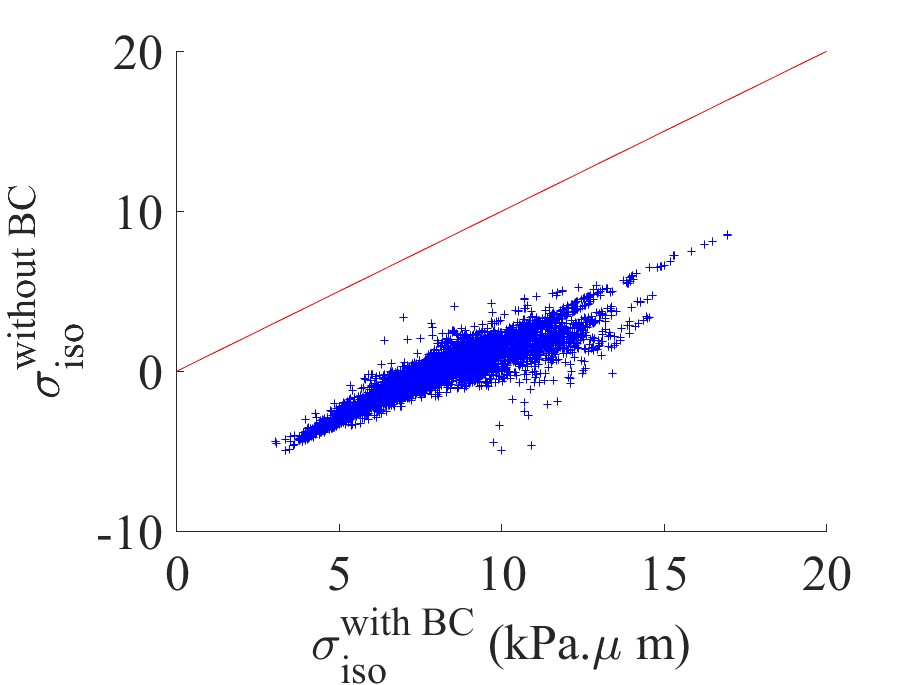}\\
  \medskip
  \textbf{g}
  \includegraphics[width=0.27\linewidth]{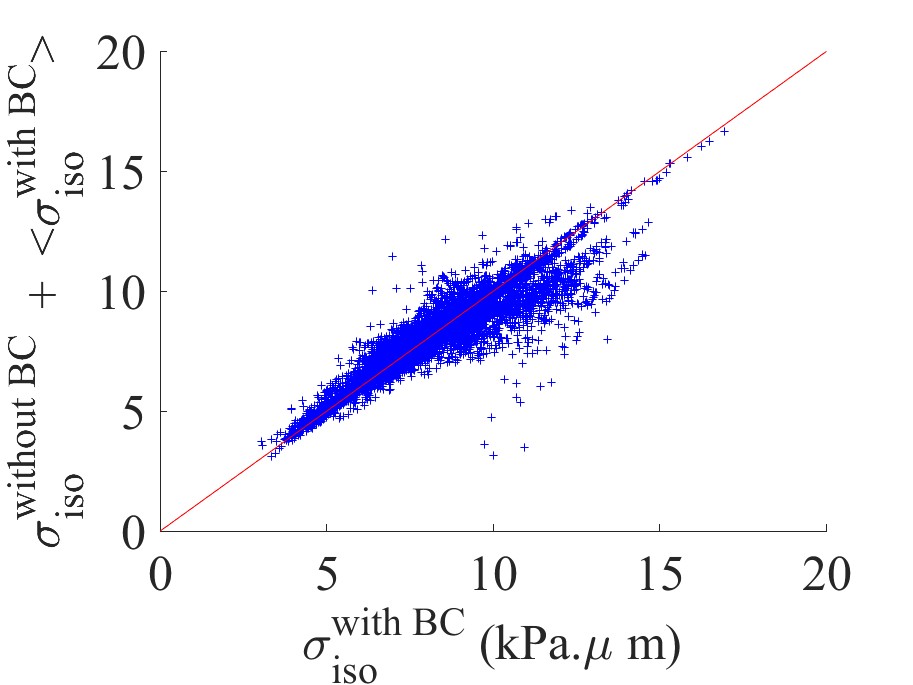} 
  \textbf{h}
  \includegraphics[width=0.27\linewidth]{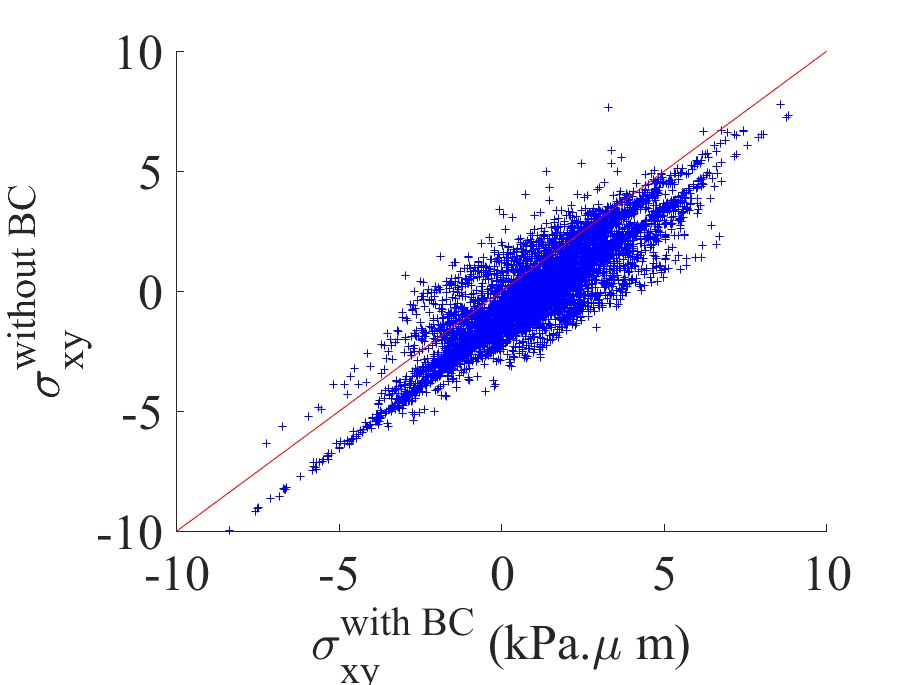} 
  \textbf{i}
  \includegraphics[width=0.27\linewidth]{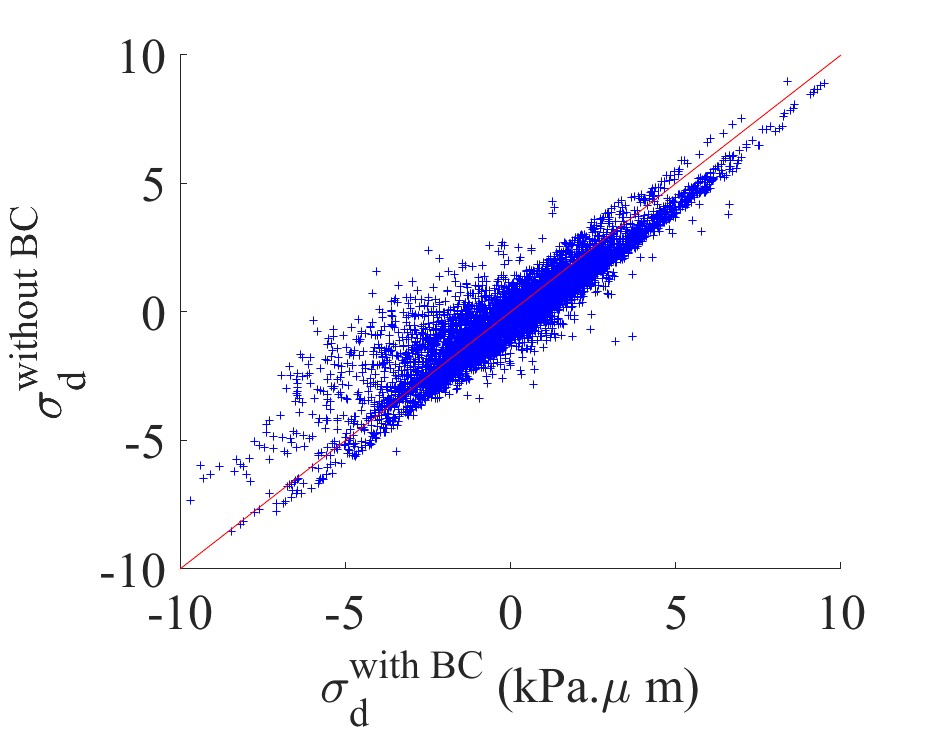} 
  \end{center}
  \caption{
  \textbf{Subdomain:} the same traction force data is used
  as in Fig.~\ref{fig:confined:MDCK}, this time focusing on the central
  subdomain of lateral extension $300 \, \mu$m, and \emph{without} imposing
  boundary conditions.
  \textbf{a, b, c)} Color maps of the inferred
isotropic stress $\sigma_{\mathrm{iso}}^{\mathrm{inf}}$ and
deviatoric stress components $\sigma_{\mathrm{xy}}^{\mathrm{inf}}$,
$\sigma_{\mathrm{d}}^{\mathrm{inf}}$ (in kPa.$\mu$m). Since the color code
is the same as in  Fig.~\ref{fig:confined:MDCK}, we see that,
contrary to deviatoric components, the absolute
value of the isotropic stress is not recovered.
\textbf{d, e)} Comparison of measured
($t^{\mathrm{true}}_x$, $t^{\mathrm{true}}_y)$
and inferred ($t^{\mathrm{inf}}_x$, $t^{\mathrm{inf}}_y$) values of
components of the traction force vectors (blue crosses),
computed from the inferred stress field
$\vec{t}^{\mathrm{inf}} = \mathrm{div} \, \sig^{\mathrm{inf}}$.
The bisectrix $y = x$ is plotted as a red line for comparison.
The coefficient of determination is $R^2_t = 1.0$.
This confirms that spatial patterns of stress are correctly
inferred without imposing boundary conditions.
\textbf{f, h, i)} Comparison of isotropic stress
 $\sigma_{\mathrm{iso}}^{\mathrm{inf}}$ and deviatoric
stress components $\sigma_{xy}^{\mathrm{inf}}$ and
$\sigma_{\mathrm{d}}^{\mathrm{inf}}$ inferred on the whole domain
with stress-free boundary conditions ($x$ axes, ``with BC'')
and inferred on the central subdomain without
imposing boundary conditions ($y$ axes, ``without BC'').
\textbf{g)} Data on the $y$ axis are shifted by
$< \sigma_{\mathrm{iso}}^{\mathrm{with BC}} >$
the mean isotropic stress computed over the subdomain
\emph{with} stress-free boundary conditions imposed
on the whole domain.
Coefficients of determination for the three panels g, h and i:
$R^2_{\mathrm{iso}}  = 0.82$,
$R^2_{\mathrm{xy}}  = 0.49$,
$R^2_{\mathrm{d}}  = 0.86$.
The spatial averages over the subdomain inferred without
boundary conditions are:
$< \sigma_{\mathrm{iso}}^{\mathrm{without BC}} > = 1.3 \pm  274.0$ Pa, 
$< \sigma_{\mathrm{xy}}^{\mathrm{without BC}} > = -2.5 \pm  566.1$ Pa and
$< \sigma_{\mathrm{d}}^{\mathrm{without BC}} > = 16.8 \pm   581.0$ Pa,
in each case consistent with a zero mean value.
Uncertainty is computed as the standard deviation.
}
\label{fig:subdomain:MDCK}
\end{figure}

When the monolayer is very large, and/or when the domain of interest is very far from 
the natural boundaries of the tissue, implementing stress-free boundary condition
may become impractical. Measuring all cell-exerted forces within a tissue is typically limited by its size, which, in our hands, may reach the centimeter scale. Such size constraints can require applying BISM on a subdomain of the whole tissue.

In this section, we evaluate
how BISM-based inference is affected by the absence of boundary conditions
(see Figs.~\ref{fig:subdomain:MDCK} and \ref{fig:subdomain:HaCaT} for
similar results obtained in MDCK and HaCaT cell monolayers, respectively):
To this end, we infer stresses from traction force data only within the (central)
subdomain of the larger, confined monolayer analysed in Figure \ref{fig:confined:MDCK}. 
Thus, the correct boundary conditions for this subdomain are unknown. Cells close
to the subdomain boundaries, yet within the subdomain, are surrounded by
other cells outside the subdomain with which they interact mechanically.
When applying BISM on the subdomain, we therefore cannot impose boundary
conditions, and modify the prior accordingly. 

We observe that spatial patterns of stress are correctly recovered 
(see Figs.~\ref{fig:subdomain:MDCK}d,e), as the likelihood function
still enforces the force balance equation \eqref{eq:forcebalance}.
We also find that the deviatoric components of the stress are correctly
reconstructed: Focusing on the case of MDCK cells, compare panels
\ref{fig:confined:MDCK}e and \ref{fig:subdomain:MDCK}b, 
then \ref{fig:confined:MDCK}f and \ref{fig:subdomain:MDCK}c,
and see panels \ref{fig:subdomain:MDCK}h and \ref{fig:subdomain:MDCK}i.
However, a discrepancy occurs for the isotropic stress
(as well as for normal stress components -- data not shown),
which is shifted by a constant when  compared to the correct result
obtained  in the full domain: Compare panels \ref{fig:confined:MDCK}d
and \ref{fig:subdomain:MDCK}a, and see panel \ref{fig:subdomain:MDCK}f.

These observations can be rationalized by noticing that the effect
of not imposing boundary conditions in the BISM algorithm is to set the mean value of all
inferred stress components to $0$ (within numerical noise).
Indeed, in Figs.~\ref{fig:subdomain:MDCK}, \ref{fig:BC0}
and \ref{fig:subdomain:HaCaT}, without using boundary conditions, 
the mean values computed over the subdomain
$< \sigma_{\mathrm{iso}}^{\mathrm{inf}} >$, $< \sigma_{\mathrm{xy}}^{\mathrm{inf}} >$ 
and $< \sigma_{\mathrm{d}}^{\mathrm{inf}} >$ are all consistent with $0$ within
error bars.
In Figs.~\ref{fig:subdomain:MDCK}g and \ref{fig:subdomain:HaCaT}g,
we shifted isotropic stresses obtained without boundary conditions
by the value of the mean isotropic stress obtained in the same
subdomain with stress-free boundary conditions: good agreement
was thus recovered.

We next tested possible consequences of foregoing stress-free
boundary conditions on other experimental conditions. 
Overall, we find that in the absence of boundary conditions,
spatial patterns of stress are still inferred correctly, 
as evidenced by values of $R^2_t$ remaining close to $1$
(see, \textit{e.g.}, Fig.~\ref{fig:BC0}adg).
We also find that spatial patterns of stress in cell islands are still
inferred correctly when less care is given to the inference than
advocated in Sec.~\ref{BC:island} that is when a larger domain
is considered, boundary conditions are not implemented, 
and spurious traction force data is taken into account, see Fig.~\ref{fig:BC0}beh anf cfi.
However, and as was the case in this section, the correct absolute values
of some stress components may be lost  (compare, \textit{e.g.}
Fig.~\ref{fig:star}d and Fig.~\ref{fig:BC0}h).

To summarize, we observe that stress-free boundary conditions is necessary 
for an absolute measure of tissue stress components with BISM.
Foregoing boundary conditions still allows to infer stress patterns
reliably, but may turn BISM-based inference into a relative estimate
of stresses, measured up to an unknown additive constant
(as is the case for Bayesian Force Inference, which by construction
lacks a reference pressure as well as a force scale). 
Based on an this analysis, we recommend to use boundary conditions 
(and to acquire traction force data for the whole domain) when stresses are 
inferred on monolayers confined by micropatterns.

However, in the absence of such boundary constraints, BISM can still provide accurate absolute stress estimates, provided the mean stress component of interest is approximately zero. This assumption is often justified in practice. Deviatoric stresses are often found to average to zero. In traction force microscopy for instance, tissue boundaries are typically inaccessible, but the tissue is confined by the culture dish walls. These tissues extend over millimeter-scale areas, thousands of times the size of individual cells, and do not experience free expansion. Under these conditions, large homeostatic tissues are observed to exhibit negligible mean isotropic stress in their central regions, enabling absolute stress estimation with BISM even without explicit boundary conditions.

\begin{figure}[!h]
\includegraphics[width=1\linewidth]{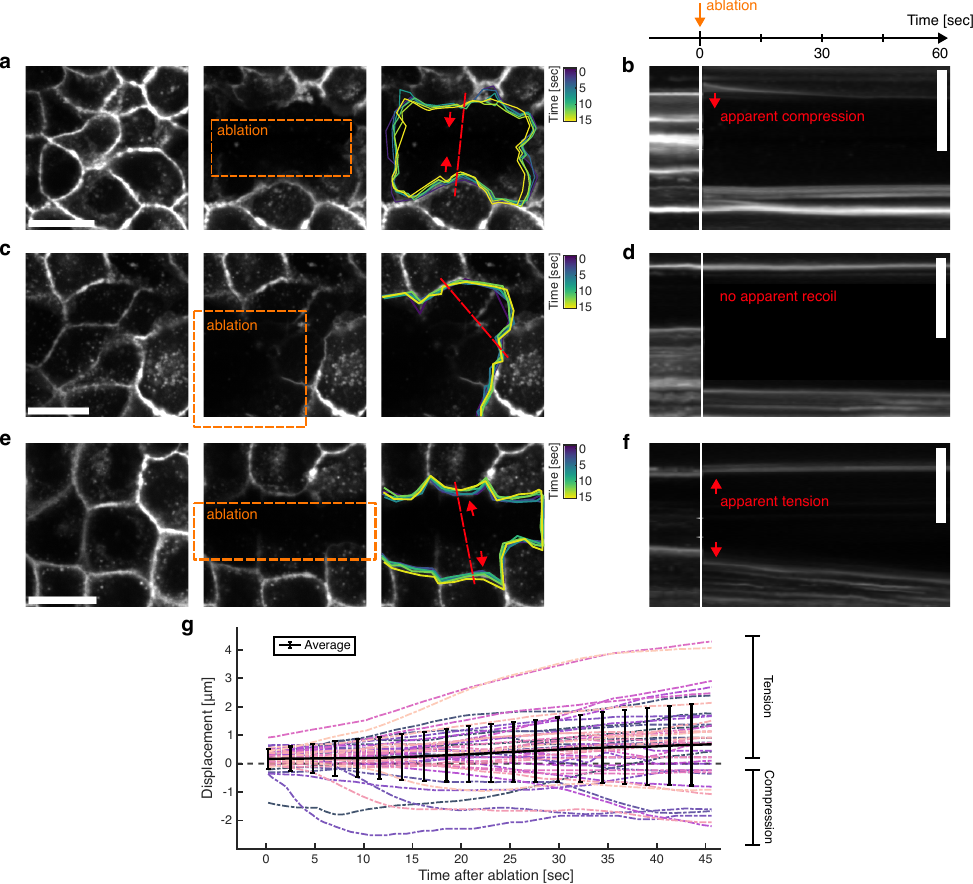}
\caption{
\textbf{Experimental validation by laser ablation of zero mean isotropic stress in a large tissue far from its boundaries.}
    Panels \textbf{a, c, e}) Left: Representative images of CAAX-GFP–expressing MDCK cells at confluence, before laser ablation. Middle : Same region immediately after ablation. The orange contour outlines the ablated region. Right: Same region, where the outline shows the evolution of the surrounding cell boundaries over time after ablation, representative of a response to laser ablation for a tissue under compression \textbf{(a)}, zero stress \textbf{(c)} or under tension tension \textbf{(e)}. The color map indicates time after ablation, in seconds. 
   Panels  \textbf{b, d, f}) Corresponding kymographs along the red line in \textbf{(a, c, e)}, showing the temporal displacement response of the surrounding tissue following laser ablation. Negative recoil illustrates a state of compression \textbf{(b)}, an absence of recoil a state of zero isotropic stress \textbf{(d)}, and positive recoil  a state of tension \textbf{(f)}, respectively. The x-axis represents time after ablation, in seconds.
  Panel  \textbf{g}) Recoil displacement over time in response to laser ablation. The dark line represents the ensemble average of the distribution of displacements, also as a function of time. Each line in gradient of pink corresponds to one ablation assay. This graph shows the variability in response, over N = 45 ablations from n = 2 independent experiments.
    [Scale bar 15 $\mu$m (a, b, c, d, e, f). Error bars show the 95 $\%$ confidence interval.]
}
\label{fig:null_laser}
\end{figure}

 We verified that this expectation is correct in MDCK monolayers cultured at confluence in 35-mm wide dishes. To assess the mechanical state of such tissues, we performed laser ablation experiments, which provide a direct evaluation of tissue isotropic stress by observing the direction of recoil following ablation. If the ablated region expands, then the tissue is under tension. If it contracts, the tissue is under compression. By ablating random positions within the inner tissue region, we observed that the likelihoods of expansion and contraction events are approximately equal ($N = 45$, $n_+ = 18$, $n_- = 14$, defined as half one standard deviation away from the mean). Moreover, in many locations, no measurable recoil occurred after ablation, suggesting a local isotropic stress close to zero (Fig.~\ref{fig:null_laser}a–f). On average, the recoil displacement was close to zero ($\Delta x = 0.17 \pm 0.36 \, \mu$m, Fig.~\ref{fig:null_laser}g), validating our expectation that the mean isotropic stress is zero. Owing to the large distance from the physical boundaries, cells located near the center of the dish can be considered as screened from boundary conditions.
 Taken together, since our laser ablation experiments revealed regions of tension, compression and negligible isotropic stress, they are qualitatively in agreement with the stress maps generated by BISM, where those features are observed as well.

Along these lines, this approach has been validated in prior studies, particularly for identifying compressive and tensile stress patterns around topological defects where the average isotropic stress vanishes (see \cite{Saw2017}, and \cite{leToquin2024}, where similar continuum models were used). These findings have been corroborated both theoretically \cite{Saw2017} and experimentally in more specific cases using laser ablation \cite{Sonam2023, Schoenit2025}, which will be discussed in the next section.

\section{Heterogeneous tissues}
\label{sec:hetero}

Biological tissues are often heterogeneous and can consist of different cell types, \textit{e.g.} stem cells and specialized, differentiated cells. Furthermore, heterogeneity can emerge in disease, when cancerous, pathogen- or viral-infected cells accumulate within a tissue. How cells within such heterogeneous tissues interact on a mechanical level remains largely unknown. This knowledge gap could be addressed by stress inference. BISM presents an ideal approach, because unlike other methods, it does not require assumptions on tissue rheology, which is generally not expected to be homogeneous in tissues composed of different cell types. 

Mixing different cell types \textit{in vitro} can be used as simple models to study physiological processes like cell sorting, boundary formation or cell competition.
To understand the role of local stresses in such experimental contexts, we applied BISM to cell mixtures of normal MDCK cells (WT) and cells where the cell-cell adhesion protein E-cadherin was removed (KO) (Fig.~\ref{fig:mixed}a). This difference in cell-cell adhesion leads to cell sorting and eventually to competition between the two cell types \cite{Schoenit2025}. 
Since studying tissue-scale processes like sorting and cell competition can strongly be affected by confining cells in 2D, we performed stress inference on a subdomain and without imposing boundary conditions. BISM stress maps then revealed that cellular differences of the competing cell types translated to a striking differences in their mechanical state, with WT cells being mainly under compression and KO cells being under tension (Fig.~\ref{fig:mixed}a,b; \cite{Schoenit2025}). Therefore, we investigate other validation approaches.

\subsection{Validation with laser ablation}
\label{sec:hetero:val}

BISM has been employed in a range of contexts,
with or without boundary conditions, 
to estimate the internal isotropic stress of epithelial monolayers
\cite{Sonam2023, Schoenit2025, balasubramaniam2024different, gupta2023mechanical}, seen
as a quantifier of the mechanical state of the system. 
Complementing the experimental and computational validations presented above,
a more invasive technique, laser ablation, has also been employed \cite{Schoenit2025, Sonam2023}. 

\begin{figure}[!t]
\includegraphics[width=1\linewidth]{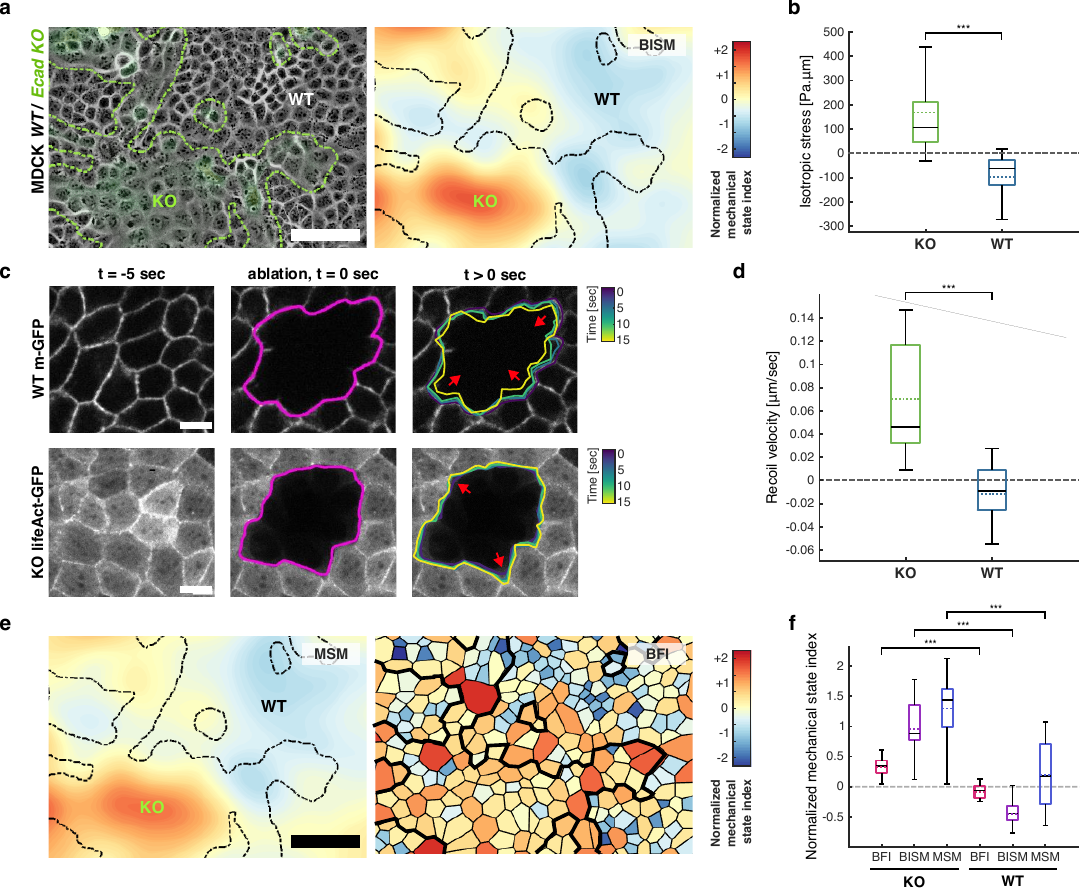}
\caption{
\textbf{Experimental validation by laser ablation and comparison with other force inference methods:} 
    \textbf{a}) Left: Representative phase contrast image of WT cells mixed with LifeAct-GFP–labeled E-cad KO cells (green). Right: Corresponding stress map of the inferred mechanical state using Bayesian Inversion Stress Microscopy (BISM).
    \textbf{b}) Boxplot of the average isotropic stress within the tissue comparing the areas occupied by KO or WT cells. n = 19 positions.
    \textbf{c}) Representative images of CAAX-GFP–labeled WT cells mixed with LifeAct-GFP–labeled E-cad KO cells, before (left) and after (middle) laser ablation. The pink contour outlines the ablated region. Immediately after ablation (right), the outline shows the evolution of the surrounding cells in time, representative of a reaction to compression (WT) or to tension (KO). The color gradient is time, in seconds.
    \textbf{d}) Boxplot of recoil velocities measured 5s immediately after laser ablation. n = 15 (WT) and n = 13 (E-cad KO) ; N = 2 independent experiments. 
    \textbf{e}) Left: Stress map corresponding to \textbf{a} inferred using Monolayer Stress Microscopy (MSM)
    Right: Cell segmentation and color-coded stress corresponding to \textbf{a} using shape based Bayesian Force Inference (BFI).
    \textbf{f}) Boxplot of median normalized mechanical state index inferred using BFI, BISM (without boundary conditions) or MSM, for WT region and E-cad KO region within the co-culture. n = 19 positions for each condition. 
    [Scale bar 70 $\mu$m (a, e), 10 $\mu$m (c). P-value from Mann-Whitney U-test (b,d)]
}
\label{fig:mixed}
\end{figure}

Laser ablation of groups of WT or E-cad KO cells experimentally confirmed their different mechanical \textit{i.e.} compression and tension, respectively (Fig.~\ref{fig:mixed}c,d; \cite{Schoenit2025}). This demonstrates that those mechanical states were correctly inferred by BISM without knowing the right boundary conditions. 
The mixed tissue represent a simple system of mechanical compartmentalisation with distinct mechanical states of the two cell populations.

\subsection{Comparison with other inference methods}
\label{sec:hetero:comp}

Other non-invasive inference methods are monolayer stress microscopy (MSM)
\cite{Tambe2013}, 
and cell shape–based Bayesian Force inference (BFI). Different BFI techniques are available \cite{Borges2024} and we used the one introduced by Ishihara and Sugimura \cite{Ishihara2012}, which was validated by laser ablation  \cite{Ishihara2012, Ishihara2013, Kong2019}.
To compute stresses using MSM (Fig.~\ref{fig:mixed}e), we used the same traction force data as used for BISM in Fig.~\ref{fig:mixed}a,b. We additionally segmented the corresponding phase-contrast images to apply BFI (Fig.~\ref{fig:mixed}e).

As discussed in Sec.~\ref{BC:unconfined}, BISM may fail to recover the absolute value 
of isotropic stress when stress-free boundary conditions cannot be imposed.
MSM may fail when the rheology of the tissue is more complex than elastic \cite{Nier2016}.
A cell-shape based method such as BFI cannot measure absolute stress \cite{Ishihara2012}.
BFI reports line tensions and relative pressure for single cells, MSM and BISM report 
continuous stress fields. Batchelor's stress tensor allows to compute tissue-scale stresses 
from the pressures and tensions inferred by BFI \cite{Ishihara2012,Ishihara2013}, whereas 
the spatial resolution of both BISM and MSM is limited only by that of TFM, which may
reach down to sub-cellular scales. 
Since the pressure differences reported by BFI and the isotropic stress differences 
reported by MSM and BISM
contain similar information about the spatial pressure fluctuations within the tissue, they may be biologically interpreted in a similar way.

To enable meaningful comparison across methods, we focused on the simplest qualitative 
measure: the normalized mechanical state index $M$, defined as a dimensionless, but signed, metric of isotropic mechanical state:
\begin{equation}
M(x,y) = \frac{\sigma^{p}_{\mathrm{iso}}(x,y)}{\widetilde{\sigma ^p}}
\end{equation}
at each spatial position $(x,y)$, with $\sigma^{p}_{\mathrm{iso}}$ being the considered isotropic stress equivalent parameter (isotropic stress for BISM or MSM, negative relative inferred pressure for BFI) and $\widetilde{\sigma ^p}$ defined as the median of the corresponding absolute value of the isotropic stress equivalent parameter within the field of view for each cell type.
This allows to assess, if the different methods can report the mechanical state of the cells validated by laser ablation (Fig.~\ref{fig:mixed}d).

Our analysis revealed that all the inference methods used report the correct change in mechanical state (Fig.~\ref{fig:mixed}f). BFI is a relative measurement technique, as it detects only changes in pressure rather than absolute values. Therefore, the physical meaning of positive or negative values is lost. What should be considered instead is the magnitude of the difference between E-cad KO and WT regions, which is consistent with the expected transition from tension to compression. While BFI can not definitively indicate if the monolayer is under tension or compression in absolute terms, it clearly captures the biphasic mechanical behavior revealed by the laser ablation experiments. In the case of MSM, although it does not indicate that WT cells are under compression on average, it does capture a differential mechanical state between E-cad KO and WT regions. 
Importantly, BISM successfully detected both the difference in mechanical state and the sign: it correctly identified that WT cells are, on average, under compression, while E-cad KO cells are under tension (Fig.~\ref{fig:mixed}f).

\subsection{\textit{Ex vivo} tissue}

\begin{figure}[!t]
\includegraphics[width=1\linewidth]{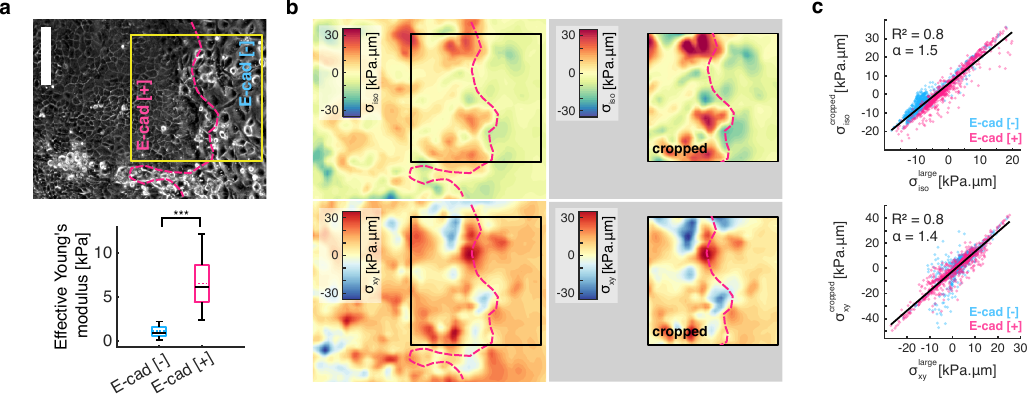}
\caption{
\textbf{Application of BISM in heterogeneous \textit{ex vivo} tissue:} 
    \textbf{a}) Top : Brightfield image showing a patient-derived metaplastic breast cancer cultivated on a flat substrate. The two cellular sub-populations (epithelial-like E-cad [+], pink; mesenchymal-like E-cad [-], blue) sort completely. Bottom :  Comparison of cell stiffness represented by their Young's modulus, measured with nano-indentation. n = 57 cells of each kind, ; N = 2 independent experiments. Data from \cite{Schoenit2025}.
    \textbf{b}) Isotropic stress field $\sigma_{\mathrm{iso}}$ and shear stress field $\sigma_{\mathrm{xy}}$ corresponding to \textbf{a}, computed either from tractions of the whole field of view (left) or a cropped region (right).
    \textbf{c}) Comparison of stresses measured in the cropped region, computed using traction forces from the whole field of view (large) or only the cropped region (cropped). Top : isotropic stress, $\sigma_{\mathrm{iso}}^{\mathrm{large}}$ vs $\sigma_{\mathrm{iso}}^{\mathrm{cropped}}$, bottom : shear stress, $\sigma_{xy}^{\mathrm{large}}$ vs $\sigma_{xy}^{\mathrm{cropped}}$. The black line represents the corresponding linear regression across all point pairs, regardless of cell type. The line coefficient $\alpha$ and the coefficient of determination $R^2$ are given. Pink indicates E-cad [+] cells, and blue indicates E-cad [–] cells. 
    [Scale bar 150 $\mu$m (a).]
}
\label{fig:mixed_tumor}
\end{figure}

To demonstrate the applicability of BISM to highly heterogeneous \textit{ex vivo} tissues, we analysed stresses within a patient-derived metaplastic breast cancer which consists of two sub-populations of cells \cite{Schoenit2025}: one displays an epithelial phenotype and strongly expresses E-cadherin (E-cad [+]), the other shows a mesenchymal phenotype, where E-cadherin is absent (E-cad [-]) and cells show only little adhesion. This heterogeneity in phenotype is reflected in different mechanical properties, which include cell-cell and cell-substrate adhesion \cite{Schoenit2025}, average cell height, and a five-fold contrast in cell stiffness (Fig.~\ref{fig:mixed_tumor}a). Those mechanical differences led to cell sorting, creating an internal tissue boundary. Cultivated on a flat substrate, the E-cad [+] cells expand, thereby pushing on the E-cad [-] cells. Thus, we expect that the E-cad [+] cells are under tension, as is common for expanding epithelial cells \cite{Vincent2015}, and apply compression on the E-cad [-] cells. Applying BISM, we observed that these population dynamics are reflected in the stress field, which indeed shows compressive stresses in the shrinking E-cad [-] population and large tensile stresses within the expanding E-cad[+] population (Fig.~\ref{fig:mixed_tumor}b, \cite{Schoenit2025}). 

Since our description is two-dimensional, the monolayer height is implicitly assumed 
to be constant. However, tissue height might vary in space and time, \textit{e.g.}
in heterogeneous tissues or when epithelial cells mature and
take a columnar shape (see Sec.~\ref{methods:height} for an explicit
treatment of height variations within the framework of BISM).
An heterogeneous tissue such as presented in Fig.~\ref{fig:mixed_tumor}
displays a well-defined, sharp interface: we thus expect that
height variations will remain small far enough from the interface.
Therefore, BISM inference away from the interface
may be applied within the usual assumption of constant height.
In the case of varying height, the height field appears in the (3D)
force balance equation as a positive multiplicative term (see Eq.~\eqref{eq:BISM:3D}).
As a consequence, a change of mean height will only
affect the amplitude of the inferred stresses, and
leave the sign of the isotropic stress unchanged.

Investigating dynamic processes at the tissue boundary with high spatial and temporal resolution might require data acquisition in a smaller field of view. Therefore, we cropped the original field of view and performed BISM on a subdomain of the tissue. Albeit we noticed that positive and negative stress values increased, we found a qualitatively good agreement between the cropped and the large field of view (Fig.~\ref{fig:mixed_tumor}c). Importantly, the mechanical compartmentalisation between the cell types indicated by the sign of the isotropic stress (E-cad [+] under tension, E-cad [-] under compression) was preserved (Fig.~\ref{fig:mixed_tumor}c). Together, these experiments highlight the power of BISM in robustly inferring stresses in heterogeneous tissues, even if cells show fundamental differences in mechanical properties, including their elastic modulus. Furthermore, experiments with \textit{ex vivo} tissue demonstrate that BISM can be in principle applied to any biological tissue, independent of the cells in contains, provided that it can be cultivated on a flat substrate and traction forces can be measured.

\section{Discussion}
\label{disc}

In this work, we reviewed and extended the domain of application of Bayesian Inversion
Stress Microscopy, with an emphasis on experimental validations and on
how various boundary conditions may affect the accuracy of BISM-based stress inference.
BISM relies on a Gaussian statistical model.
Offering high temporal and spatial resolutions, en par with the
underlying traction force measurements, BISM allows for an absolute
measure of all stress components upon the implementation of
stress-free boundary conditions.  
Importantly, the only biological requirement is that cells are in physical contact with each other, \textit{i.e.} that they form a continuous tissue. This always allows force transmission, even if cell-cell junctions as the main cellular mechanostransductive machinery \cite{Ladoux_2017} might be impaired \cite{Peyret2019} or display heterogeneity between sub-populations \cite{Schoenit2025}.

Many experimental conditions investigated with BISM are characterized by a state of
mechanical tension. In the quasi-1D case studied in Sec.~\ref{BC:front},
the force balance equation reduces to 
\begin{equation}
    \label{eq:1D}
    \frac{\mathrm{d} \sxx}{\mathrm{d} x} = t_x\,.
\end{equation}
Assuming that the tissue occupies the half-space to the right of a boundary at $x = 0$,
Eq.~\eqref{eq:1D}  may be integrated from the boundary at $x = 0$ towards the inside of the monolayer as:
\begin{equation}
    \label{eq:1D:int}
    \sxx(x) - \sxx(0) = \int_0^x t_x(x') \, \mathrm{d}x'\,.
\end{equation}
The stress-free boundary condition at $x = 0$ reads $\sxx(0) = 0$.
For inward pointing traction forces $t_x > 0$, Eq.~\eqref{eq:1D:int} shows that this tissue
is in  a state of tension $\sxx > 0$. 
This simple 1D example, which may be generalized to the 2D case, allows to explain why 
inward-pointing traction force vectors lead to a tissue under tension,   with a positive 
isotropic stress that may build up towards the inside.
For a given geometry, larger traction force amplitudes will lead to a larger tension. 
 
Most experimental studies
using BISM-based stress inference have so far focused on the
measurement and interpretation of isotropic stress.
However, deviatoric stress components (see Table~\ref{tab1}) are also estimated
reliably by BISM, and deserve more attention as a complementary
assessment of the mechanical state of the tissue.
Since it makes no assumption on the rheology of the tissue,
BISM represents a possible gateway into tissue rheology
in physiological conditions.
As it relies on a (Gaussian) statistical assumption,
we conversely do not recommend using BISM to probe the statistical
properties of tissue stress. 

When the mean isotropic stress is zero, as it is the case for large MDCK monolayers,
BISM accurately reports absolute isotropic stresses 
even when boundary conditions are unknown.
If this is not the case, foregoing stress-free boundary conditions means that 
the measure of tissue stress components becomes relative, \textit{i.e.} valid
up to an unknown additive constant.
For tissues fundamentally different to MDCK monolayers, and in cases where the absolute value of the isotropic stress are desired, we advise to probe the potential existence of such an additive constant by, \textit{e.g.}, performing  laser ablation.
Our comparative study of heterogeneous tissue stress
confirmed that BISM is a robust and reliable method for inferring stresses in heterogeneous epithelial tissues. BISM was able to detect the changes in mechanical state between WT and E-cad KO regions, and to recover the sign of the isotropic stress in agreement with the ground truth provided by laser ablation. Thus, BISM captures both the qualitative and quantitative aspects of the mechanical environment within the tissue. Its non-invasive nature, combined with its ability to compute the whole stress tensor, makes BISM a powerful tool for probing tissue mechanics in complex biological systems.

Admittedly, our approach when extending the scope of BISM from confined, rectangular
domains to various geometries and boundary conditions has been
mostly empirical, using numerical and experimental validations
in a step-wise manner. However a number of the above observations
call for a theoretical explanation of the influence of boundary
conditions on Bayesian inversion accuracy. 

Since BISM solves a 2D problem, the 3D structure of the tissue
is overlooked by construction. Therefore, only 2D stresses are inferred. Since stresses are effectively averaged over the cell height, potential cell-internal stress differences, \textit{e.g.} tension in the apical and compression in the basal part, can currently not be resolved.
On a planar substrate, normal components of the stress may be
inferred from 2.5D TFM data, which also measures the component
of traction forces normal to the substrate \cite{DelanoeAyari2023}. Inferring such stress components might be of particular relevance when studying out-of-plane processes, like orthogonal/asymmetric cell division, cell extrusion or the formation of multilayered tissue. 
The 3D geometry of tissues like the intestine can be reproduced by microfabrication approaches \cite{Xi2022, Gjorevski2022}. Thus, albeit experimentally currently limited by the lack of appropriate TFM protocols, an extension of the stress inference to curved substrates would be of great interest.
These important questions will be addressed by future developments.

\section{Conclusion}
\label{conclusion}

Bayesian Inversion Stress Microscopy allows to reliably infer
the mechanical stress tensor of a cell monolayer from
the traction forces it exerts on its substrate.
Two requirements for its use are the relevance of a continuum description
of the monolayer and the presence of a flat substrate. If these two are met,
BISM can be readily applied to tissues with different boundary conditions,
which makes it now possible to access absolute stresses in previously
inaccessible experimental systems. Importantly, BISM is not limited to
epithelia, but can be applied to any tissue which might consist of
multiple different cell types, \textit{e.g.} mixtures of epithelial
and mesenchymal cells. 
We hope that applying BISM to such tissues will help uncovering of the role of stresses in regulating tissue interactions related to morphogenesis or disease progression to answer outstanding biological questions.
For example, stresses remain largely unexplored in complex collective migration scenarios occurring during morphogenesis or invasion, where cell populations migrate through areas occupied by other cells \cite{friedl2009}. As this process is accompanied by cell deformations \cite{Barriga2019}, one generally expects local cell mechanics to be strongly affected, and perhaps to govern tissue interactions. 
Going forward, it will be exciting to see how BISM can continue to contribute to unravel the role of stresses in regulating cellular behaviour and shaping tissue form and function.

\section{Methods}
\label{methods}

\subsection{Experimental methods}
\label{methods:exp}

\subsubsection{Substrate preparation}
Soft silicone substrates for traction force microscopy were prepared as described before \cite{Nier2016, Saw2017, Peyret2019, Schoenit2025}. To obtain an elastic modulus of 15 kPa, the two components CY52-276A and CY52-276B polydimethylsiloxane (PDMS, Dow Corning Toray) were mixed in a weight ratio of 1:1, poured on a plastic dish to obtain a flat layer and cured at 80\textsuperscript{o}C for 2h. The stiffness of the substrate was verified using nanoindentation.

The substrate was silanized  using a solution of 5$\%$ (3-aminopropyl)triethoxysilane (Sigma-Aldrich) diluted at 10$\%$ in absolute ethanol for 10 min. After washing 3x with ethanol, the sample was dried at 80$^\circ$C. Then, 200 nm red carboxylated fluorescent beads (FluoSpheres, Invitrogen) were diluted at a 2:1,000 ratio in water, sonicated for 10 min, filtered using a $0.22 \,\mu$m filter and incubated on the substrates for 15 min, protected from light. After a final 3x wash with water, the substrate was dried at 80$^\circ$C. Before cell seeding, it was coated with 50 $\mu \mathrm{g.ml^{-1}}$ fibronectin (Sigma) for 45 min and washed three times with PBS.

\subsubsection{Micropatterning}
PDMS stamps for micropatterning were prepared as described previously \cite{Nier2016, Saw2017, Peyret2019, Schoenit2025}. Moulds of the desired pattern were obtained using standard lithography methods. PDMS (SYLGARD 184, Dow Corning) was prepared by mixing the base with a curing agent at a ratio of 1:10, poured over the mould, degassed and then cured at 80$^\circ$C for 2 h. Stamps were peeled of the mould and stored, protected from light and humidity. On utilization, the stamp surface was activated using plasma cleaning to make it hydrophilic, and a mixture of Cy3-conjugated fibronectin and regular fibronectin ($50 \,\mu \mathrm{g.ml^{-1}}$, Sigma) was then incubated covering the whole surface for 45 min. Then, the surface was cleaned using a gentle air flow. The stamps were gently pressed against the bottom of the PDMS substrate for about 1 min. After rinsing, the integrity of the patterns was verified using epifluorescence microscopy (Nikon). Patterns were then incubated with a solution of 2\% Pluronic F-127 (Sigma) for 1 h to passivate the areas outside the patterns.

\subsubsection{Cell culture and sample preparation}
HaCaT, MDCK-II (ATCC CCL-34), MDCK-II CAAX GFP, and MDCK-II E-cadherin knockout LifeAct-EGFP (clone B6P6) were cultured and maintained under standard conditions, as described previously \cite{Nier2016, Saw2017, Peyret2019, Schoenit2025, balasubramaniam2024different}. In brief, they were cultured at 37$^\circ$C with 5\% CO\textsubscript{2}, grown in DMEM (GlutaMax, Life Technologies) supplemented with 10\% FBS (Life Technologies) and 1\% penecillin-streptomycin and passaged every 2-3 days using 0.005\% Trypsin (Merck).

For experiments on fibronectin-coated micropatterns, a sufficient number of cells was seeded to directly cover the pattern. After letting the cells adhere for a few hours, the sample was rinsed to remove non-adhering cells. For cell competition experiments, MDCK WT and E-cadherin KO were mixed in a 1:1 ratio and seeded to obtain a confluent monolayer.

\subsection{Ex vivo culture of tumor-patient-derived xenografts}
Breast-cancer-patient-derived xenografts were obtained from triple-negative metaplastic breast tumours (HBCx-60 and HBCx-90) and generated as described previously \cite{Schoenit2025}. Tissues were surgically removed and cut into small pieces and digested in RPMI 1640 medium supplemented with 4 $\mathrm{mg.ml^{-1}}$ collagenase (Sigma-Aldrich) in 10 mM HEPES, 5\% FBS, penicillin–streptomycin (1×) and glutamine (1×) for 1h at 37$^\circ$C on a rotating wheel at 150 rpm. Tumouroids were pelleted at 400g for 10 min. Then, the tumouroids were incubated for 3-5 min at room temperature in DMEM/F-12/DNase (2 U $\mu \mathrm{l}^{-1}$). Tumouroids were pelleted again at 400g for 10 min. To remove the fibroblasts, tumouroids were washed with DMEM/F-12 medium and centrifuged at 400g for 3s at room temperature until the supernatant was clear. Tumouroids were resuspended in DMEM with 10 mM HEPES, 5\% FBS, 5 $\mu \mathrm{g.ml^{-1}}$ insulin, 10 $\mathrm{ng.ml^{-1}}$ cholera toxin, 1 $\mathrm{mg.ml^{-1}}$ hydrocortisone, penicillin–streptomycin (1×) and glutamine (1×). Hundreds of tumouroids were plated on a fibronectin-coated substrate. The tumoroids spread, leading to the formation of a monolayer and sorting of the cell types within.

\subsubsection{Data acquisition}
Prior to the experiment, samples were washed to remove dead cells and debris and fresh culture medium was added. The dish was transferred to a live-cell epifluorescence microscope (Biostation IM-Q, Nikon) equipped with a 10x phase-contrast air objective and an incubation chamber. Phase contrast images of the cells and fluorescent images of the beads were acquired every 10 or 15 minutes. To generate high-quality traction force data, special care was given to correct focus and brightness of the beads. It is advised to not acquire data of areas where beads are either sparse or aggregating or where the substrate is uneven. In the case of large micropatterns, the whole pattern was imaged using multiple positions, which were stitched together afterwards using ImageJ. Cells were removed at the end of the experiment by adding 500 $\mu$l of 10 $\%$ sodium dodecyl sulfate (SDS). Images of the relaxed state of the beads on the substrate were usually acquired one hour after SDS addition and serves as the reference.

\subsubsection{Traction force microscopy}
High-quality traction force data is a necessity for correct stress inference. See Teo \textit{et al.} \cite{Teo2020} for an extensive TFM protocol. Here,beads were analysed to compute the traction forces, as described previously \cite{Nier2016, Saw2017, Peyret2019, Schoenit2025}. In brief, the bead images obtained during TFM manipulation were merged with the corresponding reference bead image taken after SDS treatment. The resulting stack of images was preprocessed using the Image Stabilizer and then the Illumination correction plug-in ImageJ. Displacement field of beads was obtained using PIVlab (v.3.08), a particle image velocimetry toolbox within \texttt{matlab}, with an interrogation window of 32×32 pixels and an overlap of 50$\%$. Bead displacements were then correlated to a traction force field using Fourier transform traction cytometry, a known theoretical substrate stiffness and a regularization parameter of 9×10\textsuperscript{-9}. See Teo \textit{et al.}

\subsubsection{Laser ablation}

MDCK WT cells were seeded on a glass-bottom imaging dish and grown to confluency. To generate heterogeneous tissues, MDCK WT and E-cad KO cells were seeded at a ratio of 50:50 and grown until reaching confluency such that large islands of each cell type could be observed. 
Before wound induction, dishes were rinsed using warm PBS and provided with a fresh culture medium. Laser ablation was done using a spinning-disc CSU-X1 with a fluorescence recovery after photo-bleaching module (Yokogawa) and a ×40/1.2 water-immersion objective. Briefly, lines spanning 3–4 cells were ablated by focusing an ultraviolet laser (355 nm, pulse duration 3–5 ns, laser power 450 mW) for 1 s. For cell competition assays, holes of 3–4 cell diameters were generated within regions containing the same cell type in the mixture. Each sample was imaged during 15 s before ablation and until 3 min after ablation, using 5-s intervals. The recoil velocity was measured by manually segmenting the edge holes over time and plotting the change in displacement of the edges of the ablated region.

\subsubsection{Bayesian Force Inference}

The code used to implement Bayesian Force Inference is described in \cite{Kong2019}. Briefly, in terms of preparation and pre-processing, cell segmentation was performed using Cellpose \cite{stringer2025cellpose3} on brightfield images, and cell identities were assigned based on E-cad KO LifeAct fluorescence or m-GFP fluorescence.

\subsubsection{Monolayer Stress Microscopy}

The code used to implement Monolayer Stress Microscopy is available at \url{https://github.com/jknotbohm/Cell-Traction-Stress}, along with its associated documentation. The underlying methodology is described in \cite{Tambe2013}.

\subsection{Effect of height variation of the monolayer}
\label{methods:height}

By construction, the 2D description of the cell monolayer
overlooks possible height variations. In order to take them into account,
we define the tissue height $h(\vec{r})$
as a function of position $\vec{r}$ in the plane, and consider  
the 3D tissue stress tensor $\Sigma$, which satisfies the following
force balance equation ($i,j \in \{x,y\}$):
\begin{equation}
  \label{eq:forcebalance:3D}
\partial_j \Sigma_{ij} + \partial_z \Sigma_{iz}= 0 \,,  
\end{equation}
with the boundary conditions $\Sigma_{iz}(x,y,z = 0) = t_i(\vec{r})$,
$\Sigma_{iz}(x,y,z = h(\vec{r})) = 0$.
The 2D stress tensor $\sigma$ is deduced from $\Sigma$ by integrating
over the height:
$$
\sigma_{ij}(x,y) =  \int_0^{h(\vec{r})}  \Sigma_{ij}(x,y,z) \, \mathrm{d}z
  = h(\vec{r}) \, \hat{\Sigma}_{ij}(x,y) \,,  
$$
where $\hat{\Sigma}_{ij}$ denotes the average over $z$ of $\Sigma_{ij}$.
Integrating \eqref{eq:forcebalance:3D} over $z$ between $0$ and $h(\vec{r})$
yields $\int_0^{h(\vec{r})} \partial_j \Sigma_{ij} \, \mathrm{d}z = t_i$, or:
$$
\partial_j \sigma_{ij} - \left( \partial_j h(\vec{r}) \right)
\, \Sigma_{ij}(z = h(\vec{r})) = t_i \,.
$$
For small enough height, we may approximate the unknown quantity 
$\Sigma_{ij}(z = h(\vec{r}))$ by its mean value
$\hat{\Sigma}_{ij} = \sigma_{ij}/h$, and deduce:
$$
\partial_j \sigma_{ij} - \frac{\partial_j h}{h} \, \sigma_{ij} = t_i \,.
$$
This equation reduces to Eq.~\eqref{eq:forcebalance} for constant height
($\partial_j h(\vec{r}) = 0$).
Since  $\sigma_{ij} = h \, \hat{\Sigma}_{ij}$, we obtain:
\begin{equation}
  \label{eq:BISM:3D}
\partial_j \hat{\Sigma}_{ij} = \frac{t_i}{h} \,,  
\end{equation}
which is formally equivalent to \eqref{eq:forcebalance}.

Provided that the monolayer  height field $h(\vec{r})$ can be measured
experimentally, applying BISM to the data ratio $\vec{t}/h$ allows to
infer the average 3D stress components $\hat{\Sigma}_{ij}$,
$i,j \in \{x,y\}$, from which the 2D stress field can be computed
$\sigma_{ij} = h \, \hat{\Sigma}_{ij}$.

\backmatter

\bmhead{Acknowledgements}

The authors gratefully acknowledge the pivotal contributions made in the early stages of this project 
by Vincent Nier and Shuji Ishihara, and would like to thank Grégoire Peyret for the data used in 
Figs.~\ref{fig:confined:MDCK}, \ref{fig:subdomain:MDCK}, \ref{fig:confined:HaCaT} and
\ref{fig:subdomain:HaCaT}. We thank the members of the 'Cell Adhesion and Mechanics' team (Institut Jacques Monod) for helpful discussions, with a special mention to Joseph d'Alessandro. We acknowledge the ImagoSeine core facility of the Institut Jacques Monod, member of the France BioImaging infrastructure supported by the French National Research Agency (ANR-24-INBS-0005 FBI BIOGEN) and GIS-IBiSA.
We thank multiple reviewers on earlier publications where BISM was applied for their critical feedback.

\section*{Declarations}

\begin{itemize}
\item Funding. This work was supported by the European Research Council (grant no. Adv-101019835 “DeadorAlive” to B.L.), LABEX Who Am I? (ANR-11-LABX-0071 to B.L. and R.-M.M.), the Alexander von Humboldt Foundation (Alexander von Humboldt Professorship to B.L.), the Ligue Contre le Cancer (Equipe labellisée 2019 to R.-M.M. and B. L.), the CNRS through 80|Prime program (to B.L.), Institut National du Cancer INCa (‘Invadocad’, PLBIO18-236) and the Agence Nationale de la Recherche (‘MechanoAdipo’ ANR-17-CE13-0012 to B.L.). A.S. received funding from the CNRS 80|Prime program and the Fondation Recherche Medicale (FDT-202404018282). L.A. received funding from the Ligue contre le Cancer. F.W. acknowledges financial support from the LabEx "Who Am I?" (grant ANR-11-LABX-0071).
\item Conflict of interest/Competing interests. We have no competing interests.
\item Ethics approval and consent to participate. Not applicable.
\item Consent for publication. Not applicable.
\item Data availability. Example data to apply BISM is provided on \url{https://github.com/ASmushyCell/TFM-BISM-analysis}. All data supporting the findings of this study are available from the corresponding author upon reasonable request. 
\item Materials availability. Not applicable.
\item Code availability. The \texttt{matlab} script for BISM is provided on \url{https://github.com/pmarcq/BISM/}.
\item Author contribution. R.-M.M., B.L. and P.M. designed the research. 
L.A. and A.S. performed experiments. F.W. performed the laser ablation experiments on confluent MDCK monolayers. C.R. provided experimental data on tumor cells. L.A., A.S. and P.M. analysed data and drafted the manuscript.
All the authors approved the final version of the manuscript.
\end{itemize}


\bibliography{bism}

\newpage
\FloatBarrier

\subsection*{Supplementary information}


\medskip
Provided the analysis is carried out correctly, the code accompanying this paper ultimately allows access to all 2D components of the stress tensor, as well as the following invariants:

\setcounter{table}{0}
\renewcommand{\tablename}{Table}
\renewcommand{\thetable}{S\arabic{table}}

\begin{table}[h]
\caption{Output of the BISM analysis pipeline provided with this article}\label{tab1}%
\begin{tabular}{@{}l p{4.5cm} l l@{}}
\toprule
Field & Description  & Dimension & Expression\\
\midrule
stress\_xx    & $xx$-component of the stress tensor   & Pa.$\mu$m  & //  \\
stress\_yy    & yy-component of the stress tensor   & Pa.$\mu$m  & //  \\
stress\_xy \footnotemark[1]   & xy-component of the stress tensor   & Pa.$\mu$m  & // \\
stress\_M  & Maximum principal stress, the larger eigenvalue of the stress tensor & Pa.$\mu$m & $\frac{\sigma_{xx} + \sigma_{yy}}{2}$ + $\sqrt{(\frac{\sigma_{xx} - \sigma_{yy}}{2})^2 + \sigma_{xy}^2}$\\
stress\_m & Minimum principal stress, the smaller eigenvalue of the stress tensor & Pa.$\mu$m & $\frac{\sigma_{xx} + \sigma_{yy}}{2}$ - $\sqrt{(\frac{\sigma_{xx} - \sigma_{yy}}{2})^2 + \sigma_{xy}^2}$\\
stress\_iso \footnotemark[2] & Isotropic stress, opposite of a pressure  & Pa.$\mu$m & $\frac{\sigma_{xx} + \sigma_{yy}}{2}$ \\
stress\_aniso \footnotemark[3] & Anisotropic stress, also called maximum shear stress & Pa.$\mu$m & $\sqrt{(\frac{\sigma_{xx} - \sigma_{yy}}{2})^2 + \sigma_{xy}^2}$ \\
sigma\_VM & von Mises equivalent stress & Pa.$\mu$m & $\sqrt{\sigma_{xx}^2 + \sigma_{yy}^2 - \sigma_{xx}\sigma{xy} +3\sigma_{xy}^2}$ \\
angle\_stress\_x \footnotemark[4] & unit $x$-component of the principal stress orientation director field & 1 & $\tan 2 \theta = \frac{2\sigma_{xy}}{(\sigma_{xx} - \sigma_{yy})}$ \\
angle\_stress\_y \footnotemark[5] & unit $y$-component of the principal stress orientation director field & 1 & $\tan 2 \theta = \frac{2\sigma_{xy}}{(\sigma_{xx} - \sigma_{yy})}$ \\
\botrule
\end{tabular}
\footnotetext{See \url{https://github.com/ASmushyCell/TFM-BISM-analysis} for a more detailed explanation of each output.}
\footnotetext[1]{stress\_yx is equal to stress\_xy as by construction, the stress tensor is symmetric.}
\footnotetext[2]{stress\_iso is derived from the first invariant $I_1$ of the stress tensor (specifically, it corresponds to half the trace of the stress tensor).}
\footnotetext[3]{stress\_aniso is derived from the second invariant $J_2$ of the deviatoric stress tensor (specifically, it corresponds to the square root of $J_2$).}
\footnotetext[4]{$\cos{\theta}$ with $\theta$, the principal stress direction.}
\footnotetext[5]{$\sin{\theta}$ with $\theta$, the principal stress direction.}
\end{table}

\setcounter{figure}{0}
\renewcommand{\figurename}{Fig.}
\renewcommand{\thefigure}{S\arabic{figure}}

\begin{figure}[!b]
    \begin{center}
      \hspace*{0.2cm}\textbf{a}\hspace*{0.4cm}
      \includegraphics[width=0.21\linewidth]{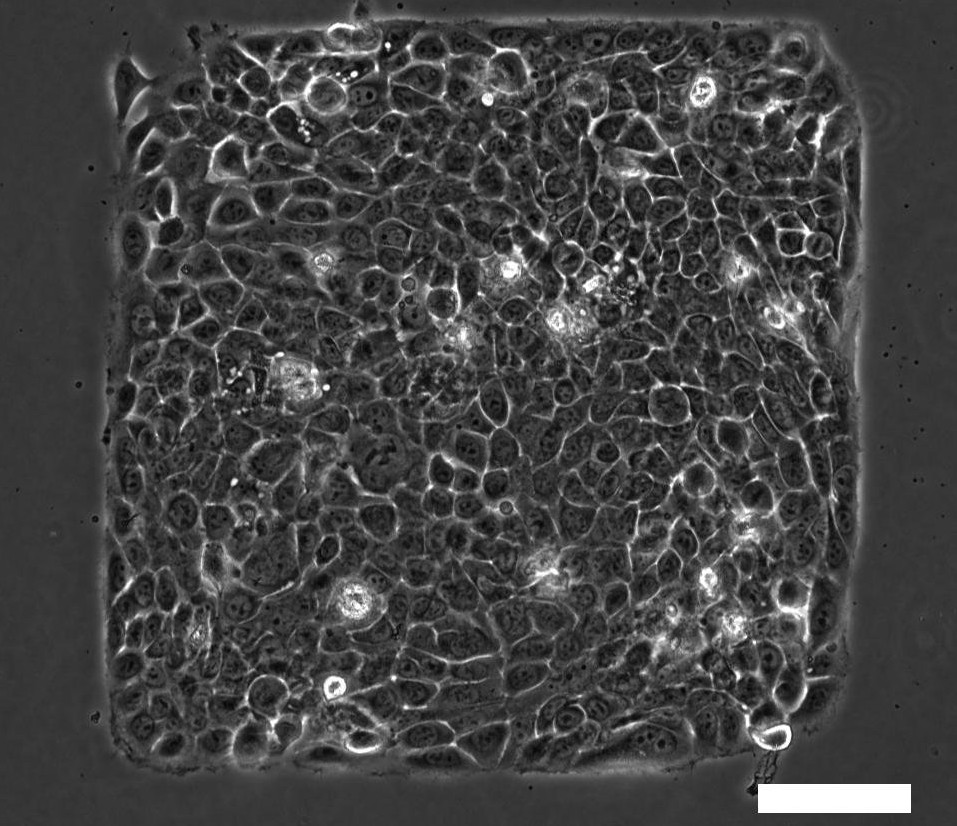}
   \hspace*{0.15cm} \textbf{b}
  \includegraphics[width=0.27\linewidth]{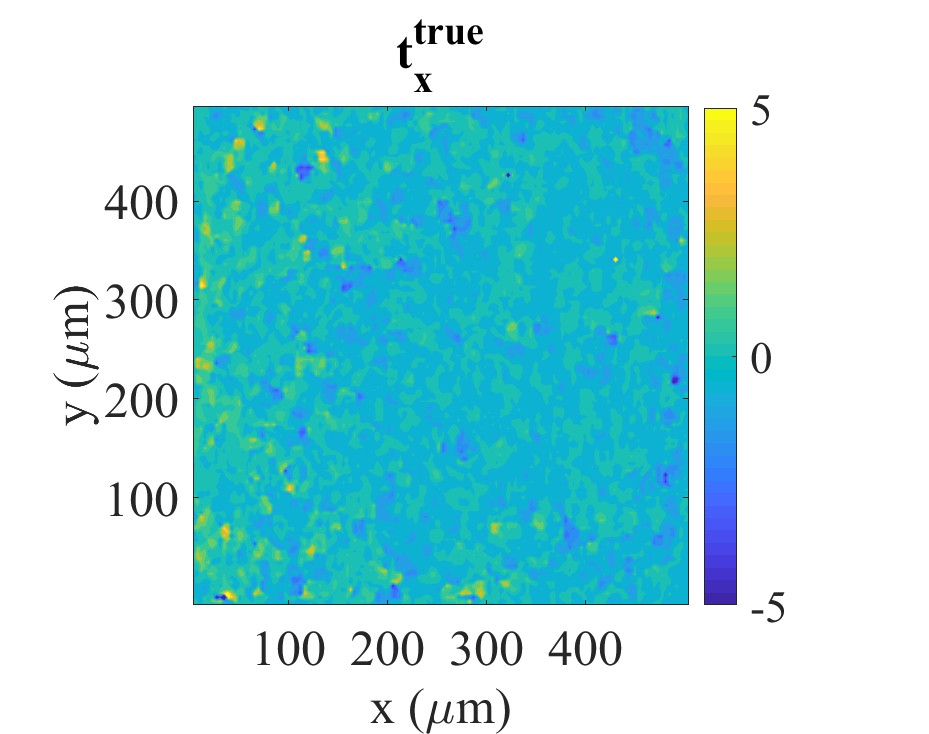} 
  \textbf{c}
  \includegraphics[width=0.27\linewidth]{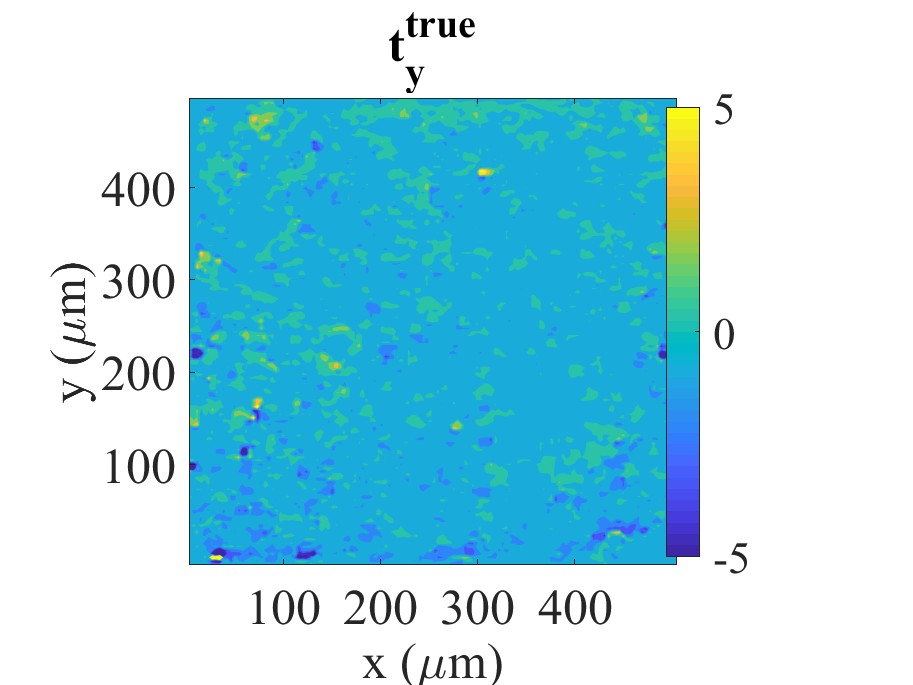}\\
  \medskip
  \textbf{d}
  \includegraphics[width=0.27\linewidth]{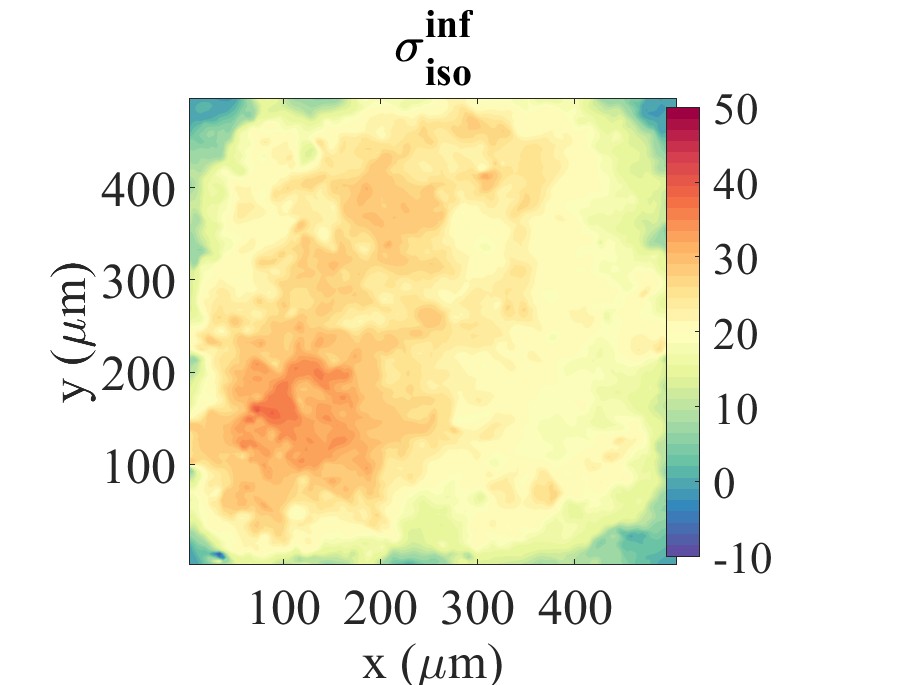} 
  \textbf{e}
  \includegraphics[width=0.27\linewidth]{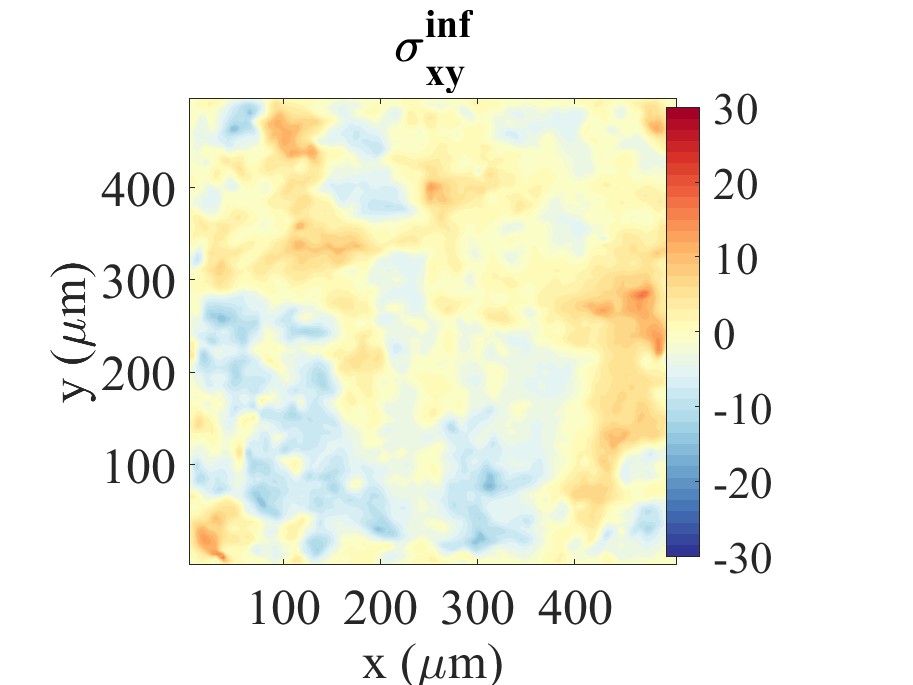}
  \textbf{f}
  \includegraphics[width=0.27\linewidth]{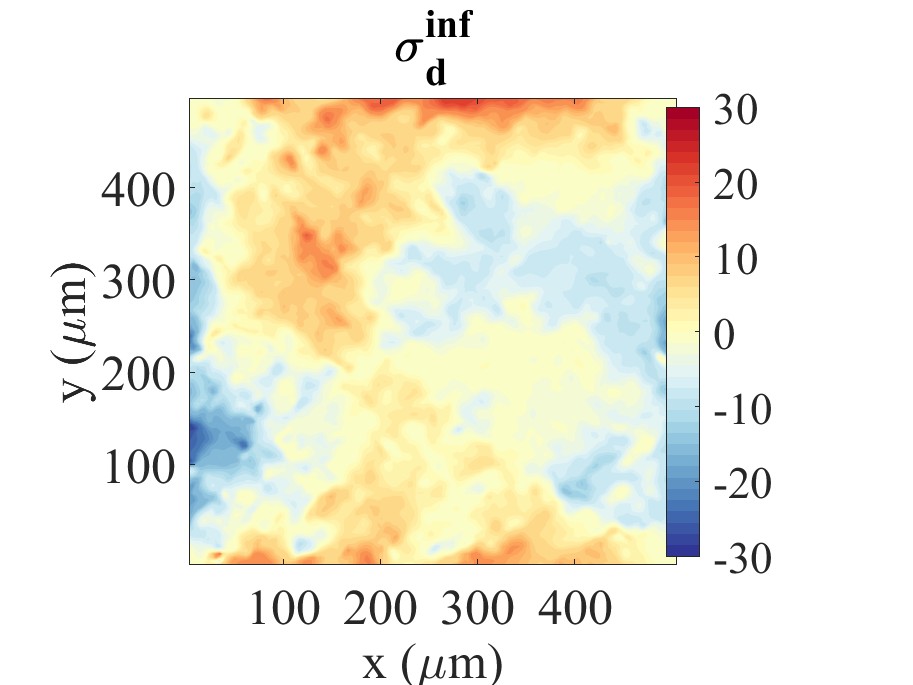} \\
  \smallskip
    \textbf{g}
  \includegraphics[width=0.27\linewidth]{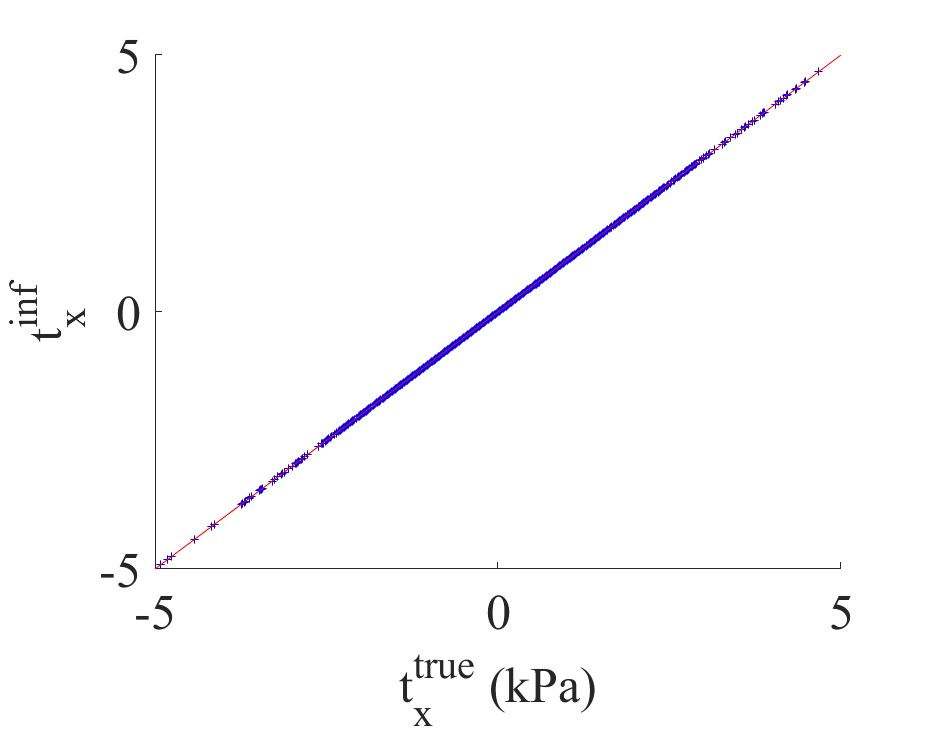}
  \textbf{h}
  \includegraphics[width=0.27\linewidth]{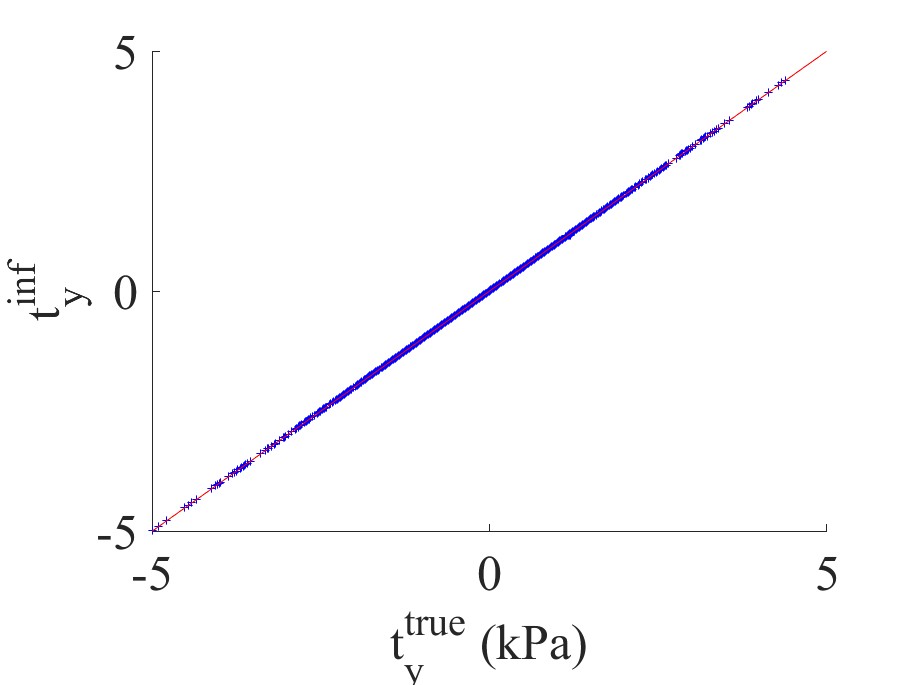}   
  \end{center}
\caption{
\textbf{Stress inference in a confined system:} 
HaCaT cell monolayer in a square domain of lateral extension
$L = 500 \,\mu$m \cite{Peyret2019}.
\textbf{a)} Phase contrast image of the confining domain.
\textbf{b, c)} Color maps of the
components of the traction force field
$t_x^{\mathrm{true}}$, $t_y^{\mathrm{true}}$ measured by TFM (in kPa).
\textbf{d, e, f)} Color maps of the inferred
isotropic stress $\sigma_{\mathrm{iso}}^{\mathrm{inf}}$ and
deviatoric stress components $\sigma_{\mathrm{xy}}^{\mathrm{inf}}$,
$\sigma_{\mathrm{d}}^{\mathrm{inf}}$ (in kPa.$\mu$m).
Here the zero-stress condition is imposed on the four boundaries.
\textbf{g, h)} Comparison of measured
($t^{\mathrm{true}}_x$, $t^{\mathrm{true}}_y)$
and inferred ($t^{\mathrm{inf}}_x$, $t^{\mathrm{inf}}_y$) values of
components of the traction force vectors (blue crosses),
computed from the inferred stress field
$\vec{t}^{\mathrm{inf}} = \mathrm{div} \, \sig^{\mathrm{inf}}$.
The bisectrix $y = x$ is plotted as a red line for comparison.
The coefficient of determination is $R^2_t = 1.0$.
We find excellent agreement between stress averages computed 
in the cell domain and the true values obtained from moments 
of the traction force field, thus confirming that BISM provides
an absolute measurement of stresses in confined domains: compare 
$  \langle \sigma_{\mathrm{iso}}^{\mathrm{true}} \rangle =  21.64$ kPa
and
$  \langle \sigma_{\mathrm{iso}}^{\mathrm{inf}} \rangle = 21.60$ kPa;
then $  \langle \sigma_{\mathrm{xy}}^{\mathrm{true}} \rangle = -895$ Pa
and
$  \langle \sigma_{\mathrm{xy}}^{\mathrm{inf}} \rangle = -896$ Pa;
and $  \langle \sigma_{\mathrm{d}}^{\mathrm{true}} \rangle = -58$ Pa
and
$  \langle \sigma_{\mathrm{d}}^{\mathrm{inf}} \rangle = -48$ Pa.
[Scale bar: $100 \,\mu$m (a).] 
}
\label{fig:confined:HaCaT}
\end{figure}

\begin{figure}[!t]
  \begin{center}
      \hspace*{-0.05cm}\textbf{a}\hspace*{0.6cm}
      \includegraphics[width=0.18\linewidth]{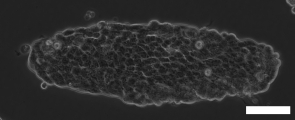}
   \hspace*{0.5cm} \textbf{b}
  \includegraphics[width=0.27\linewidth]{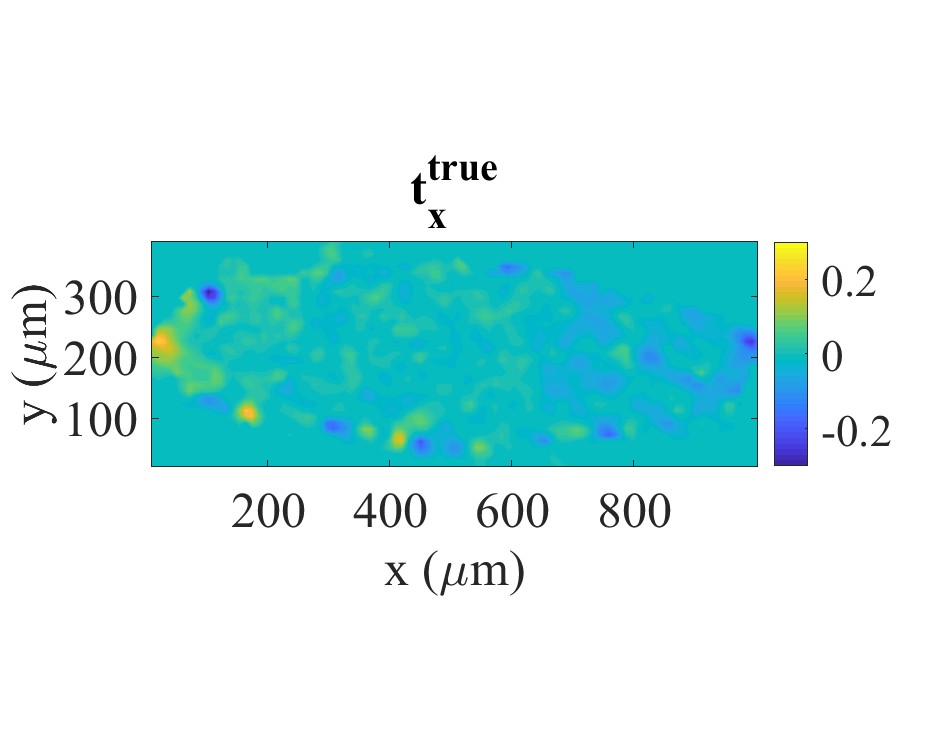} 
  \textbf{c}
  \includegraphics[width=0.27\linewidth]{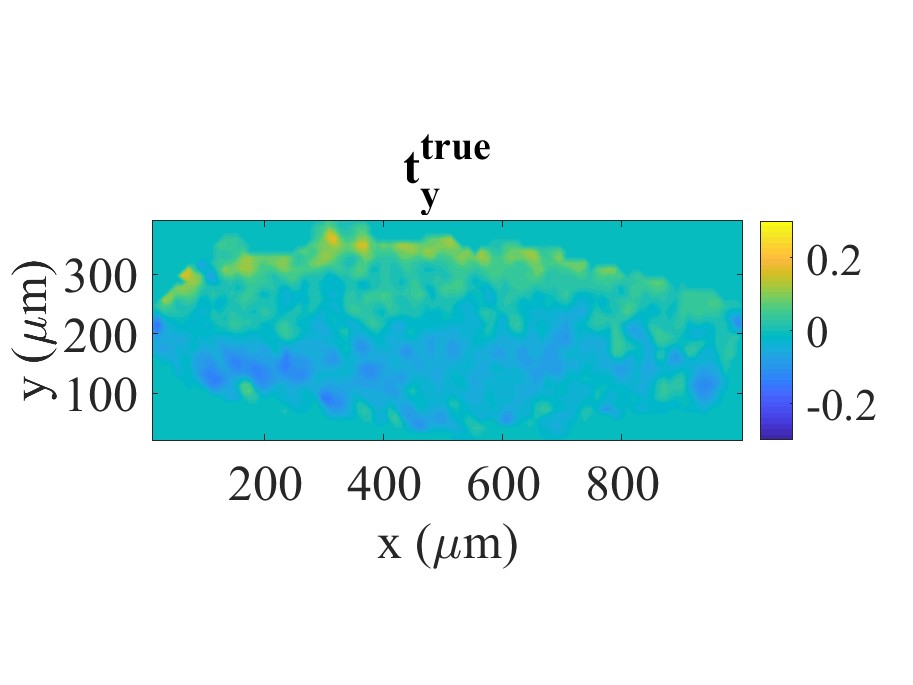}\\
  \medskip
  \textbf{d}
  \includegraphics[width=0.27\linewidth]{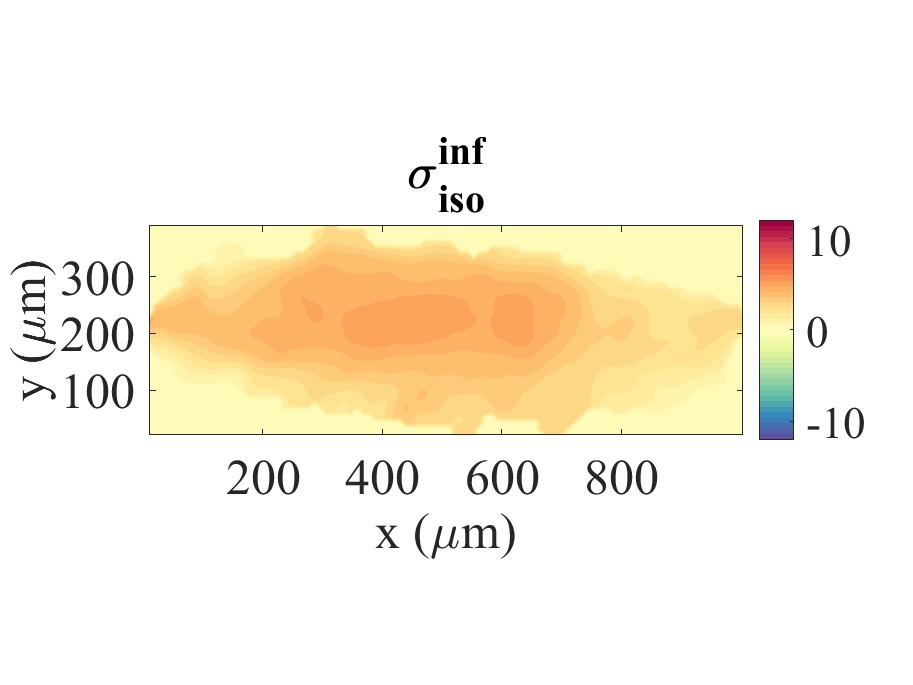} 
  \textbf{e}
  \includegraphics[width=0.27\linewidth]{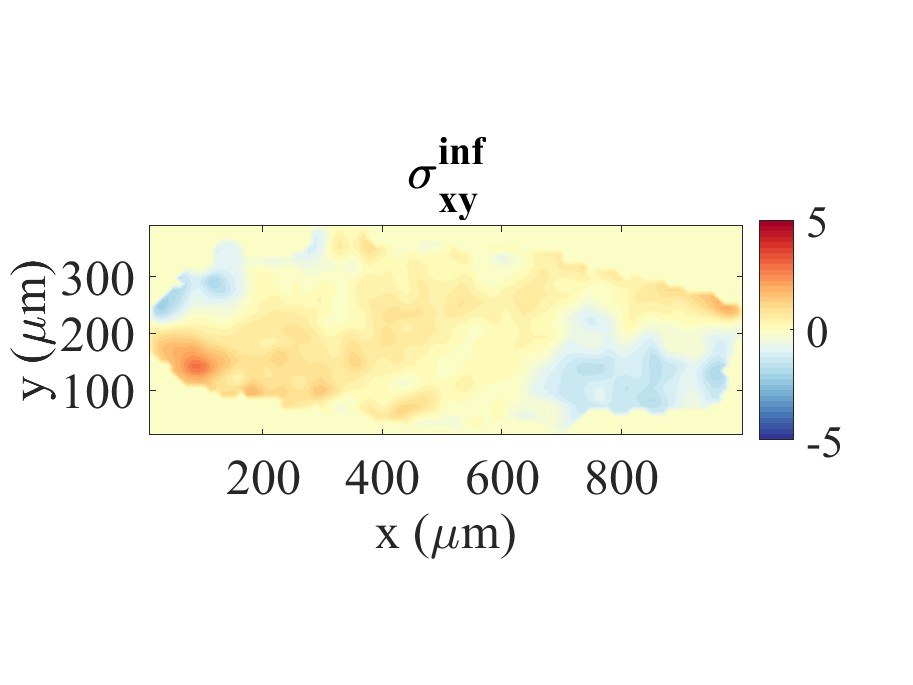}
  \textbf{f}
  \includegraphics[width=0.27\linewidth]{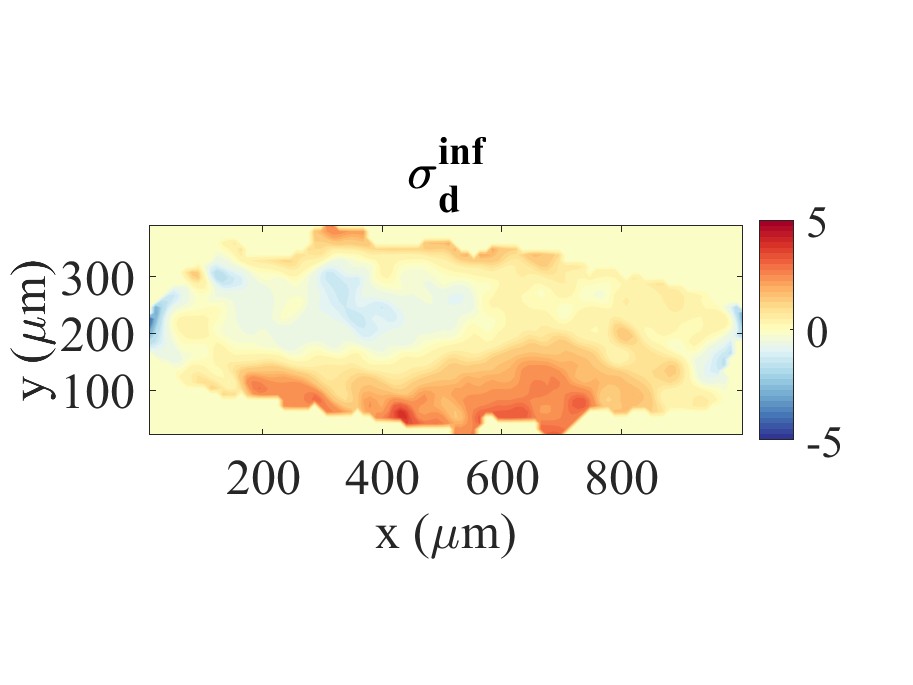} \\
  \smallskip
    \textbf{g}
  \includegraphics[width=0.27\linewidth]{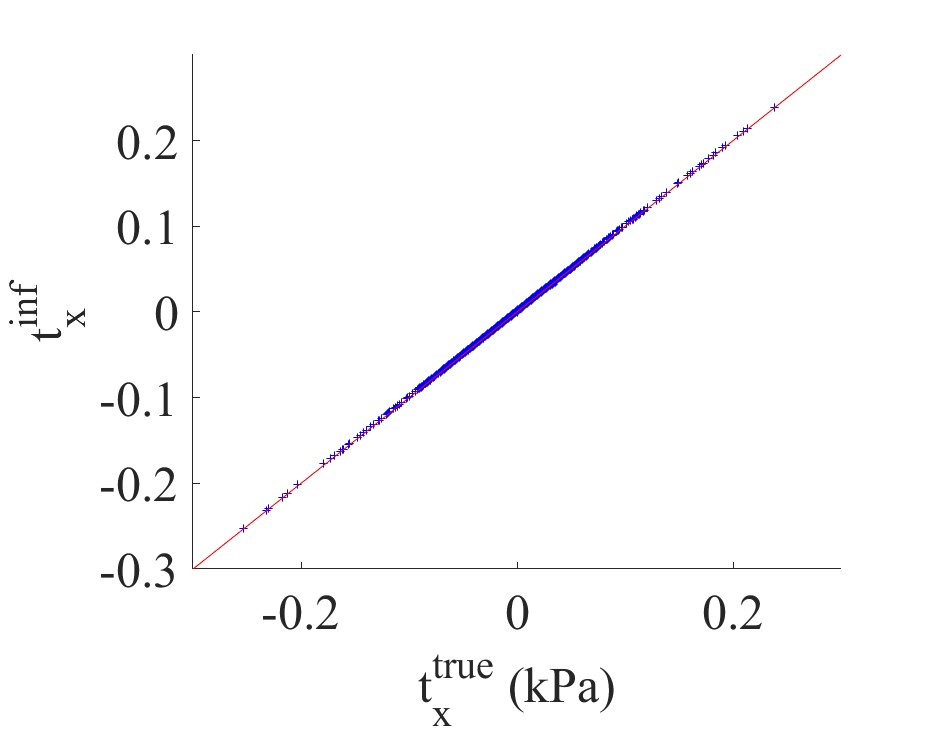}
  \textbf{h}
  \includegraphics[width=0.27\linewidth]{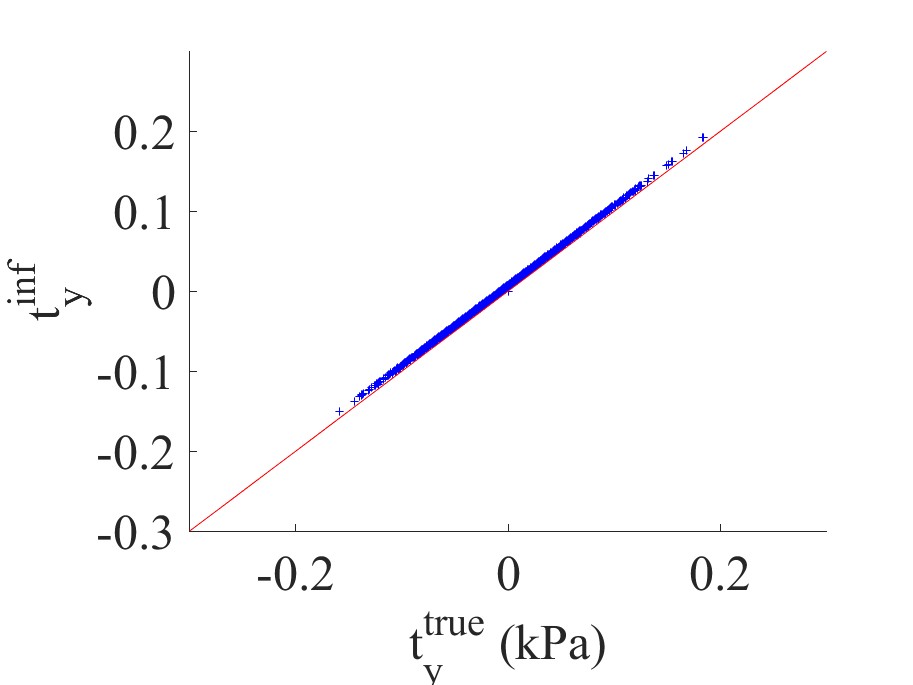}   
  \end{center}
  \caption{
\textbf{Stress inference in ellipse-shaped tissue:} 
MDCK cell monolayer.
\textbf{(a)} Phase contrast image of the cell island.
\textbf{(b, c)} Color maps of the
components of the traction force field
$t_x^{\mathrm{true}}$, $t_y^{\mathrm{true}}$ measured by TFM (in kPa).
\textbf{(c, d, e)} Color maps of the inferred
isotropic stress $\sigma_{\mathrm{iso}}^{\mathrm{inf}}$ and
deviatoric stress components $\sigma_{xy}^{\mathrm{inf}}$,
$\sigma_{d}^{\mathrm{inf}}$ (in kPa.$\mu$m).
Here the zero-normal stress condition
$\sigma_{ij} \, n_j=0$ is imposed on the four boundaries of the smallest
rectangular domain encompassing the cell domain.
\textbf{(g, h)} Comparison of measured
($t^{\mathrm{true}}_x$, $t^{\mathrm{true}}_y)$
and inferred ($t^{\mathrm{inf}}_x$, $t^{\mathrm{inf}}_y$) values of
components of the traction force vectors (blue crosses),
computed from the inferred stress field
$\vec{t}^{\mathrm{inf}} = \mathrm{div} \, \sig^{\mathrm{inf}}$.
The bisectrix $y = x$ is plotted as a red line for comparison.
The coefficient of determination is $R^2_t = 0.99$.
As in Fig.~\ref{fig:star}, we find excellent agreement between
stress averages computed in the cell domain and the true values
obtained from moments of the traction force field: compare 
$  \langle \sigma_{\mathrm{iso}}^{\mathrm{true}} \rangle = 3.91 $ kPa
and
$  \langle \sigma_{\mathrm{iso}}^{\mathrm{inf}} \rangle = 3.84$ kPa;
$  \langle \sigma_{xy}^{\mathrm{true}} \rangle = 53$ Pa
and
$  \langle \sigma_{xy}^{\mathrm{inf}} \rangle = 169$ Pa;
$  \langle \sigma_{d}^{\mathrm{true}} \rangle = 0.96$ kPa
and
$  \langle \sigma_{d}^{\mathrm{inf}} \rangle = 1.00$ kPa.
}
\label{fig:ellipse}
\end{figure}

\begin{figure}[!ht]
\includegraphics[width=1\linewidth]{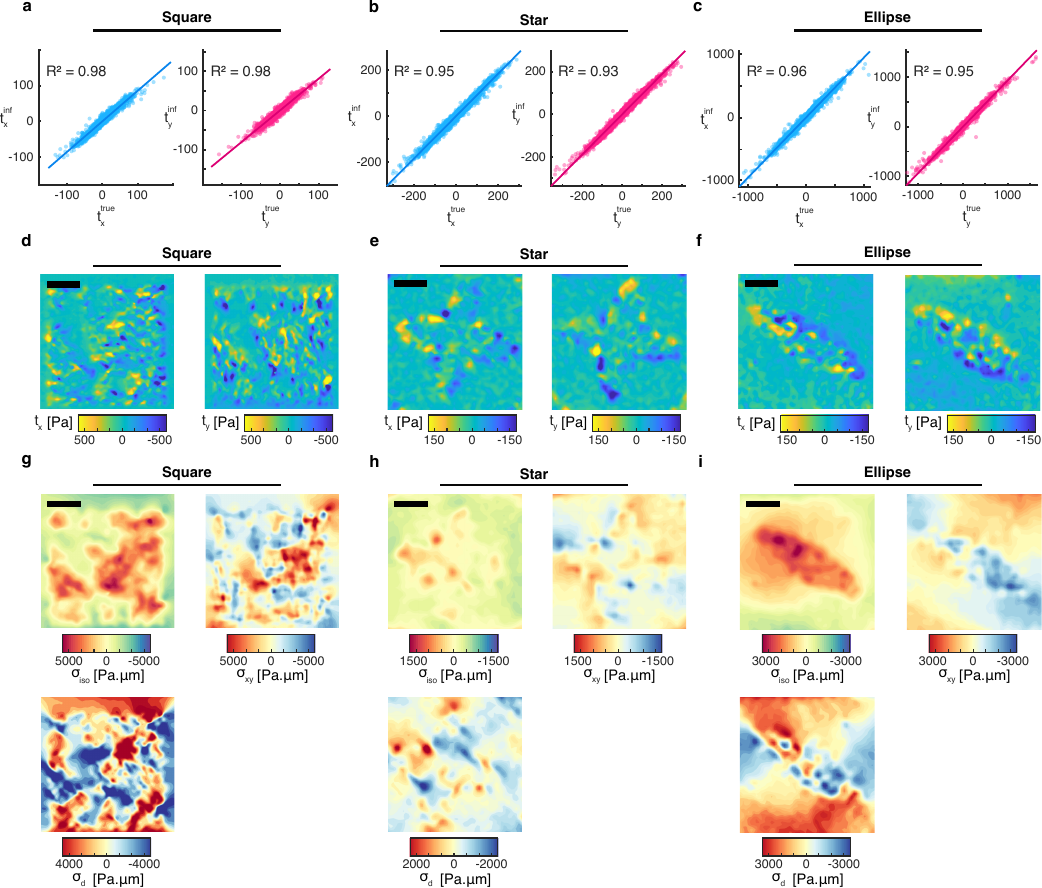}
\caption{\textbf{Stress inference in various pattern with minimum assumptions:}
    \textbf{a, b, c}) Comparison of measured
($t^{\mathrm{true}}_x$, $t^{\mathrm{true}}_y)$
and inferred ($t^{\mathrm{inf}}_x$, $t^{\mathrm{inf}}_y$) values of
components of the traction forces field,
computed from the inferred stress field
$\vec{t}^{\mathrm{inf}} = \mathrm{div} \, \sig^{\mathrm{inf}}$, for cell cultivated in a square (a), a star (b) or an ellipse (c) pattern.
    \textbf{d, e, f})  Color maps of the components of the traction force field
$t_x^{\mathrm{true}}$, $t_y^{\mathrm{true}}$ measured by TFM (in Pa), for cell cultivated in a square (d), a star (e) or an ellipse (f) pattern. 
    \textbf{g, h, i}) Color maps of the inferred
isotropic stress $\sigma_{\mathrm{iso}}^{\mathrm{inf}}$ and
shear stress components $\sigma_{xy}^{\mathrm{inf}}$ (in Pa.$\mu$m), for cell cultivated in a square (g), a star (h) or an ellipse (i) pattern.
Here no boundary condition has been imposed when computing the stress tensor. Square : $\langle \sigma_{\mathrm{iso}}^{\mathrm{all}} \rangle =$ 27.4 $\pm $ 1765 Pa, $\langle \sigma_{\mathrm{xy}}^{\mathrm{all}} \rangle =$ 1.8 $\pm $ 1566 Pa and $\langle \sigma_{\mathrm{d}}^{\mathrm{all}} \rangle =$ -11 $\pm $ 3956 Pa ; then $\langle \sigma_{\mathrm{iso}}^{\mathrm{pattern}} \rangle =$ 947 $\pm $ 1637 Pa, $\langle \sigma_{\mathrm{xy}}^{\mathrm{pattern}} \rangle =$ 111 $\pm $ 1791 Pa and $\langle \sigma_{\mathrm{d}}^{\mathrm{pattern}} \rangle =$ -489 $\pm $ 3813 Pa. Star : $\langle \sigma_{\mathrm{iso}}^{\mathrm{all}} \rangle =$ -20.2 $\pm $ 325 Pa, $\langle \sigma_{\mathrm{xy}}^{\mathrm{all}} \rangle =$ - 2.36 $\pm $ 317 Pa and $\langle \sigma_{\mathrm{d}}^{\mathrm{all}} \rangle =$ -15.1 $\pm $ 701 Pa ; then $\langle \sigma_{\mathrm{iso}}^{\mathrm{pattern}} \rangle =$ 22 $\pm $ 227 Pa, $\langle \sigma_{\mathrm{xy}}^{\mathrm{pattern}} \rangle =$ -32.7 $\pm $ 317 Pa and $\langle \sigma_{\mathrm{d}}^{\mathrm{pattern}} \rangle =$ -93.1 $\pm $ 650 Pa. Ellipse : $\langle \sigma_{\mathrm{iso}}^{\mathrm{all}} \rangle =$ 31.3 $\pm $ 948 Pa, $\langle \sigma_{\mathrm{xy}}^{\mathrm{all}} \rangle =$ - 9.6 $\pm $ 763 Pa and $\langle \sigma_{\mathrm{d}}^{\mathrm{all}} \rangle =$ 13.3 $\pm $ 1579 Pa ; then $\langle \sigma_{\mathrm{iso}}^{\mathrm{pattern}} \rangle =$ 1670 $\pm $ 631 Pa, $\langle \sigma_{\mathrm{xy}}^{\mathrm{pattern}} \rangle =$ -666 $\pm $ 718 Pa and $\langle \sigma_{\mathrm{d}}^{\mathrm{pattern}} \rangle =$ -135 $\pm $ 1308 Pa. The superscript "all" corresponds to
averages performed over the whole field of view, while "pattern" corresponds to averages performed over the 
corresponding pattern. 
 [Scale bar 150 $\mu$m (d, e, f, g, h, i). Uncertainty is computed as the standard deviation.]
     }
\label{fig:BC0}
\end{figure}

\begin{figure}[!t]
  \begin{center}
      \textbf{a.}
  \includegraphics[width=0.27\linewidth]{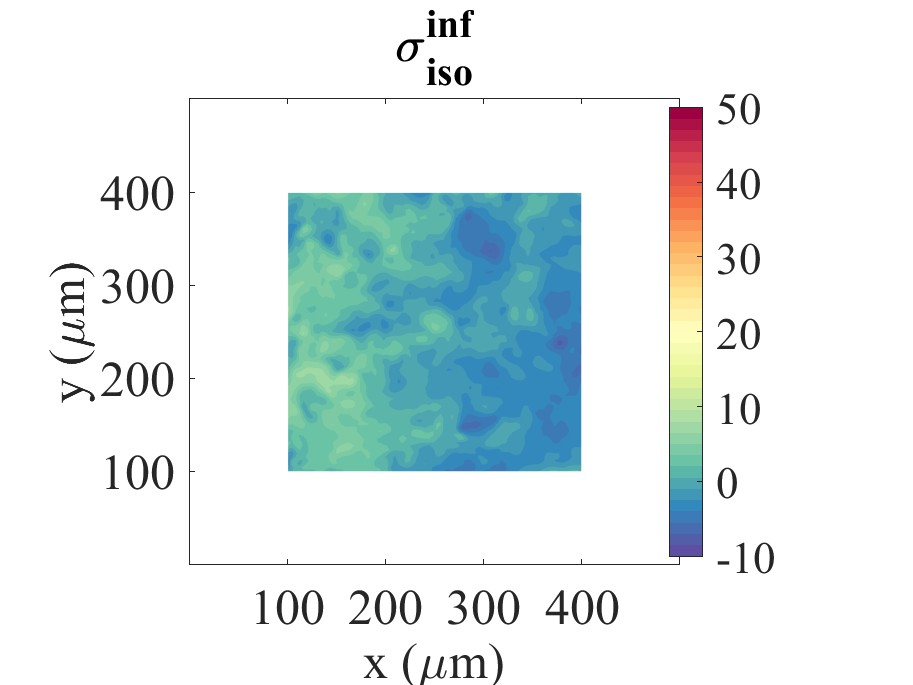} 
  \textbf{b.}
  \includegraphics[width=0.27\linewidth]{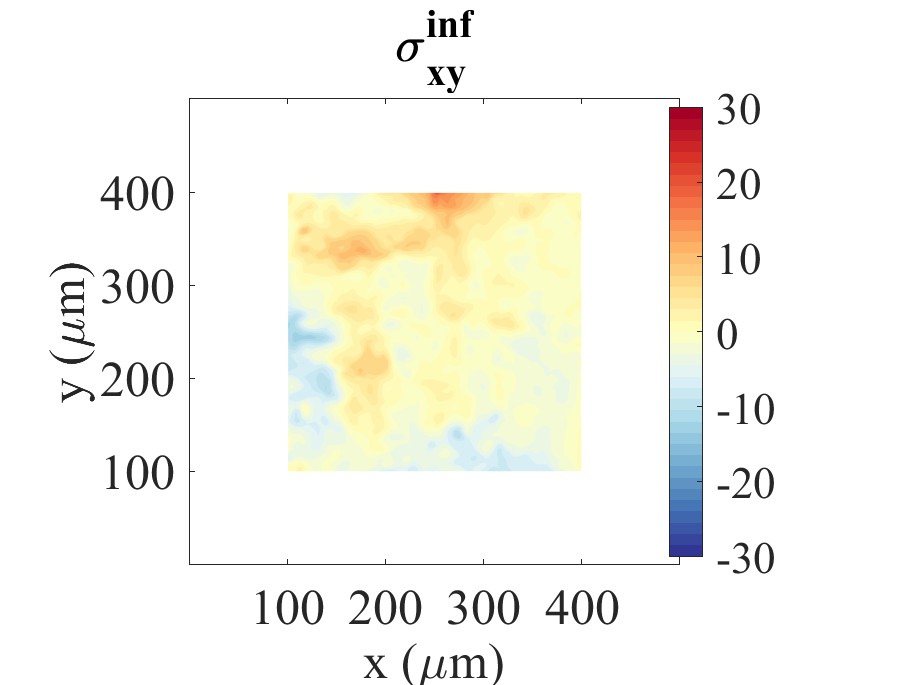} 
  \textbf{c.}
  \includegraphics[width=0.27\linewidth]{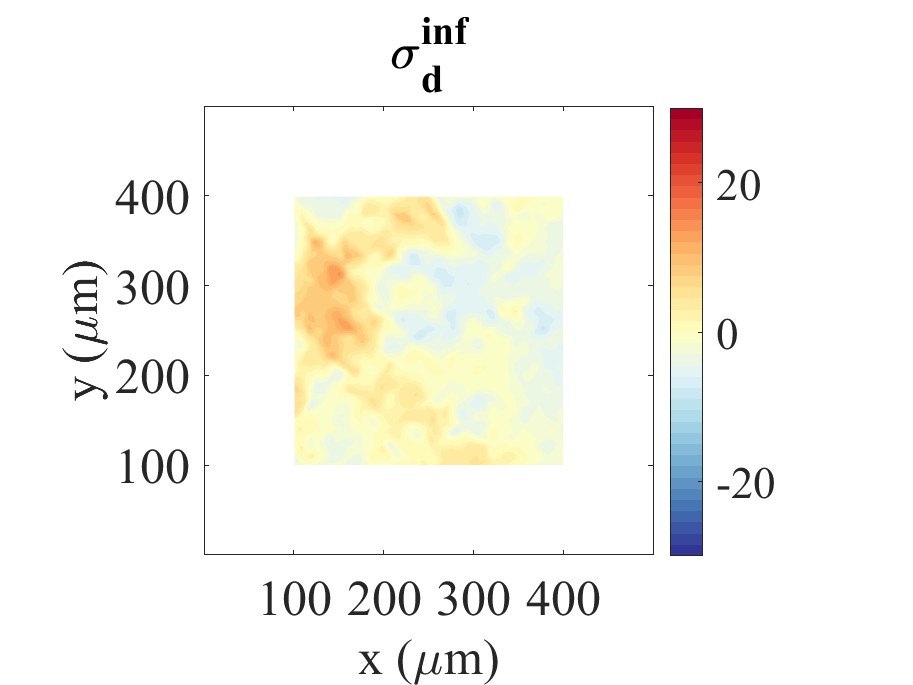}\\
  \textbf{d.}
  \includegraphics[width=0.27\linewidth]{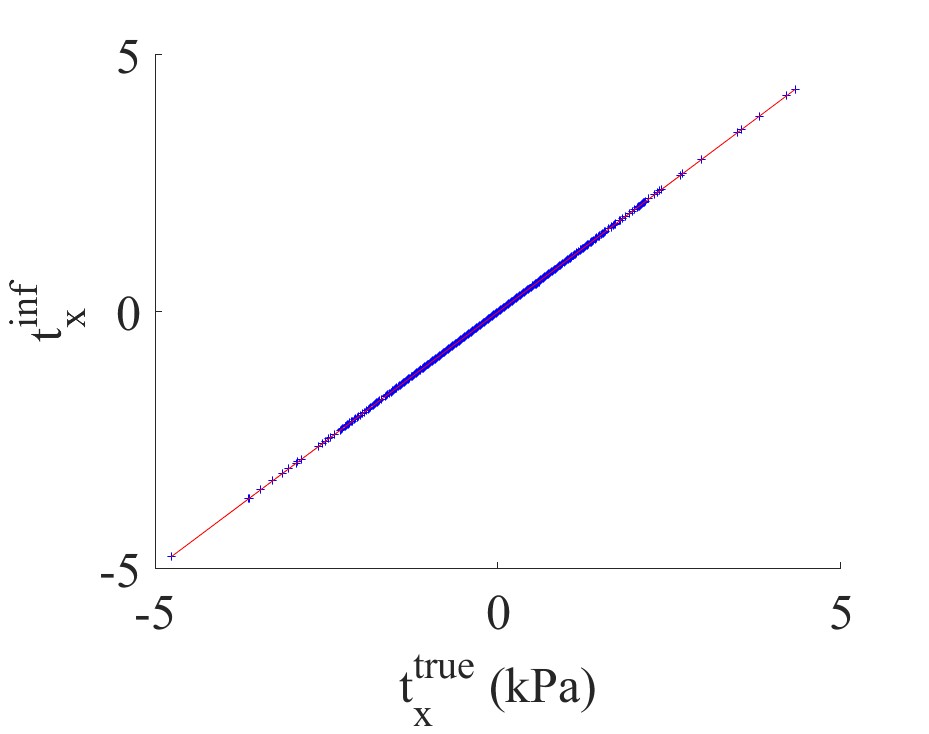}
  \textbf{e.}
  \includegraphics[width=0.27\linewidth]{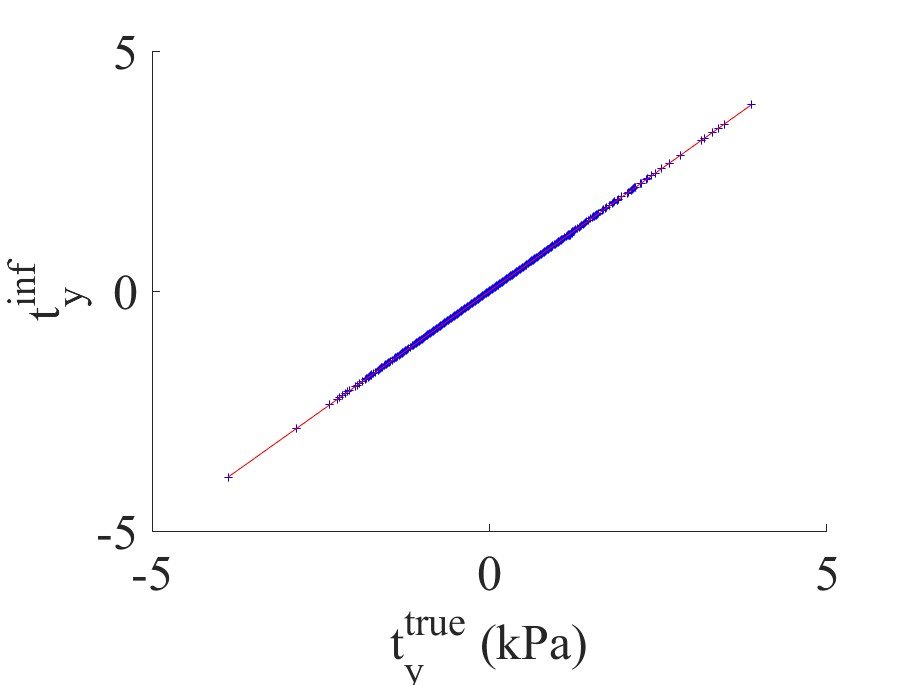}
    \textbf{f.}
  \includegraphics[width=0.27\linewidth]{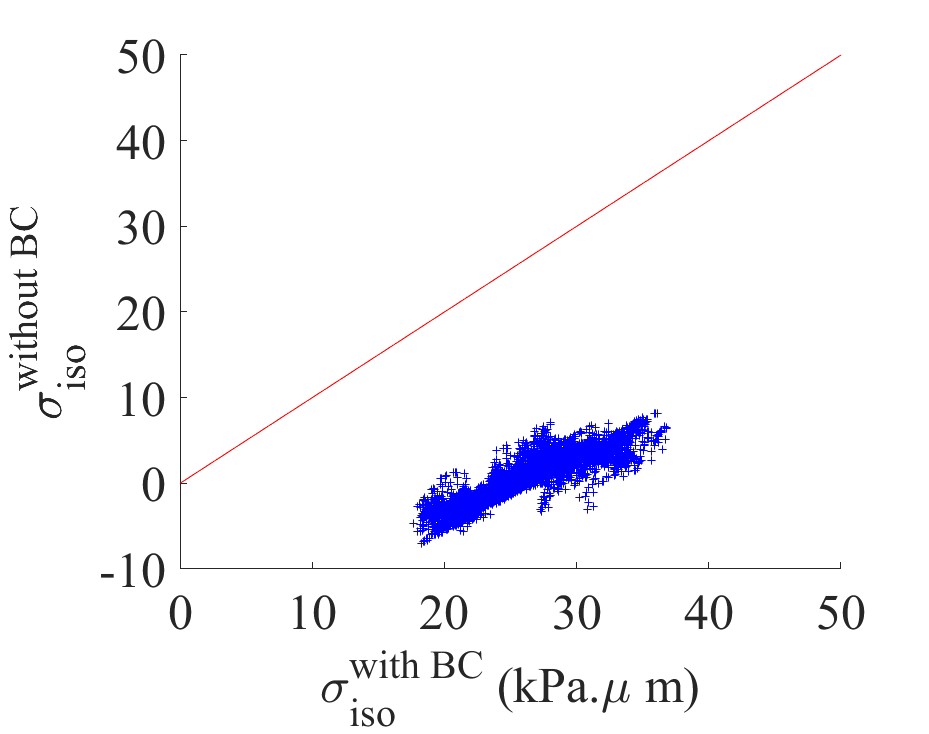}\\   
  \textbf{g.}
  \includegraphics[width=0.27\linewidth]{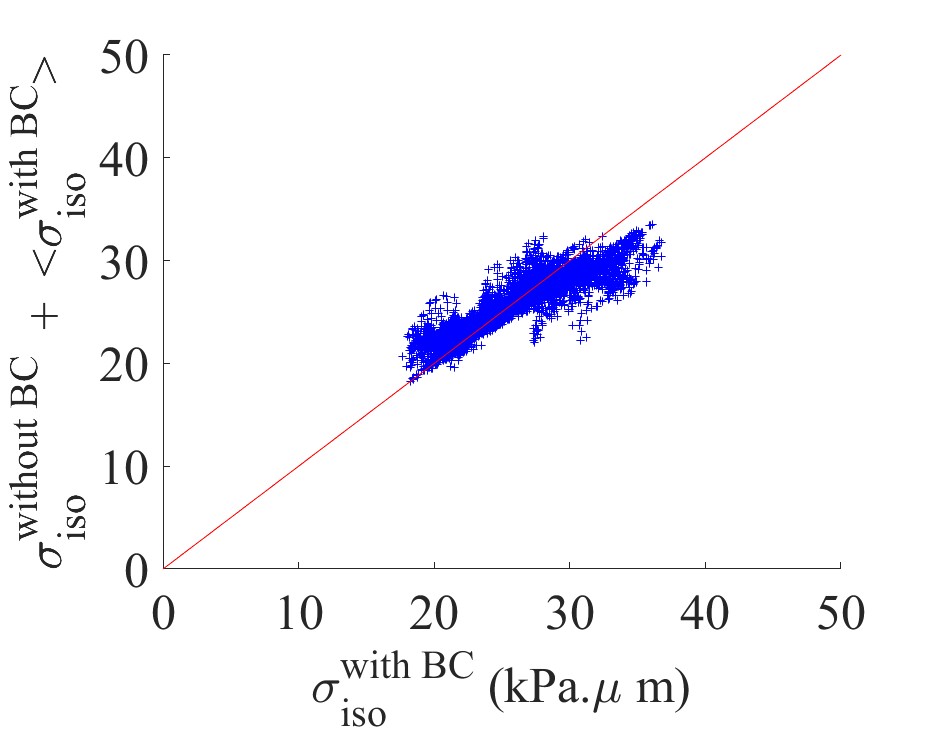}
  \textbf{h.}
  \includegraphics[width=0.27\linewidth]{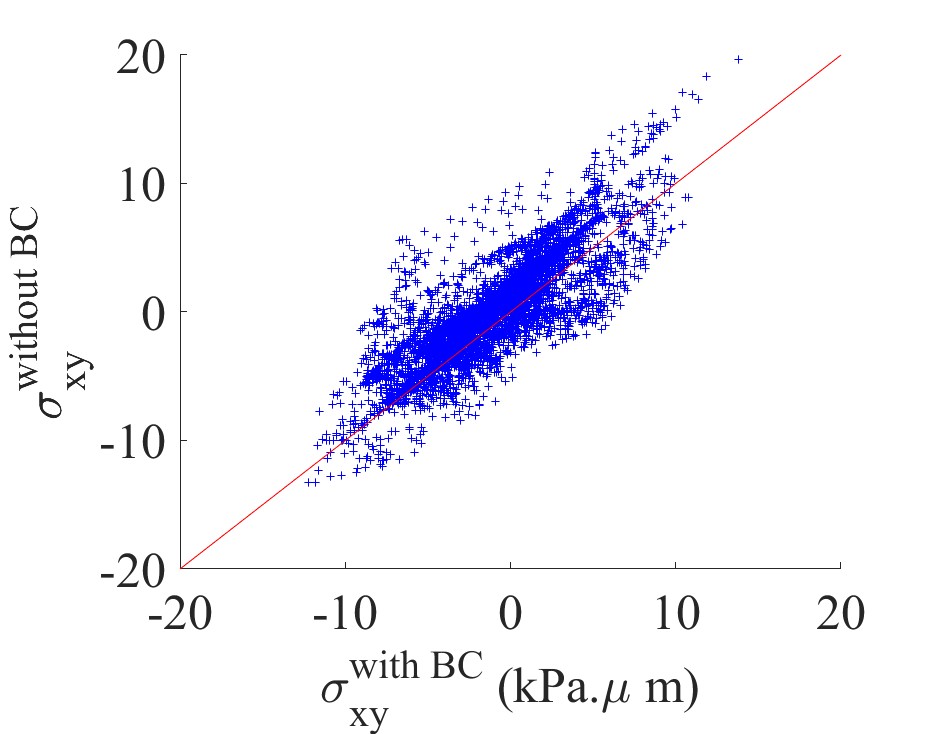}  
    \textbf{i.}
  \includegraphics[width=0.27\linewidth]{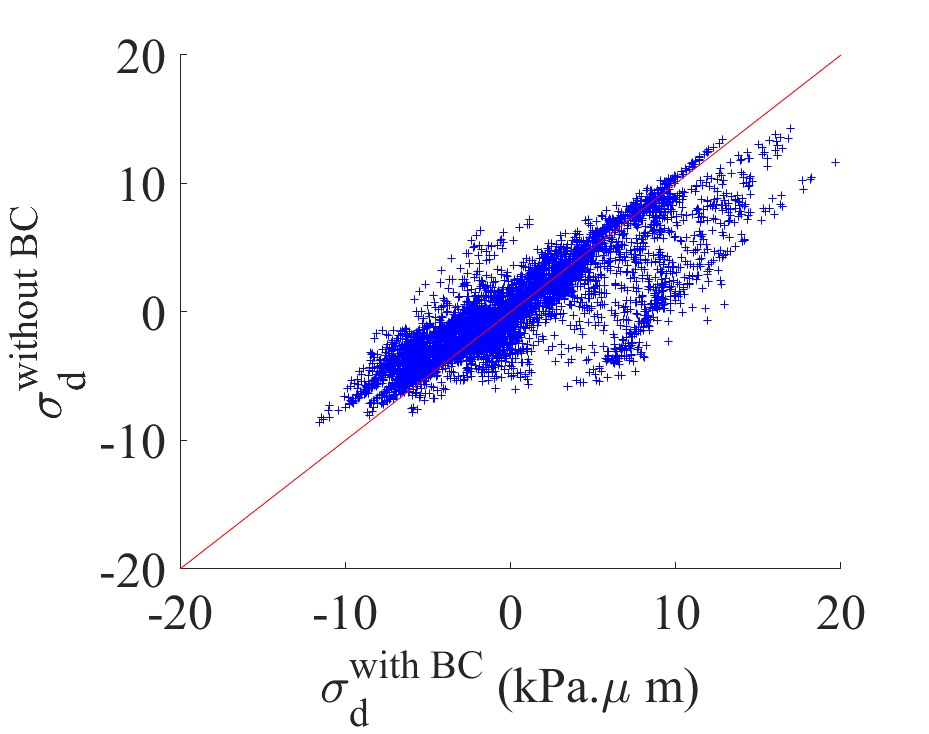}   
  \end{center}
\caption{
 \textbf{Subdomain:} the same traction force data is used
  as in Fig.~\ref{fig:confined:HaCaT}, but focusing on the central subdomain
  of lateral extension $300 \, \mu$m, and this time \textit{without} imposing
  boundary conditions.
  \textbf{a, b, c)} Color maps of the inferred
isotropic stress $\sigma_{\mathrm{iso}}^{\mathrm{inf}}$ and
deviatoric stress components $\sigma_{\mathrm{xy}}^{\mathrm{inf}}$,
$\sigma_{\mathrm{d}}^{\mathrm{inf}}$ (in kPa.$\mu$m). Since the color code
is the same as in  Fig.~\ref{fig:confined:HaCaT}, we see that,
contrary to deviatoric components, the absolute
value of the isotropic stress is not recovered.
\textbf{d, e)} Comparison of measured
($t^{\mathrm{true}}_x$, $t^{\mathrm{true}}_y)$
and inferred ($t^{\mathrm{inf}}_x$, $t^{\mathrm{inf}}_y$) values of
components of the traction force vectors (blue crosses),
computed from the inferred stress field
$\vec{t}^{\mathrm{inf}} = \mathrm{div} \, \sig^{\mathrm{inf}}$.
The bisectrix $y = x$ is plotted as a red line for comparison.
The coefficient of determination is $R^2_t = 1.0$.
This confirms that spatial patterns of stress are correctly
inferred without imposing boundary conditions.
\textbf{f, h, i)} Comparison of isotropic stress
 $\sigma_{\mathrm{iso}}^{\mathrm{inf}}$ and deviatoric
stress components $\sigma_{xy}^{\mathrm{inf}}$ and
$\sigma_{\mathrm{d}}^{\mathrm{inf}}$ inferred on the whole domain
with stress-free boundary conditions ($x$ axes, ``with BC'')
and inferred on the central sub domain without
imposing boundary conditions ($y$ axes, ``without BC'').
\textbf{g)} Data on the $y$ axis are shifted by
$< \sigma_{\mathrm{iso}}^{\mathrm{with BC}} >$
the mean isotropic stress computed over the subdomain
\emph{with} stress-free boundary conditions imposed
on the whole domain.
Coefficients of determination for the three panels g, h and i:
$R^2_{\mathrm{iso}}  = 0.77$,
$R^2_{\mathrm{xy}}  = 0.52$,
$R^2_{\mathrm{d}}  = 0.70$.
The spatial averages over the subdomain inferred without
boundary conditions are:
$<\sigma_{\mathrm{iso}}^{\mathrm{without BC}}> = 2 \pm  361 $ Pa, 
$<\sigma_{\mathrm{xy}}^{\mathrm{without BC}}> = 10 \pm   1047$ Pa and
$<\sigma_{\mathrm{d}}^{\mathrm{without BC}}> = 9 \pm   1204$ Pa,
in each case consistent with a zero mean value.
Uncertainty is computed as the standard deviation.
}
\label{fig:subdomain:HaCaT}
\end{figure}




\end{document}